\documentclass[aps,prl,twocolumn,showpacs,10pt,superscriptaddress,preprintnumbers,nofootinbib,showkeys]{revtex4-1}
\usepackage{epsfig,amssymb,amsmath,psfrag,epstopdf,color}
\pdfoutput=1

\usepackage{graphicx}
\usepackage[margin=7pt,justification=centerlast]{caption}
\usepackage{subcaption}
\usepackage{slashed} 
\usepackage{paralist}
\usepackage{hyperref}
\usepackage[utf8]{inputenc}
\usepackage{booktabs}

\def\beq{\begin{equation}}
\def\eeq{\end{equation}}
\def\bsp#1\esp{\begin{split}#1\end{split}}
\newcommand{\be}{\begin{equation}}
\newcommand{\ee}{\end{equation}}
\newcommand{\bea}{\begin{eqnarray}}
\newcommand{\eea}{\end{eqnarray}}

\def\to{\rightarrow}

\def\ksl{\not{\hbox{\kern-2.3pt $k$}}}

\def\e{\epsilon}

\def\Ord{{\cal O}}

\def\spa#1.#2{\left\langle#1\,#2\right\rangle}
\def\spb#1.#2{\left[#1\,#2\right]}
\def\lor#1.#2{\left(#1\,#2\right)}
\def\sand#1.#2.#3{%
\left\langle\smash{#1}{\vphantom1}^{-}\right|{#2}%
\left|\smash{#3}{\vphantom1}^{-}\right\rangle}

\newcommand{\nn}{\nonumber}

\newcommand{\cI}{{\cal I}}
\newcommand{\Gcusp}{\Gamma^{\text{cusp}}}
\newcommand{\Lp}{L_\perp}
\newcommand{\zero}{{(0)}}
\newcommand{\one}{{(1)}}
\newcommand{\two}{{(2)}}
\newcommand{\three}{{(3)}}

\begin{document}

\title{Quark Transverse Parton Distribution at the Next-to-Next-to-Next-to-Leading Order}
\author{Ming-xing Luo}
\email{mingxingluo@zju.edu.cn}
\affiliation{Zhejiang Institute of Modern Physics, Department of
  Physics, Zhejiang University, Hangzhou, 310027, China\vspace{0.5ex}}
\author{Tong-Zhi Yang}
\email{yangtz@zju.edu.cn}
\affiliation{Zhejiang Institute of Modern Physics, Department of
  Physics, Zhejiang University, Hangzhou, 310027, China\vspace{0.5ex}}
\author{Hua Xing Zhu}
\email{zhuhx@zju.edu.cn}
\affiliation{Zhejiang Institute of Modern Physics, Department of
  Physics, Zhejiang University, Hangzhou, 310027, China\vspace{0.5ex}}
\author{Yu Jiao Zhu}
\email{zhuyujiao@zju.edu.cn}
\affiliation{Zhejiang Institute of Modern Physics, Department of
  Physics, Zhejiang University, Hangzhou, 310027, China\vspace{0.5ex}}

\begin{abstract}
We report a calculation of the perturbative matching coefficients for the transverse-momentum-dependent parton distribution functions for quark at the next-to-next-to-next-to-leading order in QCD, which involves calculation of non-standard Feynman integrals with rapidity divergence. We introduce a set of generalized Integration-By-Parts equations, which allows an algorithmic evaluation of such integrals using the machinery of modern Feynman integral calculation. 
\end{abstract}


\maketitle

\section{Introduction}
\label{sec:introduction}

Transverse-Momentum-Dependent~(TMD) Parton Distribution Functions~(PDFs) generalize the concept of PDFs by allowing the dependence on the intrinsic transverse momentum of struck parton, beside the conventional longitudinal momentum fraction. By probing the the intrinsic confined transverse motion of parton inside the nucleon, TMDs allow reconstruction of the 3D picture of nucleon structure, as well as probing the parton orbital motion and the spin-orbit correlations between parton and nucleon. The possibility of precision measurement of TMDs in both $pp$ and $ep$ scattering also probes fundamental aspects of QCD such as gauge invariance and (non)universality. For these reasons, TMD PDFs play increasing central role in QCD theory and phenomenology in high energy collisions~\cite{Collins:2011zzd,Angeles-Martinez:2015sea}. In light of the unprecedented precision of the future Electron Ion Collider in TMD measurements~\cite{Accardi:2012qut}, perturbative knowledge about TMD PDFs at the highest available order is  desirable, and is the main subject of this Letter.

The theoretical framework of TMD PDFs is substantially more complicated than the conventional PDFs, which is partly reflected by the existence of several different formulations of TMD factorization in the literature~\cite{Dokshitzer:1978yd,Parisi:1979se,Collins:1981uw,Collins:1984kg,Ji:2004xq,Ji:2004wu,Bozzi:2005wk,Cherednikov:2007tw,Collins:2011zzd,Becher:2010tm,Echevarria:2012pw,Chiu:2012ir}, see Refs.~\cite{Collins:2012uy,Collins:2017oxh} for a discussion of the different formulations. In this Letter we adopt the rapidity renormalization group formalism of Refs.~\cite{Chiu:2011qc,Chiu:2012ir}, which is based on Soft-Collinear Effective Theory~(SCET)~\cite{Bauer:2000ew,Bauer:2000yr,Bauer:2001yt,Bauer:2002nz,Beneke:2002ph}. In this formalism, TMD factorization involves both TMD beam function ${\cal B}$ and TMD soft function ${\cal S}$. Schematically, for Drell-Yan production at low $q_\perp$,
\begin{equation}
  \label{eq:DYfact}
  \frac{d \sigma}{d^2q_\perp} \sim \sigma_0 H(Q) \int d^2 b_\perp e^{i b_\perp \cdot  q_\perp} {\cal B} \otimes {\cal B} {\cal S} \,,
\end{equation}
where $\sigma_0$ is the born cross section for Drell-Yan process, $H(Q)$ is the quark electromagnetic form factor, and the convolution is in the longitudinal momentum fraction.
The quark TMD beam function can be written as an operator matrix element in a hadron state with momenta $P$,
\begin{multline}
  \label{eq:TMDbeam}
  \mathcal{B}_{q/N}(z,b_\perp) = 
\int \frac{db_-}{4\pi} \, e^{-i z b^- P^{+}/2} 
\\
 \langle N(P) | \bar{\chi}_n(0,b^-,b_\perp) \frac{\slashed{\bar{n}}}{2} \chi_n(0) | N(P) \rangle \, ,
\end{multline}
where $\chi_n = W_n^\dagger \xi_n$ is the gauge invariant collinear quark field~\cite{Bauer:2001ct} in SCET, constructed from collinear quark field $\xi_n$ and path-ordered collinear Wilson line $W_n(x) = {\cal P} \exp \left(i g \int_{-\infty}^0 ds\, \bar{n} \cdot A_n (x + \bar{n} s) \right)$ in some null direction $n = (n^+, n^-, n_\perp) = (2, 0, 0_\perp)$ and $\bar n = (0, 2, 0_\perp)$. $b_\perp$ plays the role of Fourier conjugate of the transverse momentum of the struck quark. The TMD soft function is a vacuum matrix element of time-ordered and anti time-ordered soft Wilson lines,
\begin{equation}
\mathcal{S}(b_\perp) = \frac{1}{N_c} \mathrm{Tr} \langle 0 | {\cal T} \big[S^\dagger_{\bar{n}} \, S_n(0,0,b_\perp)\big] \, \overline{\cal T} \big[ S_n^\dagger \, S_{\bar{n}}(0) \big] | 0 \rangle \, ,
\label{eq:TMDsoft}
\end{equation}
where $S_n(x) = {\cal P} \exp \left( i g \int_{-\infty}^0 ds\, A_s(x + sn) \right)$.

Similar to PDFs, the TMD beam function in Eq.~\eqref{eq:TMDbeam} is intrinsically non-perturbative. At leading twist it admits a light-cone operator product expansion onto the PDFs,
\begin{equation}
  \mathcal{B}_{q/N}(z,b_\perp) = \sum_i  \int_z^1 \frac{d\xi}{\xi} \, \mathcal{I}_{qi}(\xi,b_\perp) \, \phi_{i/N}(z/\xi) + \mathcal{O}(|b_\perp^2|\Lambda^2_{\text{QCD}}) \, ,
\label{eq:TMDPDFOPE}
\end{equation}
where in principle the matching coefficients ${\cal I}_{qi}$ is perturbatively calculable. The main complication is that the operator matrix element in Eqs.~\eqref{eq:TMDbeam} and \eqref{eq:TMDsoft} are not well-defined even within dimensional regularization. They suffer from the so-called rapidity divergence, which originates from gluon emission at infinite rapidity in the backward direction to the struck parton. Several regulators proposed to cure the rapidity divergence in TMD exist in the literature~\cite{Collins:1984kg,Ji:2004wu,Collins:2011zzd,Becher:2010tm,Becher:2011dz,Chiu:2012ir,Chiu:2009yx,Echevarria:2015byo,Li:2016axz,Ebert:2018gsn}, which allow to calculate the matching coefficients in QCD to Next-to-Leading Order~(NLO)~\cite{Collins:1984kg,Ji:2004wu,Becher:2010tm,Collins:2011zzd,Aybat:2011zv,GarciaEchevarria:2011rb,Chiu:2012ir}, and relative recently to Next-to-Next-to-Leading Order~(NNLO)~\cite{Catani:2011kr,Catani:2012qa,Gehrmann:2012ze,Gehrmann:2014yya,Echevarria:2016scs,Luo:2019hmp,Luo:2019bmw,Gutierrez-Reyes:2019rug}. These results provide stringent test to the TMD factorization, and facilitate some of the most precise theoretical prediction at the LHC~\cite{Chen:2018pzu,Bizon:2018foh,Cieri:2018oms,Bizon:2019zgf,Bertone:2019nxa}. 

In this Letter we shall present for the first time a calculation of the matching coefficients for quark TMD beam function at the Next-to-Next-to-Next-to-Leading Order~(N$^3$LO) within the exponential regularization scheme~\cite{Li:2016axz} in a completely analytic form. We also provide  results for the regulator-independent TMD PDFs, by combining the beam function with the N$^3$LO TMD soft function~\cite{Li:2016ctv}.
Our calculation contains all the color structures and all the partonic channels, and represents a major step towards precision TMD physics.  To facilitate this calculation, we introduce a set of generalized Integration-By-Parts equations that allows an algorithmic evaluation of rapidity divergent integrals using the powerful machinery of modern Feynman diagram calculation, which shall be explained in the next section.

\section{The Method}
\label{sec:method}

Since the matching coefficients can be calculated in short-distance perturbation theory, we can use asymptotic quark and gluon state instead of the hadron state $N(P)$ in Eq.~\eqref{eq:TMDbeam}. Furthermore, we can insert a complete state of $| X \rangle \langle X|$, which also consist of asymptotic quark and gluon states, between $\bar{\chi}_n$ and $\chi_n$. This converts the calculation of the perturbative TMD beam function to the calculation of semi-inclusive partonic cross section,
\begin{equation}
  \label{eq:beamdecomp}
  {\cal B}_{q/i}(z,b_\perp) = \sum_{L=0} \sum_{n+m=L} {\cal B}_{q/i}^{(n,m)}(z,b_\perp) \,,
\end{equation}
where ${\cal B}_{q/i}^{(n,m)}(z,b_\perp)$ represents a $n$-loop, $m$-real emission contribution for the semi-inclusive sub-process $i \to q + X$ at $\Ord(\alpha_s^{n+m})$,
\begin{multline}
\label{Eq:MomentumDefinition1}
 \mathcal{B}^{(n,m)}_{q/i}(z, b_\perp) = \int \frac{d^{d-2} \widetilde K_\perp}{|\widetilde{K}_\perp^2|^{-\e}}  e^{ - i b_\perp \cdot \widetilde K_\perp}  \widetilde{\mathcal{B}}^{(n,m)}_{q/i} (z, \widetilde K_\perp) \,,
\end{multline}
with $d = 4  - 2 \e$ the space-time dimension. We have suppressed the renormalization scale $\mu$ an rapidity scale $\nu$ in the argument for simplicity. The function $\widetilde{\mathcal{B}}^{(n,m)}_{q/i} (z,\widetilde K_\perp)$ has simple power and logarithmic dependence on $\widetilde K_\perp$, which can be integrated easily~\cite{Luo:2019hmp,Luo:2019bmw}. The function $\widetilde{\mathcal{B}}^{(n,m)}_{q/i} (z,\widetilde K_\perp)$ can be calculated as
\begin{align}
\label{Eq:MomentumDefinitionM}
\widetilde{\mathcal{B}}&^{(n,m)}_{q/i}(z, \widetilde K_\perp) =  \Big[ \lim_{\tau \to 0}  \frac{2  \int d^d K}{V_{d-2}}  e^{ -b_0 \tau \frac{P \cdot K}{P^+} } \delta( K^+ - P^+ (1-z) )  \nn \\ &  \times  \delta( \widetilde K_\perp^2 - K_\perp^2 ) \mu^{2 \e (n+m)} \prod_{j=0}^n \frac{\int d^d l_j}{ (2 \pi)^d} \prod_{r=0}^m \frac{\int d^d k_r}{(2 \pi)^{d-1}}  \delta_+(k_r^2) \nn \\
&\times \delta^{(d)}( K - \sum_{r=1}^m k_r) |\overline{\mathrm{Sp}}_{q\leftarrow i}(P,\{l\},\{k\})|^2\Big]\bigg|_{\tau \to  1/\nu } \,,
\end{align}
where $ e^{ -b_0 \tau \frac{P \cdot K}{P^+} }$ is the exponential regulator~\cite{Li:2016axz} in the collinear sector~\cite{Luo:2019hmp}.
$|\overline{\mathrm{Sp}}_{q\leftarrow i}(P,\{l\},\{k\})|^2$ is the spin and color averaged squared splitting amplitudes for $i \to q + X$ with $n$ loops~(counting at the squared amplitude level) and $m$ real emission. $K^\mu$ is the total four momentum of $X$. $b_0 = 2 e^{-2 \gamma_E}$ is a conventional factor, and $V_d = 2 \pi^{d/2}/\Gamma(d/2)$ is the volume of $d$ sphere. It is easy to see from Eq.~\eqref{Eq:MomentumDefinitionM} that the dependence on $\widetilde K_\perp$ only enters through $\ln(|\widetilde{K}_\perp^2|/\mu^2)$ or $\ln(|\widetilde{K}_\perp^2|/\nu^2)$. Note that $K_\perp^2 = K^2 - K^- K^+ = K^2  - K^- P^+ (1-z)$. Using reverse unitarity~\cite{Anastasiou:2002yz}, the delta functions in Eq.~\eqref{Eq:MomentumDefinitionM} can be considered as ``cut'' propagators. Therefore, $\widetilde{\mathcal{B}}^{(n,m)}_{q/i} (z,\widetilde K_\perp)$ can be regarded as a $(n+m)$ loop Feynman integral in the algebraic sense, if not for the exponential regulator. The main new idea of this Letter is the introduction of a set of generalized Integration-By-Parts~(IBP) equation~\cite{Chetyrkin:1981qh,Laporta:2001dd}, which take into account also the exponential regularization factor in Eq.~\eqref{Eq:MomentumDefinitionM}.

Specifically, the generalized IBP equations can be written as
\begin{align}
\label{EqIBP}
0 &\, = \int d^d q \, \frac{ \partial }{  \partial {q^\mu} } \bigg[   e^{ - b_0 \tau \frac{P \cdot K}{P^+} }   F(\{\tilde l\}) \bigg]   \nonumber \\
&\, = \left\{
\begin{array}{lr}
\int d^d q \, e^{-b_0 \tau \frac{P\cdot K}{P^+} }    \left[ -b_0 \tau \frac{P_\mu}{P^+} + \frac{ \partial }{  \partial {q^\mu}} \right] F(\{ \tilde l \}) \,, & q = K \,,
\\
\int d^d q \, e^{-b_0 \tau \frac{P\cdot K}{P^+} } \frac{ \partial }{  \partial {q^\mu}}  F(\{\tilde l \}) \,,
& q \neq K \,,
\end{array}
\right.
\end{align}
where $F(\{\tilde l\})$ is some Feynman integrand, and $\{\tilde l\} = \{ l_1, l_2, \dots, l_n, k_1, k_2, \ldots, k_{m-1}, K\}$ are the set of integration momentum. Note that we have used the delta function in Eq.~\eqref{Eq:MomentumDefinitionM} to kill one of the integration momentum, $k_m$. We generate the required IBP equations in this way using in-house \textsc{Mathematica} code and \textsc{LiteRed}~\cite{Lee:2012cn}. The obtained IBP equations and the integrals are exported to \textsc{Fire}~\cite{Smirnov:2014hma}, where all the integrals can be reduced to a set of master integrals~(MIs). 

The main task in this Letter is to calculate Eq.~\eqref{Eq:MomentumDefinitionM} at three loops, i.e. $m+n=3$. The contribution at this order can be classified by the number of loop and the number of real emission: triple-real~(RRR), double-real single-virtual~(VRR), double-virtual single-real~(VVR) and the square of real-virtual~((VR)$^2$). For VVR part, we use the results in Ref.~\cite{Duhr:2014nda}, analytically continue them to spacelike case and directly integrate over the single particle phase space. The $(\text{VR})^2$ part is relatively simple. we use the method presented in Ref.~\cite{Luo:2019hmp} to handle this part.
\begin{figure}[ht!]
  \centering
  \includegraphics[width=0.5\textwidth]{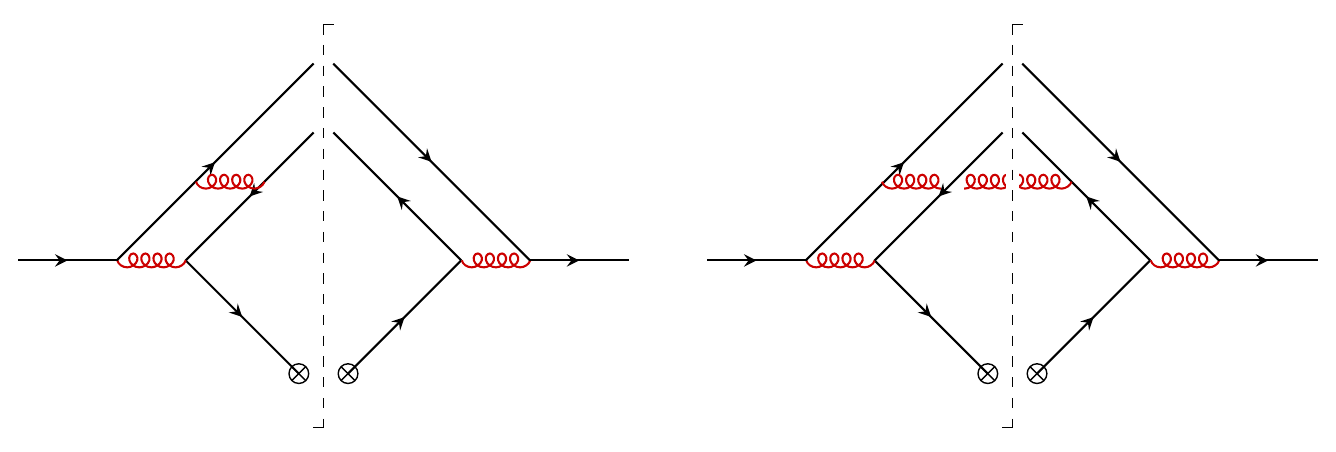}
  \caption{Representative cut diagram for the color structure $d^{ABC} d_{ABC} = (N_c^2 - 1)(N_c^2 - 4)/N_c$ from VRR~(left) and RRR~(right).}
\label{fig:d33}
\end{figure}

The main challenge of this calculation is the VRR and RRR parts. Representative cut Feynman diagrams are shown in Fig.~\ref{fig:d33}. We generate the integrand for VRR and RRR in \textsc{Qgraf}~\cite{Nogueira:1991ex}, with color/Dirac algebra and integrand manipulation aided by \textsc{Form}~\cite{Vermaseren:2000nd}, \textsc{FeynCalc}~\cite{Shtabovenko:2016sxi}, and \textsc{Apart}~\cite{Feng:2012iq}. Using the above mentioned new idea, we then pass the integrand and the generalized IBP equations to \textsc{Fire} for integral reduction. After exploring symmetry between different integral families, we find about $900$ MIs for VRR and RRR in total. To solve these MIs, we use the method of Differential Equations~(DEs)~\cite{Bern:1993kr,Gehrmann:1999as,Henn:2013pwa}.

Note that up to this stage we have kept the rapidity regulator $\tau$ finite. The resulting MIs are functions of $z$ and $\tau$, and DEs in $z$ and $\tau$ can be constructed in the standard way. The next step is to take the $\tau \to 0$ limit, as in Eq.~\eqref{Eq:MomentumDefinitionM}. We do so by expanding the DEs and the MIs around $\tau=0$ consistently, but keeping the full functional dependence on $z$. We write a MI $f_i$ as a double series in $\tau$ and $\ln\tau$,
\begin{equation}
  f_i(z, \tau, \e) \stackrel{\tau \to 0}{=} \sum_j \sum_{k=0} f_i^{(j,k)}(z,\e) \tau^j \ln^k \tau \,,
\label{eq:doubleseries}
\end{equation}
Note that for individual diagram in Feynman gauge, the result is absent of power-like singularity in $\tau$. However, application of the generalized IBPs leads to power divergence  for individual MI. We found by observation that only $j \geq -1$ is needed in Eq.~\eqref{eq:doubleseries}, by checking explicitly that terms with $j=-2$ and $j=-3$ vanish.   
In principle we can also have terms like $\tau^\e$ in the expansion. However such terms can be discarded, because we can always analytically continue to a region where $\e >0$, and take $\tau \to 0$, and then take $\e \to 0$. In practice we have checked this by enlarging the ansatz in Eq.~\eqref{eq:doubleseries} by multiplying with $\tau^{\pm n \epsilon}$ for $n = 0,\ldots, 4$. We found that when substituting the MIs into the integrand, the coefficients for $n = 1, \ldots, 4$ always vanish for each diagram. Therefore these terms are spurious at the diagram level, and we discard them in Eq.~\eqref{eq:doubleseries} to simplify our calculation. 
We then substitute the double series in Eq.~\eqref{eq:doubleseries} into the system of DEs, which are expanded in $\tau$, but with full $z$ dependence. By equating the $\tau^j \ln^k \tau$ coefficient in the DEs, we obtain a closed system of DEs in $z$ for $f_i^{(j,k)}$. By considering the double series expansion, the number of MIs are reduced to about $500$ in total for VRR and RRR.

The system of DEs in $z$ can now be solved by the standard approach. They are most conveniently solved by converting into the canonical form~\cite{Henn:2013pwa} by proper choice of MIs~\cite{Lee:2014ioa,Meyer:2016slj,Gituliar:2017vzm}. For individual VRR or RRR, the alphabet consist of five letters,
\begin{equation}
  \label{eq:alphabet}
 \{ z\,, \quad 1-z\,, \quad 1+z \,, \quad 2 - z \,, \quad z^2 - z + 1 \}\,.
\end{equation}
The DEs can be solved order-by-order in $\e$ easily in terms of Goncharov Polylogarithms. Remarkably, after summing the VRR and RRR contributions, and substituting in the boundary constants determined below, the latter two letters drop out from the sum. Therefore, Harmonic Polylogarithms~(HPLs)~\cite{Remiddi:1999ew} are sufficient to describe the final results.

The remaining task is to determine the boundary constants for the DEs. To this end we consider the threshold limit of the MIs, $z \to 1$. Following Ref.~\cite{Luo:2019hmp}, we define the so-called fully-differential beam function~\cite{Mantry:2009qz},
\begin{multline}
\label{Eq:fulldiff}
\widehat{\mathcal{B}}^{(n,m)}_{q/i}(z, K^+, K^-, K_\perp) = \prod_{j=0}^n \frac{\int d^d l_j}{ (2 \pi)^d} \prod_{r=0}^m \frac{\int d^d k_r}{(2 \pi)^{d-1}}  \delta_+(k_r^2) \\
\cdot \delta^{(d)}( K - \sum_{r=1}^m k_r)  \mu^{2 \e (n+m)}  |\overline{\mathrm{Sp}}_{q\leftarrow i}(P,\{l\},\{k\})|^2 \,.
\end{multline}
The original $\widetilde{\cal B}$ is simply obtained by integrating the $K^-$ component,
\begin{multline}
  \label{eq:relation}
  \widetilde{\cal B}^{(n,m)}_{q/i} (z, \widetilde{K}_\perp)
=
\Big[
\lim_{\tau \to 0} \frac{|\widetilde{K}_\perp^2|^{-\e}}{V_{d-2}}  \int_0^\infty\! dK^- 
e^{- b_0 \tau \frac{P \cdot K}{P^+}} 
\\
\cdot \widehat{\mathcal{B}}^{(n,m)}_{q/i}(z, P^+(1-z), K^-, \widetilde K_\perp)
\Big]\Bigg|_{\tau \to 1/\nu} \,.
\end{multline}
The advantage of having the fully-differential function is that now the threshold limit can be taken at the integrand level. For that purpose, we adopt the strategy of expansion by region~\cite{Beneke:1997zp}. For RRR, $z \to 1$ force the leading region to be $K^\mu \to 0$. VRR is more complicated. Besides $K^\mu \to 0$, we also need to consider the scaling in the loop momentum, which can be either soft or collinear~\cite{Anastasiou:2014lda}. Ultimately, expansion by region relates all the boundary constants to those computed for soft-virtual corrections to Higgs production at N$^3$LO. We have performed an independent calculation for these constants, and found agreement with those in the literature~\cite{Anastasiou:2013srw,Anastasiou:2014vaa,Li:2014afw,Li:2014bfa,Anastasiou:2014lda}.

\section{The Results}
\label{sec:results}

We are ready to combine all the ingredients and present the final results. The bare TMD beam function computed in the last section contains poles in $\e$ up to $1/\e^6$. Using the known renormalization constant and the PDF counter terms~(which contain the famous three-loop splitting kernel~\cite{Moch:2004pa,Vogt:2004mw}), we find that all the poles cancel, and finite matching coefficients can be extracted! This provides a stringent check to our calculation. For the convenience of reader, the relevant renormalization counter terms are collected in the appendix. We refer to Ref.~\cite{Luo:2019hmp,Luo:2019bmw} for the detailed renormalization procedure. 

The finite matching coefficient obey the renormalization group equation,
\begin{align}
\frac{d}{d \ln\mu} \mathcal{I}_{qi}&(z,b_\perp, P^+, \mu,\nu) = 2 \bigg[ \Gcusp(\alpha_s (\mu)) \ln\frac{\nu}{z P^{+}}  \nonumber
\\
& + \gamma^B(\alpha_s (\mu)) \bigg] \mathcal{I}_{qi}(z,b_\perp,P^+,\mu,\nu)
\nonumber 
\\
&- 2 \sum_j \mathcal{I}_{qj}(z,b_\perp,P^+,\mu,\nu) \otimes P_{ji}(z,\alpha_s (\mu) ) \,,
\label{eq:Imu}
\end{align}
where the anomalous dimension can be found in the appendix. It also obeys the rapidity evolution equation~\cite{Chiu:2012ir},
\begin{align}
\frac{d}{d \ln\nu} \mathcal{I}_{qi}&(z,b_\perp,P^+,\mu,\nu) = -2 \bigg[ \int_{\mu}^{b_0/|b_\perp|} \frac{d \bar{\mu}}{\bar{\mu}} \Gcusp(\alpha_s(\bar{\mu}))  \nonumber 
\\
&  + \gamma^R(\alpha_s(b_0/|b_\perp|)) \bigg] \mathcal{I}_{qi}(z,b_\perp,P^+,\mu,\nu) \, .
\label{eq:Inu}
\end{align}
The new results of this Letter are the initial conditions~(coefficient functions) for these equations, $I_{ij}(z) = {\cal I}_{ij}(z, b_\perp, P^+, \mu = b_0/|b_\perp|, \nu = z P^+)$. As mentioned before, the three-loop results can be written solely in terms of HPLs, which we believe is quite remarkable. The full results are too lengthy to fit in a Letter, but can be found in the appendix.~\ref{sec:beam}, along with numerical fits of to accuracy by elementary functions in appendix.~\ref{sec:numer}. We also provide computer readable files for them along with the arXiv submission. The TMD beam function depends on the rapidity regulator being used. Rapidity-regulator-independent TMD PDFs can be simply obtained using~\cite{Luo:2019hmp,Luo:2019bmw}
\begin{equation}
  \label{eq:TMDPDF}
  f_{T,ij}(z, b_\perp, P^+, \mu) = {\cal I}_{ij}(z, b_\perp, P^+, \mu, \nu) \sqrt{{\cal S}(b_\perp, \mu, \nu)} \,.
\end{equation}
The TMD soft function ${\cal S}(b_\perp, \mu, \nu)$ is also known to N$^3$LO~\cite{Li:2016ctv}. Therefore, quark TMD PDFs can also be extracted to this order. We provide computer readable file for them in the ancillary files. 

It is interesting to consider the asymptotic limit of the coefficient functions. In the threshold limit, we find the non-vanishing component to be
\begin{align}
I^{(2)}_{qq} =&  \frac{1}{(1-z)_+} \bigg[ \left(28 \zeta _3-\frac{808}{27}\right) C_A C_F+\frac{224}{27} C_F N_f T_F \bigg] \,,
\nn \\
I^{(3)}_{qq} =&  \frac{1}{(1-z)_+} \bigg[  \left(-\frac{128 \zeta
   _3}{9}-\frac{7424}{729}\right) C_F N_f^2 T_F^2   \nonumber 
   \\
   &+\biggl(-\frac{1648 \zeta _2}{81}-\frac{1808 \zeta _3}{27}+\frac{40 \zeta
   _4}{3} \nonumber 
   \\
   &+\frac{125252}{729}\biggl) C_A C_F N_f T_F+\biggl(-\frac{176}{3} \zeta _3 \zeta
   _2+\frac{6392 \zeta _2}{81} \nonumber 
   \\
   &+\frac{12328 \zeta _3}{27}+\frac{154 \zeta _4}{3}-192 \zeta
   _5-\frac{297029}{729}\biggl) C_A^2 C_F \nonumber 
   \\
   &+\left(-\frac{608 \zeta _3}{9}-32 \zeta
   _4+\frac{3422}{27}\right) C_F^2 N_f T_F  \bigg] \,,
\end{align}
where $I_{ij}^{(n)}$ is the expansion coefficient of $(\alpha_s/(4 \pi))^n$.
One can also identify the coefficient with the rapidity anomalous dimension, 
\begin{equation}
  \label{eq:largez}
  I_{qq}^{(2)}(z) = \frac{2 \gamma^R_1}{(1-z)_+} \,, \quad 
 I_{qq}^{(3)}(z) = \frac{2 \gamma^R_2}{(1-z)_+}  \,,
\end{equation}
where $\gamma^R_{1(2)}$ are the two(three)-loop rapidity anomalous dimension~\cite{Li:2016ctv,Vladimirov:2016dll}. 
This result was conjectured in Ref.~\cite{Echevarria:2016scs}, and was understood using joint $q_T$ and threshold resummation~\cite{Lustermans:2016nvk}, see also Ref.~\cite{Billis:2019vxg}. This provides another check to our calculation.

We can also consider the high energy limit $z \to 0$, which is closely related to small-$x$ physics. The leading terms are
\begin{align}
I^{(2)}_{qg} =& \frac{ 2 C_A T_F }{z} \left(  \frac{172}{27}-\frac{8 \zeta _2}{3} \right)  \,, \nonumber  
\\  
I^{(2)}_{qq} =& I^{(2)}_{qq'} = I^{(2)}_{q\bar{q}} = I^{(2)}_{q\bar{q}'} = \frac{2 C_F T_F }{z} \left(  \frac{172}{27}-\frac{8 \zeta _2}{3} \right) \,,  
\nn
\\
I^{(3)}_{qg} =& \frac{2 T_F}{z} \bigg[
   \left(\frac{208 \zeta _2}{9}+\frac{32 \zeta
   _3}{3}-\frac{17152}{243}\right) C_A^2 \ln z \nonumber 
   \\
& + \left(\frac{160 \zeta _2}{27}-\frac{32 \zeta _3}{9}-\frac{3164}{729}\right) C_A N_f
   T_F  \nonumber 
   \\
   &+\left(-16 \zeta _2+\frac{512 \zeta _3}{9}+\frac{32 \zeta
   _4}{3}-\frac{269}{9}\right) C_A C_F \nonumber 
   \\
   &+\left(\frac{12536 \zeta
   _2}{81}+\frac{1096 \zeta _3}{9}+\frac{920 \zeta _4}{9}-\frac{470494}{729}\right)
   C_A^2 \nonumber 
   \\
   &+\left(-\frac{512 \zeta _2}{27}-\frac{64 \zeta _3}{9}+\frac{40184}{729}\right) C_F
   N_f T_F \bigg] \label{eq:smallx1} \,,   
\end{align}
\begin{align}
I^{(3)}_{qq} &\, = I^{(3)}_{qq'} = I^{(3)}_{q\bar{q}} = I^{(3)}_{q\bar{q}'} \nonumber
\\
&\, = \frac{2 T_F}{z} \bigg[  \left(\frac{208 \zeta _2}{9}+\frac{32 \zeta _3}{3}-\frac{17152}{243}\right) C_A C_F \ln z \nonumber 
   \\
   &+\left(\frac{12008 \zeta _2}{81}+120 \zeta _3+\frac{920 \zeta
   _4}{9}-\frac{456266}{729}\right) C_A C_F\nonumber 
   \\
   &+\left(-\frac{32 \zeta _2}{9}-\frac{64 \zeta
   _3}{9}+\frac{16928}{729}\right) C_F N_f T_F \nonumber 
   \\
   & 
   + \left(-16 \zeta _2 +\frac{512}{9} \zeta _3 +\frac{32}{3} \zeta _4 -\frac{269}{9} \right) C_F^2 \bigg] \label{eq:smallx2}\,. 
\end{align}
There has been recent progress in understanding TMD PDFs at small $x$~\cite{Balitsky:2015qba,Marzani:2015oyb,Balitsky:2016dgz,Xiao:2017yya}~\footnote{Using the formalism of Ref.~\cite{Marzani:2015oyb}, and the anomalous dimensions in \cite{Jaroszewicz:1982gr,Catani:1994sq}, the leading logarithms for $I_{qg}^{(3)}/(2 T_F C_A^2)$ and $I_{qq}^{(3)}/(2 T_F C_A C_F)$ are predicted to be
$(208 \zeta_2/9 + 32 \zeta_3/9 - 17152/243) \ln z/z$, which differs from the analytic results in Eq.~\eqref{eq:smallx1} and \eqref{eq:smallx2} for the coefficient of $\zeta_3$. It would be interesting to understand the origin of this discrepancy.
}. Our explicit results provide useful data through Next-to-Leading Logarithmic accuracy, which can foster further progress.

\begin{figure}[ht!]
  \centering
  \includegraphics[width=0.45\textwidth]{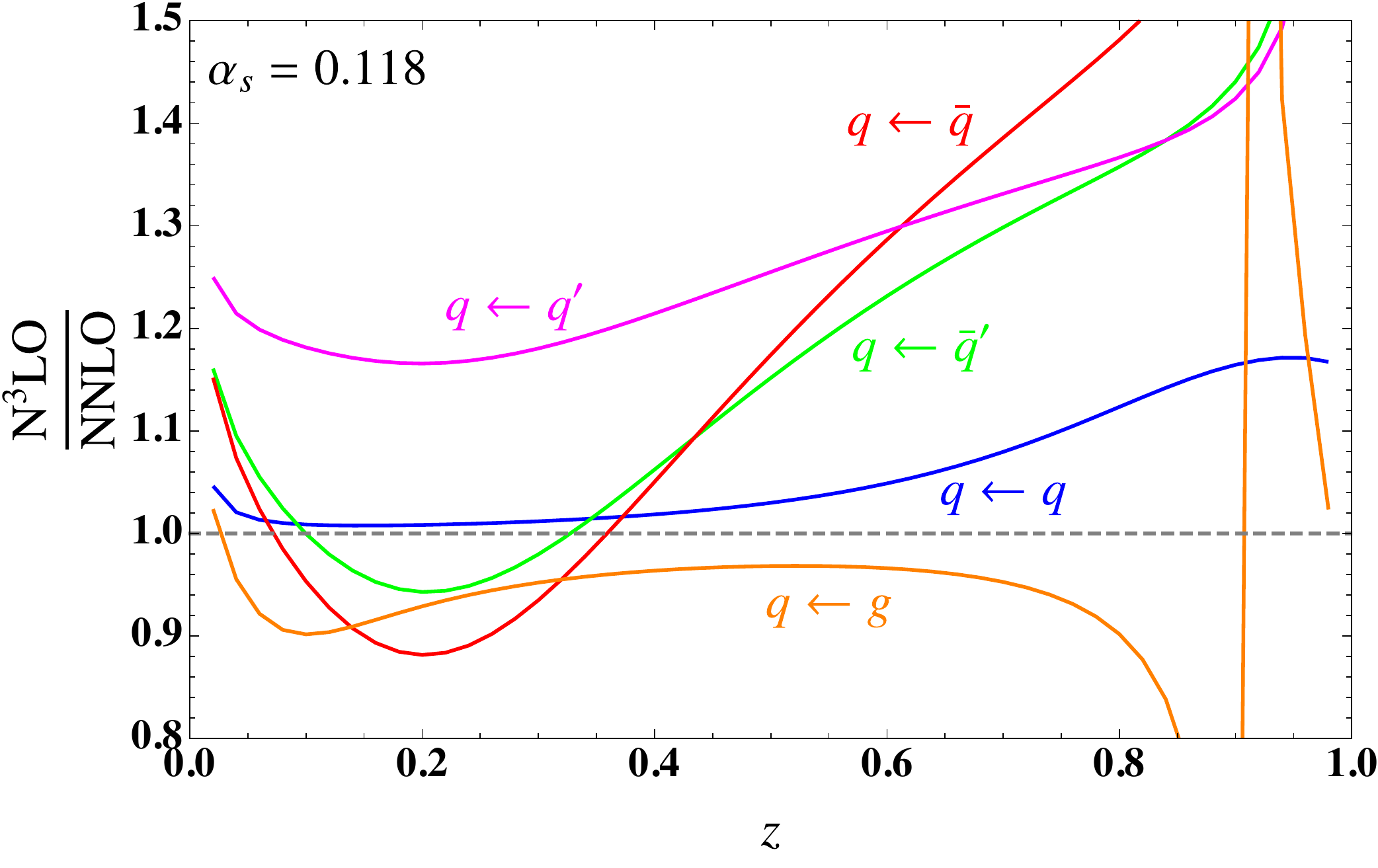}
  \caption{K factor at N$^3$LO for the coefficient functions. Numerical evaluation of HPLs are made with the \texttt{HPL} package~\cite{Maitre:2005uu}.}
  \label{fig:kfactor}
\end{figure}
\begin{figure}[ht!]
  \centering
  \includegraphics[width=0.45\textwidth]{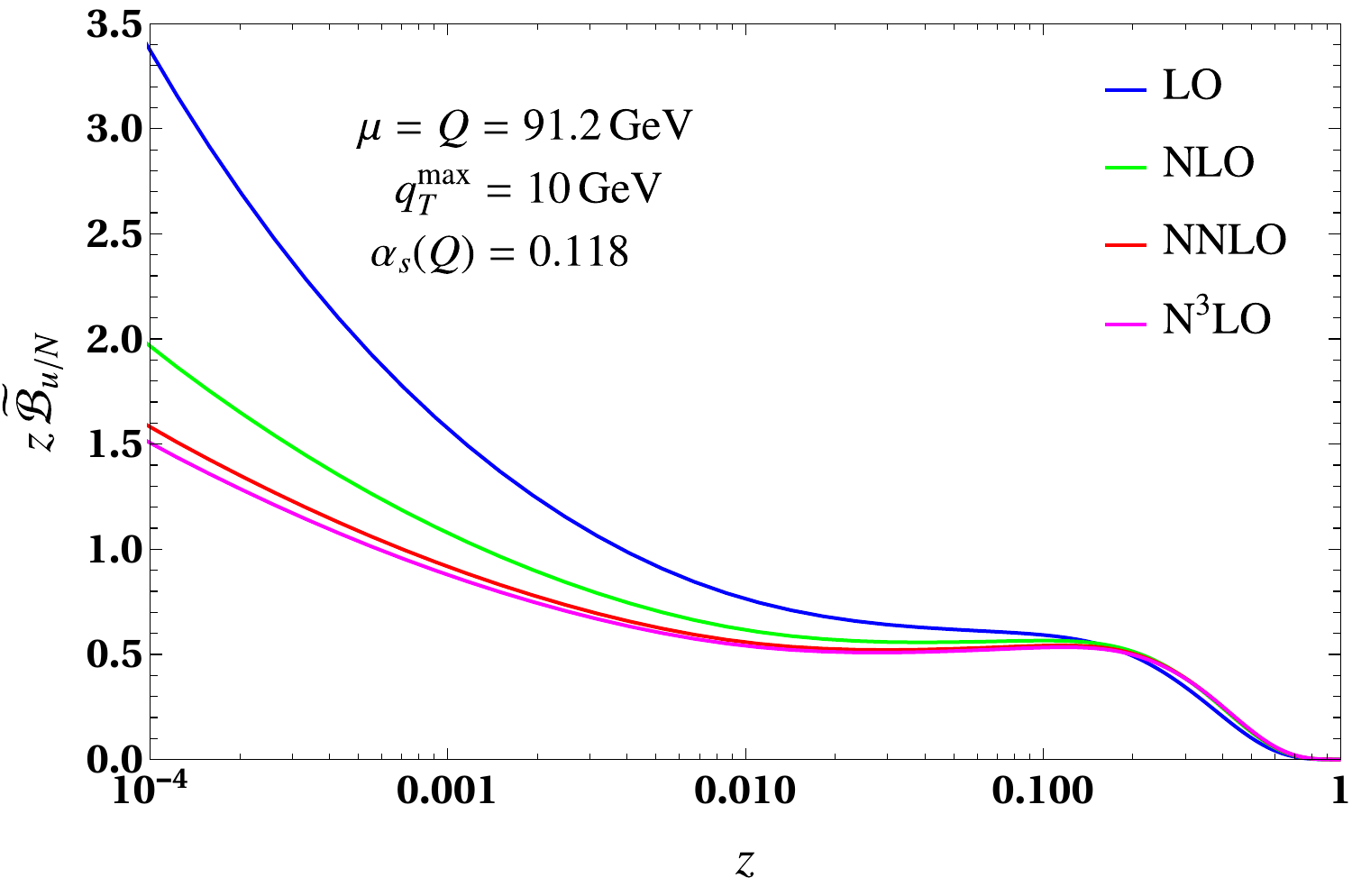}
  \caption{Integrated TMD PDFs at various perturbative order. We use  \texttt{NNPDF30\_nnlo\_as\_0118}~\cite{Ball:2014uwa} throughout the numerical calculation.}
  \label{fig:beam}
\end{figure}

To estimate the size of the three-loop corrections, we plot the ratio of N$^3$LO and NNLO coefficient functions for $0 < z <1$ in Fig.~\ref{fig:kfactor},
\begin{equation}
  \label{eq:Kfact}
  \frac{\text{N$^3$LO}}{\text{NNLO}} = 1 + \frac{ \left(\frac{\alpha_s}{4 \pi}\right)^3 I_{ij}^{(3)}(z)}{\frac{\alpha_s}{4 \pi} I_{ij}^{(1)}(z) + \left( \frac{\alpha_s}{4 \pi} \right)^2 I_{ij}^{(2)}(z) } \,.
\end{equation}
It can be seen that the three-loop corrections are non-negligible and have non-trivial shape dependence. Note that the coefficient functions are not ordinary functions of $z$, but distributions. To see the perturbative convergence of the TMD PDFs, we define an integrated version of it,
\begin{equation}
  \label{eq:integrated}
    B_{q/N}(z,q_T^{\rm max}) =\sum_i \int_0^{q_T^{\rm max}} \!\!\!\!\!\!\! dq_T \int_z^1 \frac{d\xi}{\xi} \, f_{T,qi}(\xi,q_T) \, \phi_{i/N}(z/\xi)  \, ,
\end{equation}
where $f_{T,qi}(\xi,q_T)$ is the momentum-space version of the TMD coefficients defined in Eq.~\eqref{eq:TMDPDF}, and $q_T^{\rm max}$ is a UV cutoff, below which the TMD approximation can be justified. In Fig.~\ref{fig:beam} we depict $z B_{u/N}$ with $q_T^{\rm max} = 10$ GeV at various perturbative order, where $N$ is proton. It can be seen that for moderate and small $z$, NLO and NNLO corrections are large and there no obvious perturbative convergence. On the other hand, after inclusion of the N$^3$LO corrections, the perturbative series is stabilized even for $z$ as small as $10^{-4}$. This gives us confidence that perturbative uncertainties are under good control at this order. 
%

\section{Discussion}
\label{sec:discussion}

We have calculated for the first time the quark TMD beam function and TMD PDFs at the N$^3$LO in QCD. The key new idea is the generalization of the IBP equations to Feynman integrals with the exponential regulator for rapidity divergence. We hope that the N$^3$LO results can improve our understanding about TMD factorization. There are several interesting directions to pursue in the future.

An obvious next thing to calculate is the gluon TMD at N$^3$LO. Past experience~\cite{Luo:2019bmw} shows that gluon TMD share the same set of MIs with the quark TMD, which significantly simplify the calculation. It would also be interesting to calculate TMD fragmentation function at N$^3$LO. At NNLO the timelike TMD can be obtained from spacelike ones by analytic continuation~\cite{Echevarria:2016scs,Luo:2019hmp,Luo:2019bmw}. However it is known from splitting function calculation that naive analytic continuation must be supplemented with physical constraint from reciprocity consideration~\cite{Dokshitzer:2005bf}, and still small uncertainty remains~\cite{Mitov:2006ic,Moch:2007tx,Almasy:2011eq}. It would be interesting to see if a direct calculation of timelike TMD can shed light on this problem. Our results represent the last missing ingredient for generalizing $q_T$ subtraction~\cite{Catani:2007vq} to N$^3$LO, see also Refs.~\cite{Cieri:2018oms,Billis:2019vxg}. We note that there has also been progress in calculating beam function for beam thrust~\cite{Stewart:2009yx}, where leading color contribution at N$^3$LO is available very recently~\cite{Behring:2019quf}. However, to achieve N$^3$LO accuracy based on N-jettiness subtraction~\cite{Boughezal:2015dva,Gaunt:2015pea}, N$^3$LO (beam)~thrust soft function is still missing~\cite{Kelley:2011ng,Monni:2011gb}.  Once fully differential N$^3$LO prediction with $q_T$ subtraction becomes available, it would also be important to consider the perturbative power corrections, where rapidity logarithms beyond leading power need to be understood better~\cite{Ebert:2018gsn,Moult:2019vou}, see also \cite{Cieri:2019tfv,Buonocore:2019puv}. It would also be interesting to study if the idea of generalized IBPs introduced in this Letter, perhaps with some modification, is helpful in problems where non-standard Feynman propagator are encountered, such as two-loop $q_T$ soft function for top-quark pair production~\cite{Angeles-Martinez:2018mqh,Catani:2019iny}, or thrust and hemisphere soft function~\cite{Kelley:2011ng,Monni:2011gb,Hornig:2011iu} with a step function in the integrand. Integrals with the latter form is particularly interesting since they appear frequently in precision jet substructure~\cite{Larkoski:2017jix}. Finally, given the success of the exponential regulator in perturbative calculation, it would be interesting to see if it can be applied to TMD on lattice, where rapid progress are being made recently~\cite{Ebert:2018gzl,Ebert:2019okf,Ebert:2019tvc,Ji:2019sxk,Ji:2019ewn}.

\section{Acknowledgements}
We thank V. Shtabovenko for valuable discussion, and C. Duhr, T. Gehrmann, and M. Jaquier for providing us with the results of Ref.~\cite{Duhr:2014nda} in electronic form. This work was supported in part by National Natural Science Foundation of China under contract No.~11975200 and No.~11935013, and the Zhejiang University Fundamental Research Funds for the Central Universities (2017QNA3007,2019QNA3005).

Note added: We thank Markus Ebert, Bernhard Mistlberger, and Gherardo Vita for communicating with us an independent calculation of TMD quark and gluon beam functions~\cite{1800390}, and pointing out an error in the terms proportional to the anomalous color structure $d_{ABC} d^{ABC}$ in the previous version of this paper. 

\bibliography{quarkTMDN3LO.bib}{}
\bibliographystyle{apsrev4-1}

\newpage

\onecolumngrid
\appendix*
\allowdisplaybreaks
\setcounter{secnumdepth}{2}

\section{}

In this appendix we collect all the formulas relevant to our discussion. In  Sec.~\ref{sec:beta} we collect the QCD beta functions to three loops. In Sec.~\ref{sec:AD} we collect all the relevant anomalous dimension. In Sec.~\ref{sec:soft} we collect the TMD soft function through N$^3$LO. In Sec.~\ref{sec:beam} we give the quark TMD beam function through N$^3$LO, which is one of the main results of this Letter. In Sec.~\ref{sec:numer} we give the approximate numerical fitting of the analytical results. 

\subsection{QCD Beta Function}
\label{sec:beta}

The QCD beta function is defined as
\begin{equation}
\frac{d\alpha_s}{d\ln\mu} = \beta(\alpha_s) = -2\alpha_s \sum_{n=0}^\infty \left( \frac{\alpha_s}{4 \pi} \right)^{n+1} \, \beta_n \, ,
\end{equation}
with~\cite{Baikov:2016tgj}
\begin{align}
\beta_0 &= \frac{11}{3} C_A - \frac{4}{3} T_F N_f \, , \nn
\\
\beta_1 &= \frac{34}{3} C_A^2 - \frac{20}{3} C_A T_F N_f - 4 C_F T_F N_f \, ,\nn
\\
\beta_2 &= \left(\frac{158 C_A}{27}+\frac{44 C_F}{9}\right) N_f^2 T_F^2 +\left(-\frac{205 C_A
   C_F}{9}-\frac{1415 C_A^2}{27}+2 C_F^2\right) N_f T_F  +\frac{2857 C_A^3}{54}\,.
\end{align}

\subsection{Anamolous dimension}
\label{sec:AD}

For all the anomalous dimensions entering the renormalization group equations of various TMD functions, we define the perturbative expansion in $\alpha_s$ according to
\begin{equation}
\gamma(\alpha_s) = \sum_{n=0}^\infty \left( \frac{\alpha_s}{4 \pi} \right)^{n+1} \, \gamma_n \, ,
\end{equation}
where the coefficients up to $\Ord(\alpha_s^3)$ are given by
\begin{align}
\Gcusp_{0} =& 4 C_F\,, \nn
\\
\Gcusp_{1} =&  \left(\frac{268}{9}-8 
                 \zeta_2\right) C_A C_F -\frac{80 C_F T_F N_f}{9}\,, \nn
\\
\Gcusp_{2} =&\bigg[ \left(\frac{320 \zeta _2}{9}-\frac{224 \zeta _3}{3}-\frac{1672}{27}\right) C_A
   C_F+\left(64 \zeta _3-\frac{220}{3}\right) C_F^2\bigg] N_f T_F  \nn
   \\
 &+\left(-\frac{1072 \zeta
   _2}{9}+\frac{88 \zeta _3}{3}+88 \zeta _4+\frac{490}{3}\right) C_A^2 C_F  -\frac{64}{27} C_F
   N_f^2 T_F^2\,,\nn
  \\
\gamma^B_0 =& 3C_F \, , \nn
\\
\gamma^B_1 =&  \left[  \left( \frac{3}{2} - 12\zeta_2 + 24\zeta_3 \right) C_F + \left( \frac{17}{6} + \frac{44 \zeta_2}{3} - 12\zeta_3 \right)  C_A  + \left( -\frac{2}{3} - \frac{16\zeta_2}{3} \right) T_F N_f  \right] C_F  , \nn
\\
\gamma^B_2 = &
 \bigg[ \left(-\frac{2672 \zeta _2}{27}+\frac{400 \zeta _3}{9}+4
   \zeta _4+40\right) C_A C_F+\left(\frac{40 \zeta _2}{3}-\frac{272 \zeta
   _3}{3}+\frac{232 \zeta _4}{3}-46\right) C_F^2\bigg] N_f T_F \nn 
\\   
  & +\left(16 \zeta _3
   \zeta _2-\frac{410 \zeta _2}{3}+\frac{844 \zeta _3}{3}-\frac{494 \zeta
   _4}{3}+120 \zeta _5+\frac{151}{4}\right) C_A C_F^2+\left(\frac{320 \zeta _2}{27}-\frac{64 \zeta _3}{9}-\frac{68}{9}\right) C_F N_f^2T_F^2\nn 
   \\
   &+\left(\frac{4496
   \zeta _2}{27}-\frac{1552 \zeta _3}{9}-5 \zeta _4+40 \zeta _5-\frac{1657}{36}\right) C_A^2 C_F
   +\left(-32 \zeta _3 \zeta _2+18 \zeta _2+68 \zeta _3+144 \zeta
   _4-240 \zeta _5+\frac{29}{2}\right) C_F^3\,, \nn 
   \\
\gamma^H_0 =& -3C_F \, , \nn
\\
\gamma^H_1 =& C_F \left[ C_F \left( -\frac{3}{2} + 12\zeta_2 - 24\zeta_3 \right) + C_A \left( -\frac{961}{54} - 11 \zeta_2 + 26\zeta_3 \right) + T_FN_f \left( \frac{130}{27} +4\zeta_2\right) \right] \,, \nn
\\
\gamma^H_2 = & N_f T_F \left(\left(\frac{5188 \zeta _2}{81}-\frac{1928 \zeta _3}{27}+44
   \zeta _4-\frac{17318}{729}\right) C_A C_F+\left(-\frac{52 \zeta
   _2}{3}+\frac{512 \zeta _3}{9}-\frac{280 \zeta
   _4}{3}+\frac{2953}{27}\right) C_F^2\right)\nn
   \\
 &  +\left(-16 \zeta _3 \zeta
   _2+\frac{410 \zeta _2}{3}-\frac{844 \zeta _3}{3}+\frac{494 \zeta
   _4}{3}-120 \zeta _5-\frac{151}{4}\right) C_A C_F^2+\left(-\frac{80 \zeta _2}{9}-\frac{32 \zeta
   _3}{27}+\frac{9668}{729}\right) C_F N_f^2 T_F^2\nn
   \\
  & +\left(-\frac{88}{3}
   \zeta _3 \zeta _2-\frac{7163 \zeta _2}{81}+\frac{3526 \zeta _3}{9}-83
   \zeta _4-136 \zeta _5-\frac{139345}{2916}\right) C_A^2
   C_F\nn
   \\
   &+\left(32 \zeta _3
   \zeta _2-18 \zeta _2-68 \zeta _3-144 \zeta _4+240 \zeta
   _5-\frac{29}{2}\right) C_F^3 \,, \nn \\
\gamma^S_0 =& 0 \, , \nn
\\
\gamma^S_1 =& \left[ \left( -\frac{404}{27} + \frac{11\zeta_2}{3} + 14\zeta_3 \right) C_A   + \left( \frac{112}{27} - \frac{4 \zeta_2}{3} \right)T_F N_f   \right]  C_F   \,, \nn
\\
\gamma^S_2  =&\left(-\frac{88}{3} \zeta
   _3 \zeta _2+\frac{6325 \zeta _2}{81}+\frac{658 \zeta _3}{3}-88 \zeta
   _4-96 \zeta _5-\frac{136781}{1458}\right) C_A^2 C_F
+\left(\frac{80\zeta _2}{27}-\frac{224 \zeta _3}{27}+\frac{4160}{729}\right) C_FN_f^2 T_F^2\nn
   \\
  & + \left(-\frac{2828 \zeta _2}{81}-\frac{728 \zeta _3}{27}+48 \zeta
   _4+\frac{11842}{729}\right) C_A C_F N_f T_F
   +\left(-4 \zeta _2-\frac{304 \zeta _3}{9}-16 \zeta
   _4+\frac{1711}{27}\right) C_F^2 N_f T_F\,.\nn
   \\
   \gamma^R_0 = &0 \, , \nn
\\
\gamma^R_1 = & \left[ \left( -\frac{404}{27} + 14\zeta_3 \right) C_A  +  \frac{112}{27} T_F N_f \right] C_F  \,, \nn 
\\
\gamma^R_2 =&\bigg[\left(-\frac{824 \zeta _2}{81}-\frac{904 \zeta _3}{27}+\frac{20 \zeta
   _4}{3}+\frac{62626}{729}\right) C_A N_f T_F 
   +\left(-\frac{88}{3} \zeta _3 \zeta
   _2+\frac{3196 \zeta _2}{81}+\frac{6164 \zeta _3}{27}+\frac{77 \zeta _4}{3}-96 \zeta_5\right.\nn
   \\
 & \left. -\frac{297029}{1458}\right)C_A^2 
   + \left(-\frac{304 \zeta _3}{9}-16 \zeta
   _4+\frac{1711}{27}\right)  C_F N_f T_F 
   +\left(-\frac{64 \zeta
   _3}{9}-\frac{3712}{729}\right) N_f^2 T_F^2 \bigg] C_F \,.
\end{align}
The cusp anomalous dimension $\Gamma^{\text{cusp}}$ can be found in \cite{Moch:2004pa}. The hard anomalous dimensions $\gamma^H$ can be extracted from the three-loop quark form factor~\cite{Moch:2005tm,Gehrmann:2010ue}, and can also be found in, e.g., Ref.~\cite{Becher:2009qa}. The soft anomalous dimension $\gamma^S$ are calculated explicitly through N$^3$LO in Ref.~\cite{Li:2014afw}. The beam anomalous dimension $\gamma^B$ is related to $\gamma^S$ and $\gamma^H$ through the renormalization group invariance condition $\gamma^B = \gamma^S - \gamma^H$. The rapidity anomalous dimension $\gamma^R$ can be found from \cite{Li:2016ctv,Vladimirov:2016dll}. Note that the normalization here differ from those in \cite{Li:2016ctv} by a factor of $1/2$.

\subsection{Renormalization Constants}
\label{sec:RC}

The following constants are needed for the renormalization of zero-bin subtracted~\cite{Manohar:2006nz} TMD beam function through N$^3$LO, see e.g. Ref.~\cite{Luo:2019hmp,Luo:2019bmw}. The first three-order corrections to $Z^B $ and $Z^S$ are 
\begin{align}
\label{eqZqZs}
Z^B_1 =& \frac{1}{2\epsilon} \left(2 \gamma^B_0 -\Gamma_0^{\text{cusp}} L_Q \right) \,, \nonumber \\
Z^B_2 =& \frac{1}{8 \epsilon^2} \bigg( ( \Gamma_0^{\text{cusp}} L_Q - 2 \gamma^B_0)^2 + 2 \beta_0 (  \Gamma_0^{\text{cusp}} L_Q - 2 \gamma^B_0)    \bigg) + \frac{1}{4\epsilon} \left( 2 \gamma^B_1 - \Gamma_1^{\text{cusp}} L_Q \right) \,, \nonumber \\
Z^B_3  =& \frac{1}{48 \epsilon^3}  \left( 2 \gamma^B_0 -\Gamma^{\text{cusp}}_0 L_Q \right) \biggl( 8 \beta_0^2 + 6 \beta_0 \left( -2 \gamma^B_0 + \Gamma^{\text{cusp}}_0 L_Q \right) + \left( -2 \gamma^B_0 + \Gamma^{\text{cusp}}_0 L_Q \right)^2 \biggl) + \frac{1}{6 \epsilon} \biggl(  2 \gamma^B_2 -  \Gamma^{\text{cusp}}_2 L_Q  \biggl)  \nn \\ 
&+ \frac{1}{24 \epsilon^2} \biggl( \beta_1 \left(-8 \gamma^B_0 + 4  \Gamma^{\text{cusp}}_0 L_Q \right) + \left(4 \beta_0 - 6 \gamma^B_0 + 3 \Gamma^{\text{cusp}}_0 L_Q \right) \left( -2 \gamma^B_1 + \Gamma^{\text{cusp}}_1 L_Q \right)  \biggl)  \nn \,, \\      
Z^S_1 =& \frac{1}{\epsilon^2} \Gamma^{\text{cusp}}_0  +  \frac{1}{\epsilon} \left( -2 \gamma^S_0 - \Gamma_0^{\text{cusp}} L_\nu \right) \,,\nonumber \\
Z^S_2 =& \frac{1}{2 \epsilon^4} (\Gamma^{\text{cusp}}_0)^2 - \frac{1}{4 \epsilon^3} \bigg(\Gamma^{\text{cusp}}_0 (3 \beta_0 + 8 \gamma^S_0)+4( \Gamma^{\text{cusp}}_0)^2 L_\nu\bigg)   - \frac{1}{2 \epsilon} \left( 2 \gamma^S_1 +  \Gamma^{\text{cusp}}_1 L_\nu \right) \nonumber \\
& + \frac{1}{4 \epsilon^2} \bigg(\Gamma^{\text{cusp}}_1 + 2 ( 2 \gamma^S_0 + \Gamma^{\text{cusp}}_0 L_\nu ) ( \beta_0 + 2 \gamma^S_0 + \Gamma^{\text{cusp}}_0 L_\nu) \bigg) \,, \nn \\
Z^S_3 =& \frac{ 1}{6 \epsilon^6} \left(\Gamma^{\text{cusp}}_0\right)^3  - \frac{1}{4 \epsilon^5} \left(\Gamma^{\text{cusp}}_0 \right)^2 \left(  3 \beta_0 + 4 \gamma^S_0 + 2 \Gamma^{\text{cusp}}_0 L_\nu \right) + \frac{1}{36 \epsilon^4 } \Gamma^{\text{cusp}}_0 \bigg( 22 \beta_0^2 + 45 \beta_0 \left(2 \gamma^S_0 + \Gamma^{\text{cusp}}_0 L_\nu \right)  \nn \\
&+ 9 \left( \Gamma^{\text{cusp}}_1 + 2 \left( 2 \gamma^S_0+ \Gamma^{\text{cusp}}_0 L_\nu\right)^2  \right)   \biggl)  + \frac{1}{36 \epsilon^3} \biggl( -16 \beta_1 \Gamma^{\text{cusp}}_0 - 12 \beta_0^2 \left( 2 \gamma^S_0 + \Gamma^{\text{cusp}}_0 L_\nu \right) \nn \\
& - 2 \beta_0 \left( 5 \Gamma^{\text{cusp}}_1 + 9 \left( 2 \gamma^S_0 + \Gamma^{\text{cusp}}_0 L_\nu \right)^2 \right) - 3 \bigg[ \Gamma^{\text{cusp}}_1 \left(6 \gamma^S_0 + 9 \Gamma^{\text{cusp}}_0 L_\nu \right)  \nn \\
&+ 2 \left( 8 \left( \gamma^S_0\right)^3 + 6 \Gamma^{\text{cusp}}_0 \gamma^S_1 + 12 \Gamma^{\text{cusp}}_0  \left(\gamma^S_0\right)^2 L_\nu + 6 \left(\Gamma^{\text{cusp}}_0\right)^2 \gamma^S_0 L_\nu^2 + \left( \Gamma^{\text{cusp}}_0 \right)^3 L_\nu^3 \right) \bigg]    \biggl) \nn \\
& + \frac{1}{18 \epsilon^2} \biggl(  2 \Gamma^{\text{cusp}}_2 + 3 \left( 2 \beta_1 \left( 2 \gamma^S_0 + \Gamma^{\text{cusp}}_0 L_\nu \right) + \left( 2 \beta_0 + 6 \gamma^S_0 + 3 \Gamma^{\text{cusp}}_0 L_\nu \right) \left( 2 \gamma^S_1 + \Gamma^{\text{cusp}}_1 L_\nu \right) \right)    \biggl) -  \frac{2 \gamma^S_2 + \Gamma^{\text{cusp}}_2 L_\nu}{3 \epsilon} \,.
\end{align}
Here we have defined $L_\nu = \ln (\nu^2/\mu^2)$, and $L_Q = 2 \ln ( z P^+/\nu) = 2 \ln( Q/\nu)$, where $Q$ is the Drell-Yan lepton pair invariant mass. We also need the PDF counter terms to three loops, which are given by
\begin{align}
\phi_{ij}(z, \alpha_s) =& \delta_{ij} \delta(1-z) - \frac{\alpha_s}{4 \pi} \frac{P^\zero_{ij}(z)}{\epsilon} +  \left(\frac{\alpha_s}{4 \pi}\right)^2 \bigg[ \frac{1}{2 \epsilon^2} \biggl( \sum_k P^\zero_{ik} \otimes P^\zero_{k j}(z) + \beta_0 P^\zero_{ij} (z)  \biggl) - \frac{1}{2\epsilon} P^\one_{ij}(z)  \bigg]  \nn \\
&+  \left(\frac{\alpha_s}{4 \pi}\right)^3 \bigg[  \frac{-1}{6 \epsilon^3} \biggl(  \sum_{m \,, k }P^{\zero}_{im} \otimes P^{\zero}_{mk} \otimes P^{\zero}_{k j}(z)  + 3 \beta_0 \sum_k P^{\zero}_{ik} \otimes P^{\zero}_{kj}(z) +2 \beta_0^2 P^{\zero}_{ij}(z) \biggl)  \nn \\ 
&+ \frac{1}{6 \epsilon^2} \biggl( \sum_k P^{\zero}_{ik} \otimes P^{\one}_{kj}(z) 
+ 2 \sum_k P^{\one}_{ik} \otimes P^{\zero}_{kj} (z)  + 2  \beta_0 P^{\one}_{ij}(z) +  2  \beta_1 P^{\zero}_{ij}(z) \biggl) -\frac{1}{3 \epsilon} P^\two_{ij}(z) \bigg] \,,
\end{align}
where $P_{ij}^{(n)}$ is the $(n+1)$-loop spacelike splitting function~\cite{Moch:2004pa,Vogt:2004mw}. The convolution is defined as $f \otimes g(z) \equiv \int_0^1 \! dx dy\, f(x) g(y) \delta(z - x y)$. 

\subsection{TMD Soft Function}
\label{sec:soft}

The TMD soft function is calculated through N$^3$LO in Ref.~\cite{Li:2016ctv}. With the exponential regulator~\cite{Li:2016axz}, it obeys the Non-Abelian exponentiation theorem~\cite{Gatheral:1983cz,Frenkel:1984pz},
\begin{align}
\mathcal{S}(b_\perp,\mu,\nu)=\exp\Bigg\{ \frac{\alpha_s}{4 \pi} s^{(1)}+\left(\frac{\alpha_s}{4 \pi} \right)^2 s^{(2)}+\left(\frac{\alpha_s}{4 \pi} \right)^3 s^{(3)}+ \ldots \Bigg\} \, .
\end{align}
The exponent can be written as~(we have used that $\gamma_0^R = \gamma_0^S = 0$ to simplify the expression)
\begin{align}
s^{(1)}=& -2 \zeta _2 C_F + \frac{\Gcusp_{0} \Lp^2}{2}-\Gcusp_{0} \Lp L_R  \,,\nn
\\
s^{(2)}=&\left(-\frac{67 \zeta _2}{3}+10 \zeta _4-\frac{154 \zeta_3}{9} +\frac{2428}{81}\right)C_A C_F    + 2 
   \left(\frac{10 \zeta _2}{3}+\frac{28 \zeta_3}{9}-\frac{328}{81}\right)C_F N_f T_F \nn
   \\
  & +\Lp \left(-2 \beta_0 \zeta _2
   C_F-\Gcusp_{1} L_R-2 \gamma^S_1\right)+\Lp^2
   \left(\frac{\Gcusp_{1}}{2}-\frac{\beta_0 \Gcusp_{0}
   L_R}{2}\right) +\frac{1}{6} \beta_0 \Gcusp_{0} \Lp^3  +2
   \gamma^R_1 L_R\,,\nn
   \\
   s^{(3)}=&2 \left(-\frac{80 \zeta _2 \zeta_3}{3}+\frac{275 \zeta
   _2}{9}+\frac{152 \zeta _4}{9}+\frac{224 \zeta_5}{9}+\frac{3488 \zeta_3}{81}-\frac{42727}{486}\right) C_F^2 N_f T_F +4 \left(-\frac{136 \zeta
   _2}{27}-\frac{44 \zeta _4}{27}\right.\nn
   \\
   &\left.-\frac{560 \zeta_3}{243}-\frac{256}{6561}\right)C_F N_f^2 T_F^2 
   +2\left(\frac{40 \zeta _2 \zeta_3}{9}+\frac{74530 \zeta
   _2}{729}-\frac{416 \zeta _4}{27}-\frac{184 \zeta_5}{3}+\frac{8152 \zeta_3}{81}-\frac{412765}{6561}\right) C_A C_F N_f T_F \nn
   \\
   &+ \left(\frac{1100 \zeta _2 \zeta_3}{9}-\frac{297481 \zeta _2}{729}+\frac{3649 \zeta _4}{27}-\frac{3086 \zeta
   _6}{27}+\frac{1804 \zeta_5}{9}+\frac{928 \zeta_3^2}{9}-\frac{151132 \zeta_3}{243}+\frac{5211949}{13122}\right) C_A^2 C_F \nn
   \\
&+\Lp \bigg[2 \beta_0 \left(\left(-\frac{67 \zeta _2}{3}+10 \zeta
   _4-\frac{154 \zeta_3}{9}+\frac{2428}{81}\right)C_A C_F  +2 \left(\frac{10 \zeta
   _2}{3}+\frac{28 \zeta_3}{9}  -\frac{328}{81}\right)C_F N_f T_F \right) +(4 \beta_0
   \gamma^R_1\nn
   \\
  & -\Gcusp_{2}) L_R  -2 \beta_1 \zeta _2 C_F-2 \gamma^S_2\bigg]     
   +\Lp^2 \bigg[-2 \beta_0^2 \zeta _2
   C_F+ \left(-\beta_0 \Gcusp_{1}-\frac{\beta_1
   \Gcusp_{0}}{2}\right) L_R  -2 \beta_0
   \gamma^S_1+\frac{\Gcusp_{2}}{2}\bigg]\nn
   \\
  & +\Lp^3 \bigg[-\frac{1}{3}
   \beta_0^2 \Gcusp_{0} L_R+\frac{\beta_0
   \Gcusp_{1}}{3}+\frac{\beta_1 \Gcusp_{0}}{6}\bigg]+\frac{1}{12}
   \beta_0^2 \Gcusp_{0} \Lp^4+2 \gamma^R_2 L_R\,,
\end{align}
where $L_\perp = \ln ( |b_\perp^2| \mu^2/b_0^2)$, $L_R = L_\perp + L_\nu = \ln(|b_\perp|^2 \nu^2/b_0^2)$.

\subsection{Analytical Results for the TMD Beam Function through N$^3$LO}
\label{sec:beam}

This section contains one of the main results in this Letter. The quark TMD beam functions through N$^3$LO are given by
\begin{align}
\cI^\zero_{qi}(z,b_\perp,&\,P^+,\mu,\nu) =  \delta_{qi} \delta(1-z) \, ,\nn
\\
\cI^\one_{qi}(z,b_\perp,&\,P^+,\mu,\nu) = \left( - \frac{\Gcusp_0}{2} \Lp L_Q + \gamma_0^B \Lp + \gamma_0^R L_Q \right) \delta_{qi} \delta(1-z) - P_{qi}^\zero(z) \Lp + I_{qi}^\one(z) \, , \nn
\\
\cI^\two_{qi}(z,b_\perp,&\,P^+,\mu,\nu) =  \bigg[ \frac{1}{8} \left( -\Gcusp_0 L_Q + 2\gamma^B_0 \right) \left( -\Gcusp_0 L_Q + 2\gamma^B_0 + 2\beta_0 \right) \Lp^2+ \left(  (-\Gcusp_0 L_Q + 2\gamma^B_0 + 2\beta_0) \frac{\gamma_0^R}{2} L_Q \right. \nn
\\
&\left.-\frac{\Gcusp_1}{2} L_Q + \gamma^B_1  \right) \Lp \nn
+ \frac{(\gamma_0^R)^2}{2} L_Q^2 + \gamma_1^R L_Q \bigg] \, \delta_{qi} \delta(1-z) \nn
+ \bigg( \frac{1}{2} \sum_j P^\zero_{qj} \otimes P^\zero_{ji}(z) \nn
\\
&+ \frac{P^\zero_{qi}(z)}{2} (\Gcusp_0 L_Q - 2\gamma_0^B - \beta_0) \bigg) \Lp^2 + \bigg[ -P^\one_{qi}(z) - P^\zero_{qi}(z) \gamma_0^R L_Q - \sum_j I^\one_{qj} \otimes P^\zero_{ji}(z) \nn
\\
&+ \left( -\frac{\Gcusp_0}{2} L_Q + \gamma_0^B + \beta_0 \right) I^\one_{qi}(z) \bigg] \Lp + \gamma_0^R L_Q I^\one_{qi}(z) + I^\two_{qi}(z) \, ,\nn
\\
\cI^\three_{qi}(z,b_\perp,&\,P^+,\mu,\nu) = \Lp^3\bigg[
\left(\frac{1}{2}\beta_0+\frac{1}{4}(2 \gamma^B_0-\Gcusp_0 L_Q) \right) \sum_{j}P^\zero_{qj} \otimes P^\zero_{ji}(z) -\frac{1}{6} \sum_{j\,k } P^\zero_{qj} \otimes P^\zero_{jk} \otimes P^\zero_{ki}(z) \nn
\\
&+\delta_{qi} \delta(1-z) \left( \frac{1}{6}\beta_0^2(2 \gamma^B_0-\Gcusp_0 L_Q)+\frac{1}{8} \beta_0(2 \gamma^B_0-\Gcusp_0 L_Q)^2+\frac{1}{48} (2 \gamma^B_0-\Gcusp_0 L_Q)^3 \right) \nn
\\
&+P^\zero_{qi}  \left( -\frac{1}{2} \beta_0 (2 \gamma^B_0-\Gcusp_0 L_Q)-\frac{1}{3}\beta_0^2-\frac{1}{8} (2 \gamma^B_0-\Gcusp_0 L_Q)^2 \right) \nn
\bigg]
\\
&+\Lp^2\bigg[ 
\left(-\frac{3}{2} \beta_0-\frac{1}{2}(2 \gamma^B_0-\Gcusp_0 L_Q) \right)\sum_j I^\one_{qj} \otimes P^\zero_{ji} (z) 
+\frac{1}{2} \sum_{j\,k } I^\one_{qj} \otimes P^\zero_{jk} \otimes P^\zero_{ki}(z)\nn
\\
&+\frac{1}{2} \sum_j P^\zero_{qj} \otimes P^\one_{ji}(z)
+\frac{1}{2}\sum_j P^\one_{qj} \otimes P^\zero_{ji}(z) \nn
+P^\zero_{qi}(z) \left(-\frac{1}{2}\beta_1-\frac{1}{2}(2 \gamma^B_1-\Gcusp_1 L_Q) \right)
\\
&+\delta_{qi} \delta(1-z)  \left ( \frac{1}{4} \beta_1(2 \gamma^B_0-\Gcusp_0 L_Q)+\frac{1}{2}\beta_0(2 \gamma^B_1-\Gcusp_1 L_Q)+\frac{1}{4}(2 \gamma^B_0-\Gcusp_0 L_Q)(2 \gamma^B_1-\Gcusp_1 L_Q)
 \right)  \nn
\\
&+I^\one_{qi}(z) \left( \frac{3}{4} \beta_0(2 \gamma^B_0-\Gcusp_0 L_Q)+\beta_0^2+\frac{1}{8}(2 \gamma^B_0-\Gcusp_0 L_Q)^2  \right)+P^\one_{qi}(z) \left(-\beta_0-\frac{1}{2}(2 \gamma^B_0-\Gcusp_0 L_Q) \right)
\bigg] \nn
\\
&+\Lp \bigg[
-\sum_{j } I^\one_{qj} \otimes P^\one_{ji} (z)-\sum_{j } I^\two_{qj} \otimes P^\zero_{ji} (z) -P^\zero_{qi}(z) \gamma^R_1 L_Q-P^\two_{qi}(z) 
+\delta_{qi} \delta(1-z)  \biggl( 2 \beta_0  \gamma_1^R L_Q \nn
\\
& +\frac{1}{2}\gamma_1^R (2 \gamma^B_0-\Gcusp_0 L_Q)L_Q+\frac{1}{2} (2\gamma_2^B-\Gcusp_2 L_Q) \biggl )+I^\one_{qi} (z)\left( \beta_1+\frac{1}{2}(2 \gamma^B_1-\Gcusp_1 L_Q) \right) \nn
\\
&+I^\two_{qi} (z)\left(2 \beta_0+\frac{1}{2}(2 \gamma^B_0-\Gcusp_0 L_Q) \right)\bigg]+\delta_{qi} \delta(1-z) \gamma^R_2 L_Q+I^\one_{qi}(z) \gamma^R_1 L_Q+I^\three_{qi} (z)\,,
\end{align}
where in $\cI_{qi}^{\three}$ we have used $\gamma_0^R = 0$ to simplify the expression. 

The scale independent coefficient functions at NLO are
\begin{align}
I_{qq}^\one(z) &= 2 C_F (1 - z) \,, \nn
\\
I_{qg}^\one(z) &= 4 T_F z (1 - z) \,,\nn
\\
I_{qq'}^\one(z)& =I_{q\bar{q}} ^\one(z)=0\,.
\end{align}
At NNLO they are given by
\begin{align}
\label{eq:twoloopCoef}
I^{(2)}_{q\bar{q}'}(z)=& I^{(2)}_{qq'}(z) =  + C_F T_F \bigg\{-\frac{8}{3} \frac{1-z}{z} \left(2 z^2-z+2\right) \left(H_{1,0}+\zeta
   _2\right)-\frac{2}{3} \left(8 z^2+3 z+3\right) H_{0,0}+4 (z+1) H_{0,0,0} \nn 
   \\
   &+\frac{4}{9}
    \left(32 z^2-30 z+21\right) H_0 +\frac{2 (1-z) \left(136 z^2-143 z+172\right)}{27
   z}\bigg\}  \,, \nn
   \\
   I^{(2)}_{q\bar{q}}(z) = & +I^{(2)}_{qq'}(z)+ C_F \left(C_A-2 C_F\right) \bigg\{+4 (1-z)
   H_{1,0}+4 (z+1) H_{-1,0}+ (-11 z-3) H_0 -2 (z-3) \zeta _2 \nn 
\\
&-15 (1-z)- \frac{2 \left(z^2+1\right)}{z+1}  \bigg[4 H_{-2,0}-2
   H_{2,0}-4 H_{-1,-1,0}+2 H_{-1,0,0}-H_{0,0,0}-2 H_{-1} \zeta _2+\zeta _3\bigg]    
   \bigg\} \,, \nn
   \\
   I^{(2)}_{qq}(z) = & +I^{(2)}_{qq'}(z)+ \frac{1}{(1-z)_+} \bigg[ \left(28 \zeta _3-\frac{808}{27}\right) C_A C_F+\frac{224}{27} N_f  C_F
   T_F \bigg]  \nn
\\   
   &+ C_A C_F \bigg\{\frac{1}{1-z} \bigg[   2 \left(z^2-13\right) \zeta _3-2 \left(z^2+1\right)
   \left(2 H_{1,2}+2 H_{2,0}+H_{0,0,0}+2 H_{1,1,0}\right)   \bigg] \nn 
\\
& +\frac{1}{3}
   \frac{1}{1-z} \left(z^2-12 z-11\right) H_{0,0}-2 (1-z) \left(2 H_{1,0}+3 \zeta
   _2\right)-\frac{2}{9} \frac{1}{1-z}  \left(83 z^2-36 z+29\right)H_0 -2 z  H_1 \nn 
\\
& +\frac{8
   (z+100)}{27}\bigg\}
   +C_F N_f T_F \bigg\{\frac{4}{3} \frac{z^2+1}{1-z} 
   H_{0,0}+\frac{20}{9} \frac{z^2+1}{1-z}   H_0 -\frac{4}{27} (19
   z+37)\bigg\} \nn 
   \\
   &
   +C_F^2 \bigg\{4 \frac{1+z^2}{1-z}  \left(2 H_{1,2}+2
   H_{2,0}+H_{2,1}-H_{1,0,0}+2 H_{1,1,0}+6 \zeta _3\right)-2 \frac{1}{1-z} \left(2 z^2-2
   z-3\right) H_{0,0} \nn 
   \\
   &+4 (1-z) \left(3 H_{1,0}+H_2+2 \zeta _2\right)+2 (z+1) H_{0,0,0}+2
   \frac{1}{1-z} \left(16 z^2-13 z+5\right) H_0 +2 z H_1-22 (1-z)\bigg\} \,, \nn
   \\
   I^{(2)}_{qg}(z) = & +C_A T_F \bigg\{-\frac{8}{3} \frac{1-z}{z} \left(11 z^2-z+2\right) H_{1,0}+4 \left(2 z^2+2
   z+1\right) \left(2 H_{-2,0}-2 H_{-1,-1,0}+H_{-1,0,0}-H_{-1} \zeta _2\right) \nn 
   \\
   &-\frac{2}{3}
   \left(44 z^2-12 z+3\right) H_{0,0}+4 \left(2 z^2-2 z+1\right)
   \left(H_{1,2}+H_{1,1,0}-H_{1,1,1}\right)-8 z \left(2 H_{2,0}-\zeta _3\right) \nn 
   \\
   &+8 z (z+1)
   H_{-1,0}-8 (1-z) z H_{1,1}+4 (2 z+1) H_{0,0,0}+\frac{8}{3} \frac{1}{z} \left(11
   z^3-9 z^2+3 z-2\right) \zeta _2 \nn 
   \\
   & +\frac{4}{9} \left(68 z^2-30 z+21\right)  H_0 +2  z (4
   z-3) H_1-\frac{2 \left(298 z^3-387 z^2+315 z-172\right)}{27 z}\bigg\} \nn
   \\
   &
   +C_F T_F \bigg\{4
   \left(2 z^2-2 z+1\right) \left(H_{2,1}-H_{1,0,0}+H_{1,1,1}+7 \zeta _3\right)+\left(-8
   z^2+12 z+1\right) H_{0,0} \nn 
   \\
   &-2 \left(4 z^2-2 z+1\right) H_{0,0,0}+8 (1-z) z
   \left(H_{1,0}+H_{1,1}+H_2-\zeta _2\right)+ \left(-8 z^2+15 z+8\right) H_0  \nn
   \\
   &-2  (4 z-3)z  H_1 -72 z^2+75 z-13\bigg\} \,.     
\end{align}

For the three-loop coefficient functions we decompose them into different color structures,
\begin{align}
 I^{(3)}_{qq'} (z) &= I^*_{qq'}(z) + \frac{d^{ABC} d_{ABC} \, T_F}{16 N_c } I_{d33}(z),   \nonumber \\
 I^{(3)}_{q\bar{q}'}(z) &= I^*_{qq'}(z) - \frac{d^{ABC} d_{ABC} \, T_F}{16 N_c} I_{d33}(z),   \nonumber \\  
I^{(3)}_{qq}(z) &=  I^*_{qq}(z) + I^{(3)}_{qq'}(z) ,   \nonumber  \\
I^{(3)}_{q\bar{q}}(z) &=   C_F \left(C_A-2 C_F\right) I^*_{q\bar{q}}(z) +I^{(3)}_{q\bar{q}'}(z) \,,
\label{eq:Idecomp}
\end{align}
where $d^{ABC} d_{ABC} = 4 \mathrm{Tr}[T^A \{T^B , T^C \} ] \mathrm{Tr}[T^A \{ T^B,  T^C\}] = (N_c^2-1)(N_c^2-4)/N_c$. Note the sign difference in the coefficient of $d^{ABC} d_{ABC}$ in the first and second line of Eq.~\eqref{eq:Idecomp}, which is due to the reversal of fermion line in the anomalous triangle diagram. These anomalous terms do not simply add up to zero because the PDFs for $q'$ and $\bar{q}'$ are different. Remarkably, we find that all the results at this order can be written in terms of HPLs only. We stress that this is a non-trivial statement, since symbol letter such as $2-z$ and $z^2 - z + 1$ appear in the results for VRR and RRR, but cancel out in the sum.  The fully analytical results of individual color structure are given~\footnote{There was an error in $I_{d33}(z)$ in the previous version of this paper, which has been corrected in the current version. 
We thank Markus Ebert, Bernhard Mistlberger, and Gherardo Vita for pointing out this error.}
\begin{align}
I_{d33}(z)&\, = -\frac{256}{3} \frac{1}{z} \zeta _2 H_{-1,-1} (z+1)^3-\frac{7264}{27} H_{-1,0}
   (z+1)+\frac{1312}{9} H_{2,1} (z+1)-64 \left(H_{-1,-2,0}+H_{-1,2,0}\right)
   (z+1) \nonumber \\
   & +\frac{64(z+1)}{3}   \bigg(-5 H_2 \zeta _3-6 H_{-1,0} \zeta _3-\frac{3}{2} H_{-1} \zeta
   _4-6 \zeta _2 H_{-1,2}-13 \zeta _2 H_{2,0}-8 \zeta _2 H_{2,1}+9 H_{2,3}+3 H_{4,1}+3
   \zeta _2 H_{-1,0,0} \nonumber \\
   &+3 H_{-1,3,0}+9 \zeta _2 H_{0,0,0}-4 H_{2,-2,0}-9 H_{2,1,2}-2
   H_{2,2,0}+6 H_{-1,2,0,0}-6 H_{-1,2,1,0}+H_{2,0,0,0}-9 H_{2,1,0,0}-15 H_{2,1,1,0}\bigg) \nonumber
    \\
& +\frac{33280 (1-z)}{27}+\frac{16}{27} (369 z-388) H_0+\frac{16}{27} (2243 z-144)
   \frac{1-z}{z} H_1+\frac{32}{27} (773 z+116) H_2-\frac{32}{27} \left(232 z^2+273
   z+222\right) H_3 \nonumber 
   \\
   &+\frac{224}{9} \left(8 z^2+3 z+3\right) H_4-\frac{64}{3} (9 z+7)
   H_5-\frac{32}{27} (359 z+757) \zeta _2+\frac{64}{3} \left(20 z^2+z+27\right) H_{-2}
   \zeta _2 \nonumber 
   \\
   & +\frac{128}{27} \left(116 z^2+93 z+60\right) H_0 \zeta _2-\frac{32}{27}
   \left(116 z^2+83 z+116\right) \frac{1-z}{z} H_1 \zeta _2+\frac{64}{3} (9 z-1) H_3
   \zeta _2+\frac{64}{27} \left(290 z^2-141 z+522\right) \zeta _3 \nonumber 
   \\
   &-\frac{64}{3} (5 z-7)
   \zeta _2 \zeta _3-\frac{16}{3} (5 z+39) H_0 \zeta _4 -\frac{64}{3} (15 z+53) \zeta
   _5-\frac{128}{3} (5 z-1) H_{-4,0}-\frac{64}{9} \left(32 z^2-15 z+39\right)
   H_{-3,0}\nonumber 
   \\
   & +\frac{64}{27} \left(116 z^2+198 z+99\right) H_{-2,0}-\frac{64}{3} \left(16
   z^2+z+27\right) H_{-2,2}  +\frac{64}{27} (271 z-227) H_{0,0}-\frac{32}{9} (1-z) \left(153
   H_{1,0}-91 H_{1,1}\right)\nonumber 
   \\
   & +\frac{128}{3} (4 z-5) H_{2,0}+\frac{64}{3} \left(8 z^2+3
   z+6\right) H_{2,2}-\frac{32}{9} \left(16 z^2+27 z-3\right) H_{3,1}-\frac{64}{3} (13
   z-5) H_{-3,0,0}+\frac{512}{3} z^2 H_{-2,-1,0}\nonumber 
   \\
   & -\frac{32}{9} \left(88 z^2-99 z+87\right)
   H_{-2,0,0}+\frac{1}{27} \frac{1+z}{z} \bigg[ -192 \left(58 z^2-19 z+58\right) H_{-1}
   \zeta _2+64 \left(116 z^2-101 z+116\right) H_{-1,2}\nonumber 
   \\
   & -256 \left(29 z^2+22 z+29\right)
   H_{-1,-1,0}+32 \left(232 z^2+65 z+232\right) H_{-1,0,0}\bigg] -\frac{32}{27} \left(232
   z^2+261 z+336\right) H_{0,0,0}\nonumber 
   \\
   & -128 (1-z) H_{1,-2,0}-\frac{64}{9} (1-z) \left(27
   H_{1,2}+46 H_{1,0,0}+72 H_{1,1,0}\right)+\frac{1}{9} \frac{1}{1+z} \bigg[96 \left(8
   z^3+11 z^2+6 z+9\right) H_2 \zeta _2 \nonumber 
   \\
   & -96 \left(24 z^3+40 z^2+34 z+21\right) H_{0,0}
   \zeta _2-64 \left(44 z^3+26 z^2-24 z-15\right) H_0 \zeta _3+8 \left(152 z^3-421 z^2-606
   z-15\right) \zeta _4 \nonumber 
   \\
   &
   +96 \left(16 z^3+9 z^2+8 z+12\right) H_{3,0}+64 \left(8 z^3+5
   z^2-6\right) H_{2,0,0}+96 \left(32 z^3+35 z^2+12 z+15\right)
   H_{2,1,0}\bigg]  \nonumber 
   \\
   &
    -\frac{128}{3} (2 z-1) H_{3,0,0}-\frac{64}{3} (1-z) \bigg(2 H_{-3}
   \zeta _2+4 H_{-2,-1} \zeta _2-2 H_{-2} \zeta _3+2 H_{-3,2}-H_{-2,3}-2 H_{3,2}+8
   H_{-3,-1,0}\nonumber 
   \\
   & +2 H_{-2,-2,0}-2 H_{-2,-1,2}  +2 H_{-2,2,0}+5 H_{3,1,0}+4 H_{-2,-1,-1,0}+11
   H_{-2,-1,0,0}-8 H_{-2,0,0,0}\bigg)\nonumber 
   \\
   &  +\frac{1}{9} \frac{1+z}{z} \bigg[s  64 \left(4 z^2+11
   z+4\right) H_{-1} \zeta _3-96 \left(8 z^2+z+8\right) H_{-1,3}+64 \left(16 z^2+17
   z+16\right) H_{-1,-1,2} \nonumber 
   \\
   & +128 \left(2 z^2+z+2\right) \left(3 \zeta _2 H_{-1,0}+2
   H_{-1,2,1}\right)+128 \left(4 z^2-7 z+4\right) H_{-1,-1,-1,0}+96 \left(8 z^2-29
   z+8\right) H_{-1,-1,0,0}\nonumber 
   \\
   & -512 \left(z^2-4 z+1\right) H_{-1,0,0,0}\bigg] +\frac{64}{3}
   \left(8 z^2+3\right) H_{0,0,0,0}+\frac{1}{9} \frac{1-z}{z} \bigg[ -160 \left(16
   z^2+z+16\right) H_1 \zeta _3\nonumber 
   \\
   & -32 \left(80 z^2+77 z+80\right) \zeta _2 H_{1,0}-128
   \left(14 z^2+11 z+14\right) \zeta _2 H_{1,1}+32 \left(32 z^2+65 z+32\right) H_{1,3}\nonumber 
   \\
   & -96
   \left(16 z^2+19 z+16\right) H_{1,1,2}-192 \left(8 z^2-z+8\right) H_{1,2,0}+32 \left(32
   z^2-7 z+32\right) H_{1,0,0,0} \nonumber 
   \\
   &-32 \left(16 z^2+73 z+16\right) H_{1,1,0,0}-96 \left(32
   z^2+29 z+32\right) H_{1,1,1,0}\bigg]-\frac{128}{3} \left(3 H_{0,0,0,0,0}-13 \zeta _3
   H_{0,0}\right) \,, 
\end{align}

\begin{align}
I^*_{qq'}(z) &\, = C_F N_f T_F^2 \bigg\{ \frac{1}{27}\frac{1-z}{z} \bigg[ 16 \left(4 z^2+7 z+4\right)
   H_{1,1,1}-192 \left(2 z^2-z+2\right) \left(H_{1,2}-H_{1,0,0}+2 H_{1,1,0}+H_1 \zeta
   _2\right) \bigg] \nonumber \\
   &+\frac{1}{81}\frac{1-z}{z} \bigg[ 16 \left(38 z^2-79 z+38\right)
   H_{1,1}-576 \left(z^2-z+1\right) H_{1,0}\bigg]-\frac{32}{27} \left(6 z^2-5 z-5\right)
   \left(-H_{2,1}-2 H_0 \zeta _2+2 H_3\right)  \nonumber \\
   & -\frac{32}{81} \left(18 z^2+38 z+137\right)
   H_{0,0}+\frac{16}{27} \left(24 z^2-49 z-49\right) H_{0,0,0}+\frac{16}{9} (z+1)
   \biggl(-8 \zeta _2 H_{0,0}-4 H_{3,1}+2 H_{2,1,1} \nonumber \\
   & -10 H_{0,0,0,0}-4 H_0 \zeta _3+8 H_4-9
   \zeta _4\biggl)+\frac{32}{81}\frac{1-z}{z} H_1 \left(75 z^2+19
   z+75\right) -\frac{32}{81}\frac{1}{z} \left(39 z^3+10 z^2+64 z+18\right) \zeta
   _2  \nonumber \\
   & +\frac{32}{27}\frac{1}{z} \left(18 z^3-23 z^2+13 z-12\right) \zeta
   _3+\frac{16}{243} H_0 \left(534 z^2-337 z-661\right)+\frac{32}{81} H_2 \left(57 z^2-26
   z+100\right)  \nonumber \\
   & +\frac{32 (1-z) \left(1571 z^2-2689 z+1058\right)}{729 z}\bigg\} + C_F^2 T_F \bigg\{ -\frac{2 (1-z) \left(7512 z^2+56 z+2421\right)}{81 z}\nonumber \\
   & +\frac{2}{81}
   \left(5720 z^2-8308 z+2707\right) H_0-\frac{2 (1-z)}{81z} \left(7088 z^2-5805 z+4343\right)
    H_1-\frac{4}{81} \left(2882 z^2+75 z-1524\right) H_2 \nonumber \\
   &+\frac{8}{9}
   \left(104 z^2-55 z+229\right) H_3-\frac{16}{9} \left(40 z^2-27 z-24\right)
   H_4+\frac{4}{81 z } \left(2774 z^3+687 z^2-1380 z-648\right)  \zeta
   _2 \nonumber \\
   &-\frac{16}{9} \left(60 z^2-26 z+149\right) H_0 \zeta _2-\frac{16}{3} \left(4 z^2+7
   z+14\right) H_2 \zeta _2-\frac{4}{27 z } \left(1528 z^3-999 z^2-720 z-768\right) \zeta _3 \nonumber \\
   &+\frac{32}{9} \left(38 z^2+81 z+42\right) H_0 \zeta
   _3-\frac{4}{9z } \left(52 z^3+111 z^2+486 z-48\right) \zeta
   _4-\frac{2}{81} \left(5864 z^2+9273 z+2733\right) H_{0,0} \nonumber \\
   &+\frac{16}{9} \left(52 z^2-36
   z-15\right) \zeta _2 H_{0,0}+\frac{1}{81}\frac{1-z}{z} \bigg[ 8 \left(362 z^2+161
   z+101\right) H_{1,0}-4 \left(406 z^2+2557 z+361\right) H_{1,1}\bigg] \nonumber \\
   &+\frac{16}{27}
   \left(68 z^2+123 z-15\right) H_{2,0}-\frac{4}{27} \left(184 z^2+423 z+612\right)
   H_{2,1}+\frac{16}{3} (10 z+11) H_{2,2}-\frac{16}{9} \left(4 z^2+45 z+12\right)
   H_{3,0} \nonumber \\
   & +\frac{16}{9} \left(16 z^2+9 z-6\right) H_{3,1}+\frac{4}{9} \left(376 z^2-167
   z+257\right) H_{0,0,0}+\frac{1}{27}\frac{1-z}{z} \bigg[ 8 \left(8 z^2-493 z+8\right)
   H_1 \zeta _2 \nonumber \\
   &+16 \left(20 z^2+203 z+20\right) H_{1,2}-16 \left(26 z^2-160 z-1\right)
   H_{1,0,0}-8 \left(8 z^2+209 z+8\right) H_{1,1,0} \nonumber \\
   & +12 \left(40 z^2+47 z+4\right)
   H_{1,1,1} \bigg]+\frac{16}{3} (z-2) (4 z-5) H_{2,0,0}-\frac{16}{3} \left(4 z^2+2
   z+7\right) H_{2,1,0} \nonumber \\
   & +\frac{16}{9} \left(4 z^2+21 z+9\right) H_{2,1,1}-\frac{8}{9}
   \left(140 z^2-15 z+66\right) H_{0,0,0,0} +\frac{1}{9} \frac{1-z}{z} \bigg[     96 \left(4
   z^2-11 z+4\right) H_1 \zeta _3 \nonumber \\
   & -16 \left(20 z^2-19 z+20\right) H_{1,3}+64 \left(5 z^2+2
   z+5\right) H_{1,1,2}+16 \left(8 z^2-13 z+8\right) \left(2 \zeta _2
   H_{1,1}+H_{1,2,0}\right) \nonumber \\
   & -8 \left(52 z^2-17 z+52\right) H_{1,0,0,0}+32 \left(2 z^2+17
   z+2\right) H_{1,1,0,0}+16 \left(32 z^2-25 z+32\right) \left(\zeta _2
   H_{1,0}+H_{1,1,1,0} \right) \nonumber \\
   & -16 \left(4 z^2+7 z+4\right) H_{1,1,1,1}\bigg] +\frac{16}{3}
   (z+1) \biggl( 14 H_5-12 H_2 \zeta _3+7 \zeta _2 \zeta _3-14 H_0 \zeta _4-36 \zeta _5-18
   \zeta _3 H_{0,0}-2 \zeta _2 H_{2,0} \nonumber \\
   & -4 \zeta _2 H_{2,1}+2 H_{2,3}+4 H_{4,0}-6
   H_{4,1}-14 \zeta _2 H_{0,0,0}+4 H_{2,1,2}-2 H_{2,2,0}+4 H_{3,0,0}-2
   H_{3,1,0}-H_{3,1,1}-H_{2,0,0,0} \nonumber \\
   &+8 H_{2,1,0,0}-2 H_{2,1,1,0}-2 H_{2,1,1,1}+12
   H_{0,0,0,0,0}\biggl) \bigg\} +   C_A C_F T_F \bigg\{ -\frac{4 (1-z) \left(356788 z^2-36824 z+228133\right)}{729
   z} \nonumber \\
   &-\frac{2}{243z } \left(214384 z^3+34691 z^2+143678 z+17152\right)
   H_0+\frac{2(1-z)}{243 z } \left(15520 z^2-3245 z+14944\right)  H_1 \nonumber \\& +\frac{4}{27} \left(224 z^2-451 z-724\right) H_3+\frac{4}{9} \left(8 z^2-91 z-109\right)
   H_4-\frac{8}{3} \left(20 z^2-7 z+27\right) H_{-2} \zeta _2 \nonumber \\
   & +\frac{1}{81}  \frac{1}{z}
   \bigg[ 4 \left(1000 z^3+2969 z^2-2068 z+1256\right) H_2-8 \left(3932 z^3-2936 z^2+1099
   z-3002\right) \zeta _2\bigg]  \nonumber \\
   &-\frac{8}{3} (27 z+11) \zeta _2 \zeta _3+\frac{2}{3} (5
   z-201) H_0 \zeta _4+32 (9 z-1) \zeta _5+\frac{16}{3} (13 z-25) H_{-4,0}+\frac{8}{9}
   \left(112 z^2-99 z+87\right) H_{-3,0} \nonumber \\
   & -\frac{8}{27} \left(544 z^2+93 z+576\right)
   H_{-2,0}+\frac{8}{3} \left(16 z^2+z+27\right) H_{-2,2}+\frac{4}{27} \left(828 z^2-77
   z+906\right)\frac{z+1}{z} H_{-1,0} \nonumber \\
   & +\frac{4}{81} \left(16868 z^2+3176 z+14795\right)
   H_{0,0}-\frac{208}{3} (z+2) \zeta _3 H_{0,0}-20 (1-z) \zeta _2 H_{1,0} \nonumber \\
   & +\frac{1}{81}
  \frac{1-z}{z} \bigg[  4 \left(6272 z^2+272 z+4481\right) H_{1,0}+4 \left(184 z^2+2098
   z+463\right) H_{1,1}\bigg]  \nonumber \\
   & +\frac{1}{27}\frac{1}{z} \bigg[ 8 \left(254 z^3+731
   z^2+776 z+156\right) H_0 \zeta _2-4 \left(700 z^3-4100 z^2+2203 z-1620\right) \zeta
   _3 \nonumber \\
   & +24 \left(124 z^3+123 z^2+144 z+52\right) H_{2,0}-4 \left(40 z^3-488 z^2-371
   z-116\right) H_{2,1}    \bigg]+\frac{4}{9} \left(8 z^2+35 z+59\right)
   H_{3,1} \nonumber \\
   & -\frac{16}{3} (3 z+5) H_{3,2}-64 z H_{4,0}+\frac{8}{3} (29 z-37)
   H_{-3,0,0}-\frac{64}{3} (z-2) z H_{-2,-1,0}+\frac{4}{9} \left(184 z^2-315 z+159\right)
   H_{-2,0,0} \nonumber \\
   & +\frac{1}{27}\frac{z+1}{z} \bigg[      16 \left(95 z^2-35 z+68\right) H_{-1}
   \zeta _2-8 \left(116 z^2-101 z+116\right) H_{-1,2}+16 \left(74 z^2+31 z+20\right)
   H_{-1,-1,0} \nonumber \\
   & -4 \left(592 z^2-25 z+484\right) H_{-1,0,0}  \bigg]  +8 (z+1)
   H_{-1,2,0}-\frac{4}{27} \left(1758 z^2-587 z+1261\right) H_{0,0,0} \nonumber \\
   & +\frac{1}{27}
  \frac{1-z}{z} \bigg[    4 \left(748 z^2+475 z+748\right) H_1 \zeta _2+8 (2 z-5) (67 z-16)
   H_{1,2}-24 \left(249 z^2-2 z+249\right) H_{1,0,0} \nonumber \\
   & +16 \left(257 z^2+11 z+230\right)
   H_{1,1,0}-4 \left(76 z^2+103 z-32\right) H_{1,1,1}  \bigg]  \nonumber \\
   & +\frac{1}{9}\frac{1}{z (z+1)}
   \bigg[       12 \left(24 z^4+87 z^3+120 z^2+79 z+16\right) H_2 \zeta _2+8 \left(136 z^4+47
   z^3-136 z^2-14 z+24\right) H_0 \zeta _3 \nonumber \\
   & +\left(672 z^4+4935 z^3+4956 z^2+2551
   z+1840\right) \zeta _4-8 \left(28 z^4+109 z^3+141 z^2+141 z+72\right)
   H_{2,0,0}    \bigg] \nonumber \\
   &  +\frac{1}{9}\frac{1}{z+1} \bigg[   -4 \left(8 z^3-2 z^2-56 z-37\right)
   \zeta _2 H_{0,0}+12 \left(47 z^2+26 z-24\right) H_{3,0} \nonumber \\
   &  +4 \left(112 z^3+265 z^2+294
   z+159\right) H_{2,1,0} \bigg]+\frac{1}{9z } \bigg[   48 \left(4 z^3-7 z^2-6
   z-4\right) H_{2,2}-8 \left(4 z^3+35 z^2+11 z-4\right) H_{2,1,1} \bigg] \nonumber \\
   &  -\frac{16}{3}
   (11 z-6) H_{3,0,0}+\frac{8}{3} (3 z-11) H_{3,1,0}-\frac{8}{3} (1-z) \biggl(  H_5-2 H_{-3}
   \zeta _2+7 H_3 \zeta _2-2 H_{-2} \zeta _3-2 H_{-3,2}  -8 \zeta _2 H_{-2,0} \nonumber \\
   &     +H_{-2,3}-8
   H_{-3,-1,0}-10 H_{-2,-2,0}+2 H_{-2,-1,2}-2 H_{-2,2,0}+4 H_{-2,-1,-1,0}-23
   H_{-2,-1,0,0}+16 H_{-2,0,0,0}   \biggl) \nonumber \\
   &  +\frac{1}{9}\frac{z+1}{z} \bigg[      -24 \left(4
   z^2-z+4\right) H_{-1} \zeta _3-16 \left(14 z^2-11 z+14\right) \zeta _2 H_{-1,0}+12
   \left(8 z^2+z+8\right) H_{-1,3} \nonumber \\
   & -8 \left(16 z^2-37 z+16\right) H_{-1,-2,0}-8 \left(16
   z^2+17 z+16\right) H_{-1,-1,2}-16 \left(2 z^2+z+2\right) \left(2 H_{-1,2,1}-5 \zeta _2
   H_{-1,-1}\right) \nonumber \\
   & +16 \left(4 z^2-7 z+4\right) H_{-1,-1,-1,0}-36 \left(8 z^2-19
   z+8\right) H_{-1,-1,0,0}+96 (z-2) (2 z-1) H_{-1,0,0,0}    \bigg] \nonumber \\
   & +\frac{8}{9} (25 z+121)
   H_{0,0,0,0}+\frac{1}{9}\frac{1-z}{z} \bigg[     -32 \left(z^2-5 z+1\right) \zeta _2
   H_{1,1}-4 \left(64 z^2-5 z+64\right) H_{1,3} \nonumber \\
   & +16 \left(8 z^2+5 z+8\right) H_{1,-2,0}-4
   \left(32 z^2+29 z+32\right) H_{1,1,2}-24 \left(16 z^2+13 z+16\right) H_{1,2,0} \nonumber \\
   & +4
   \left(104 z^2+119 z+104\right) H_{1,0,0,0}-28 \left(16 z^2+19 z+16\right)
   H_{1,1,0,0}-4 \left(32 z^2-25 z+32\right) \left(H_{1,1,1,0}-H_1 \zeta _3\right) \nonumber \\
   & +16
   \left(4 z^2+7 z+4\right) H_{1,1,1,1}      \bigg] -\frac{8}{3} (z+1) \biggl(    H_2 \zeta _3+6
   H_{-1,0} \zeta _3+\frac{3}{2} H_{-1} \zeta _4+6 \zeta _2 H_{-1,2}+5 \zeta _2 H_{2,0}-4
   \zeta _2 H_{2,1}  \nonumber \\
   & +3 H_{2,3}+H_{4,1}-3 \zeta _2 H_{-1,0,0}-3 H_{-1,3,0}-\zeta _2
   H_{0,0,0}-4 H_{2,-2,0}+5 H_{2,1,2}+14 H_{2,2,0}-2 H_{3,1,1}-6 H_{-1,2,0,0} \nonumber \\
   & +6
   H_{-1,2,1,0}-19 H_{2,0,0,0}+21 H_{2,1,0,0}-H_{2,1,1,0}-4 H_{2,1,1,1}  \biggl)+16 (3 z-4)
   H_{0,0,0,0,0}\bigg\} \,. 
\end{align}

\begin{align}
I^{*}_{q\bar{q}} (z) &\,= N_f T_F  \bigg\{  \frac{1}{81}\frac{1}{z+1} \bigg[  448 \left(z^2-6 z+1\right) H_{-1,0}+8
   \left(23 z^2+54 z-193\right) H_{0,0}+16 \left(137 z^2-18 z-43\right) \zeta
   _2\bigg]  \nonumber \\
   & +\frac{1}{27}\frac{1}{z+1} \bigg[      32 \left(53 z^2+18 z+35\right) H_{-2,0}-32\left(7 z^2-6 z+7\right) H_{-1,2}-48 \left(13 z^2+6 z+13\right) H_{2,0}  \nonumber \\
   & -192 \left(7
   z^2+4 z+7\right) H_{-1,-1,0}+16 \left(49 z^2+18 z+49\right) H_{-1,0,0}-16 \left(43
   z^2+36 z+43\right) H_{0,0,0}  \nonumber \\
   & -64 H_{-1} \left(7 z^2+9 z+7\right) \zeta _2-16
   \left(5 z^2+18 z+23\right) \zeta _2  H_0 +16 H_3 \left(19 z^2+18
   z+19\right)     \bigg]  \nonumber \\
   & +\frac{16}{9}\frac{1}{z+1} \left(z^2+1\right) \biggl( -4 \zeta _2
   H_{-1,-1}+4 \zeta _2 H_{-1,0}-4 \zeta _2 H_{0,0}+20 H_{-3,0}-14 H_{-2,2}-8 H_{-1,3}+6
   H_{2,2}-9 H_{3,0}  \nonumber \\
   & -2 H_{3,1}-20 H_{-2,-1,0}+17 H_{-2,0,0}-20 H_{-1,-2,0}+12
   H_{-1,-1,2}+6 H_{-1,2,0}+4 H_{-1,2,1}-9 H_{2,0,0}+12 H_{2,1,0}  \nonumber \\
   & +16 H_{-1,-1,-1,0}-14
   H_{-1,-1,0,0}+11 H_{-1,0,0,0}-8 H_{0,0,0,0}+4 H_{-2} \zeta _2+6 H_2 \zeta _2+2 H_{-1}
   \zeta _3-H_0 \zeta _3+5 H_4+6 \zeta _4 \biggl )  \nonumber \\
   & +\frac{16}{3} (1-z) \left(2 H_{1,2}-3
   H_{1,0,0}+4 H_{1,1,0}+2 H_1 \zeta _2\right)-\frac{32}{9} (z+1) \left(H_{2,1}-\zeta
   _3\right)-\frac{64}{9} (1-z) \left(H_{1,0}+H_{1,1}\right)  \nonumber \\
   & -\frac{592}{27} 
   (1-z) H_1+\frac{32}{81} H_0(47 z+11)-\frac{16}{27} H_2*(29 z-7)+\frac{392
   (1-z)}{81} \bigg\} +       C_A \bigg\{  \frac{10832 (1-z)  }{27} H_1  \nonumber \\
   &  -\frac{8   (1-z)}{9} \left(17 H_{1,0}-70 H_{1,1}\right) -\frac{4   (1-z) }{3} \left(-47 H_1 \zeta _2+86 H_{1,2}+41 H_{1,0,0}+67 H_{1,1,0}\right) +\frac{16 (1-z)}{3} H_{-2,-1,-1,0}   \nonumber \\
   &  -\frac{16 (1-z) }{3} \biggl(  -22 H_1 \zeta _3-20 \zeta
   _2 H_{1,0}-9 \zeta _2 H_{1,1}+11 H_{1,3}+8 H_{1,-2,0}-13 H_{1,1,2}-13 H_{1,2,0}+11
   H_{1,0,0,0} \nonumber\\
   &  -H_{1,1,0,0}-20 H_{1,1,1,0} \biggl) +\frac{22292
   (1-z)}{81}-\frac{4}{81} \left(3065 z^2+2287 z-859\right)\frac{1}{z+1}
   H_0+\frac{4}{27} (1744 z+439) H_2 \nonumber\\
   &  -\frac{4}{81 z (z+1)} \left(162 z^4+1619 z^3-723 z^2+1457
   z+81\right) H_{-1,0}  \nonumber \\
   & +\frac{1}{81} \frac{1}{1+z} \bigg[  2 \left(324 z^3+6287 z^2+4590 z+2183\right) H_{0,0}-2 \left(324 z^3+9473 z^2+12294 z+6701\right) \zeta _2    \bigg]  \nonumber \\
   &  +\frac{280}{9} (z+1) H_{2,1}+\frac{1}{27}\frac{1}{z+1}  \bigg[     4
   \left(73 z^2+81 z-296\right) H_3+4 \left(1133 z^2+2142 z+1205\right) H_{-1} \zeta _2  \nonumber \\
   & -4
   \left(689 z^2+738 z-22\right) H_0 \zeta _2-6 \left(41 z^2+592 z+515\right) \zeta _3-4
   \left(1415 z^2+486 z+983\right) H_{-2,0}  \nonumber \\
   & -8 \left(259 z^2+822 z+259\right) H_{-1,2}+12
   \left(185 z^2+33 z+116\right) H_{2,0}+24 \left(205 z^2+166 z+229\right) H_{-1,-1,0}  \nonumber \\
   & -8
   \left(709 z^2+864 z+709\right) H_{-1,0,0}+4 \left(1307 z^2+972 z+470\right)
   H_{0,0,0}      \bigg] +\frac{1}{9}\frac{1}{z+1} \bigg[       -4 \left(193 z^2+186 z+103\right)
   H_4  \nonumber \\
   & -4 \left(533 z^2+513 z+41\right) H_{-2} \zeta _2-12 \left(z^2-28 z-3\right) H_2
   \zeta _2-4 \left(91 z^2+141 z+67\right) H_{-1} \zeta _3  \nonumber \\
   & +4 \left(35 z^2+21 z-19\right)
   H_0 \zeta _3+3 \left(145 z^2+249 z+279\right) \zeta _4-8 \left(95 z^2-39 z+59\right)
   H_{-3,0}  \nonumber \\
   & +8 \left(314 z^2+261 z+101\right) H_{-2,2}+4 \left(197 z^2+309 z+200\right)
   \zeta _2 H_{-1,-1}-4 \left(32 z^2+15 z+62\right) \zeta _2 H_{-1,0}  \nonumber \\
   & +64 \left(10 z^2+9
   z+10\right) H_{-1,3}+4 \left(134 z^2+129 z+65\right) \zeta _2 H_{0,0}-24 \left(53
   z^2+58 z+27\right) H_{2,2}  \nonumber \\
   & -12 \left(11 z^2+62 z-15\right) H_{3,0}+8 \left(47 z^2+48
   z+23\right) H_{3,1}+8 \left(95 z^2+9 z+161\right) H_{-2,-1,0}  \nonumber \\
   & -4 \left(31 z^2-168
   z+184\right) H_{-2,0,0}+8 \left(134 z^2+39 z+98\right) H_{-1,-2,0}-48 \left(23 z^2+24
   z+23\right) H_{-1,-1,2}  \nonumber \\
   & -24 \left(15 z^2+8 z+6\right) H_{-1,2,0}-16 \left(35 z^2+48
   z+35\right) H_{-1,2,1}  -12 \left(15 z^2+68 z+11\right) H_{2,0,0}  \nonumber \\
   & -12 \left(131 z^2+112
   z+51\right) H_{2,1,0}-8 \left(79 z^2-21 z+76\right) H_{-1,-1,-1,0}+44 \left(17 z^2+3
   z+14\right) H_{-1,-1,0,0}  \nonumber \\
   & -8 \left(59 z^2-30 z+50\right) H_{-1,0,0,0}+4 \left(67 z^2-72
   z+64\right) H_{0,0,0,0}             \bigg]   +\frac{1}{3}\frac{1}{z+1} \bigg[              -4 \left(23
   z^2+25\right) H_5  \nonumber \\
   & -16 \left(30 z^2+29\right) H_{-3} \zeta _2-8 \left(105 z^2+107\right)
   H_{-2} \zeta _3+16 \left(8 z^2+5\right) \zeta _2 \zeta _3+\left(197 z^2+231\right) H_0
   \zeta _4+8 \left(11 z^2+30\right) \zeta _5  \nonumber \\
   & +8 \left(13 z^2+7\right) H_{-4,0}+8 \left(71
   z^2+73\right) H_{-3,2}+32 \left(37 z^2+38\right) \zeta _2 H_{-2,-1}+4 \left(147
   z^2+145\right) H_{-2,3}  \nonumber \\
   & +4 \left(37 z^2+11\right) \zeta _3 H_{0,0}-8 \left(15
   z^2+17\right) H_{3,2}+16 \left(11 z^2+15\right) H_{-3,-1,0}+4 \left(59 z^2+41\right)
   H_{-3,0,0}  \nonumber \\
   & +8 \left(9 z^2+11\right) H_{-2,-2,0}-8 \left(149 z^2+151\right)
   H_{-2,-1,2}-8 \left(9 z^2+7\right) H_{-2,2,0}-4 \left(29 z^2+35\right) H_{3,0,0}  \nonumber \\
   &-4\left(21 z^2+11\right) \left(H_3 \zeta _2+H_{3,1,0}\right)-4 \left(133 z^2+111\right)
   H_{-2,-1,0,0}+8 \left(33 z^2+25\right) H_{-2,0,0,0} \nonumber \\
   & +8 \left(z^2+1\right) \biggl(44 H_2
   \zeta _3+113 H_{-1,-1} \zeta _3-22 H_{-1,0} \zeta _3-34 H_{-1} \zeta _4-94 \zeta _2
   H_{-2,0}+138 \zeta _2 H_{-1,-2}+2 \zeta _2 H_{-1,2}  \nonumber \\
   & +36 H_{-1,4}+40 \zeta _2 H_{2,0}+18
   \zeta _2 H_{2,1}-22 H_{2,3}-6 H_{4,0}+4 H_{4,1}-16 H_{-2,2,1}-20 H_{-1,-3,0}-150
   H_{-1,-2,2}  \nonumber \\
   & -168 \zeta _2 H_{-1,-1,-1}+95 \zeta _2 H_{-1,-1,0}-84 H_{-1,-1,3}-34 \zeta
   _2 H_{-1,0,0}+58 H_{-1,2,2}+31 H_{-1,3,0}-16 H_{-1,3,1}  \nonumber \\
   & +14 \zeta _2 H_{0,0,0}-16
   H_{2,-2,0}+26 H_{2,1,2}+26 H_{2,2,0}-24 H_{-1,-2,-1,0}-57 H_{-1,-2,0,0}-18
   H_{-1,-1,-2,0}  \nonumber \\
   & +168 H_{-1,-1,-1,2}+14 H_{-1,-1,2,0}+32 H_{-1,-1,2,1}+34 H_{-1,2,0,0}+56
   H_{-1,2,1,0}-22 H_{2,0,0,0}+2 H_{2,1,0,0}  \nonumber \\
   & +40 H_{2,1,1,0}+60 H_{-1,-1,-1,0,0}-38
   H_{-1,-1,0,0,0}+16 H_{-1,0,0,0,0} \biggl)-12 \left(5 z^2+3\right)
   H_{0,0,0,0,0}   \bigg] \bigg\}   \nonumber \\
   &    + C_F \bigg\{  -\frac{1532 (1-z) }{3} H_1 -80 (1-z) \left(H_{1,0}+H_{1,1}\right) -\frac{16 (1-z) }{3}
   H_{-2,2,0} +\frac{4 (1-z) }{3} \biggl(-129 H_1 \zeta _2+101 H_{1,2} \nonumber \\
   &+174 H_{1,0,0}+15
   H_{1,1,0}\biggl) +\frac{16(1-z)}{3} \biggl(-47 H_1 \zeta _3-38 \zeta _2 H_{1,0}-16
   \zeta _2 H_{1,1}+24 H_{1,3}+2 H_{1,-2,0}-19 H_{1,1,2}  \nonumber \\
   & -13 H_{1,2,0}+21 H_{1,0,0,0}+6
   H_{1,1,0,0}-32 H_{1,1,1,0}\biggl)+\frac{1294 (1-z)}{3}+\frac{4}{3} \left(638
   z^2+703 z+62\right)\frac{1}{z+1} H_0  \nonumber \\
   & -\frac{4}{3} (205 z+103) H_2+\frac{8}{3} \left(2
   z^4-27 z^3-65 z^2-29 z+1\right)\frac{1}{z (z+1)} H_{-1,0}  \nonumber \\
   & +\frac{1}{3}\frac{1}{z+1}
   \bigg[   2 \left(8 z^3+451 z^2+484 z+53\right) \zeta _2-2 \left(8 z^3+671 z^2+670
   z+19\right) H_{0,0}  \bigg]-\frac{4}{3} (56 z+17) H_{2,0}  \nonumber \\
   & -\frac{8}{3} (11 z+19)
   H_{2,1}-\frac{8}{3} (43 z+51) H_{-1,-1,0}+\frac{176}{3} (z+1) H_{-1,2,0}+\frac{1}{3}
  \frac{1}{z+1} \bigg[    -4 \left(85 z^2+87 z-30\right) H_3  \nonumber \\
   & -4 \left(141 z^2+354
   z+149\right) H_{-1} \zeta _2+4 \left(160 z^2+151 z-30\right) H_0 \zeta _2+2 \left(395
   z^2+368 z-123\right) \zeta _3  \nonumber \\
   & +4 \left(167 z^2+162 z+27\right) H_{-2,0}+8 \left(49
   z^2+130 z+49\right) H_{-1,2}+16 \left(45 z^2+82 z+45\right) H_{-1,0,0}  \nonumber \\
   & -4 \left(187
   z^2+177 z+33\right) H_{0,0,0}     \bigg] +\frac{16}{3} (13 z+25) H_{2,0,0}+\frac{8}{3} (9
   z+11) H_{-1,-1,-1,0}  \nonumber \\
   & +\frac{1}{3}\frac{1}{z+1} \bigg[       8 \left(43 z^2+50 z+28\right)
   H_4+4 \left(271 z^2+220 z+9\right) H_{-2} \zeta _2-4 \left(27 z^2+68 z+47\right) H_2
   \zeta _2  \nonumber \\
   & +8 \left(27 z^2+32 z+26\right) H_{-1} \zeta _3-4 \left(109 z^2+140 z+49\right)
   H_0 \zeta _3-2 \left(56 z^2+510 z+631\right) \zeta _4  \nonumber \\
   & -8 \left(13 z^2+22 z-18\right)
   H_{-3,0}-8 \left(135 z^2+118 z+28\right) H_{-2,2}-4 \left(99 z^2+124 z+97\right) \zeta
   _2 H_{-1,-1}  \nonumber \\
   & +4 \left(29 z^2+32 z+33\right) \zeta _2 H_{-1,0}-16 \left(11 z^2+10
   z+11\right) H_{-1,3}-4 \left(93 z^2+104 z+44\right) \zeta _2 H_{0,0}  \nonumber \\
   & +4 \left(125
   z^2+156 z+73\right) H_{2,2}+4 \left(29 z^2+44 z-12\right) H_{3,0}-8 \left(25 z^2+32
   z+13\right) H_{3,1}  \nonumber \\
   & +8 \left(z^2-16 z-47\right) H_{-2,-1,0}-4 \left(108 z^2+98
   z-61\right) H_{-2,0,0}-8 \left(37 z^2+56 z+37\right) H_{-1,-2,0}  \nonumber \\
   & +144 \left(3 z^2+4
   z+3\right) H_{-1,-1,2}+48 \left(5 z^2+8 z+5\right) H_{-1,2,1}+4 \left(131 z^2+128
   z+27\right) H_{2,1,0}  \nonumber \\
   & -4 \left(37 z^2+36 z+35\right) H_{-1,-1,0,0}+8 \left(13 z^2-4
   z+10\right) H_{-1,0,0,0}+12 \left(16 z^2+26 z-5\right) H_{0,0,0,0}         \bigg]  \nonumber \\
   & +\frac{1}{3}
  \frac{1}{z+1} \bigg[       8 \left(29 z^2+31\right) H_5+32 \left(34 z^2+33\right) H_{-3}
   \zeta _2+16 H_3 \zeta _2+8 \left(31 z^2+29\right) \zeta _2 \zeta _3-2 \left(353
   z^2+267\right) H_0 \zeta _4  \nonumber \\
   & -4 \left(313 z^2+231\right) \zeta _5-16 \left(29
   z^2+7\right) H_{-4,0}-32 \left(31 z^2+32\right) H_{-3,2}-16 \left(119 z^2+121\right)
   \zeta _2 H_{-2,-1}  \nonumber \\
   & +64 \left(22 z^2+21\right) \zeta _2 H_{-2,0}-16 \left(69
   z^2+68\right) H_{-2,3}-8 \left(57 z^2+31\right) \zeta _3 H_{0,0}+16 \left(15
   z^2+17\right) H_{3,2}  \nonumber \\
   & +16 \left(11 z^2+5\right) H_{4,0}-16 \left(59 z^2+38\right)
   H_{-3,0,0}+96 \left(2 z^2+1\right) H_{-2,-2,0}+32 \left(58 z^2+59\right)
   H_{-2,-1,2}  \nonumber \\
   & +56 \left(9 z^2+7\right) H_{3,0,0}+8 \left(11 z^2+13\right) H_{3,1,0}+272
   \left(5 z^2+4\right) H_{-2,-1,0,0}-24 \left(33 z^2+25\right) H_{-2,0,0,0}  \nonumber \\
   & -8
   \left(z^2+1\right) \biggl(    -169 H_{-2} \zeta _3+94 H_2 \zeta _3+177 H_{-1,-1} \zeta
   _3-90 H_{-1,0} \zeta _3-62 H_{-1} \zeta _4+234 \zeta _2 H_{-1,-2}-10 \zeta _2
   H_{-1,2}  \nonumber \\
   & +66 H_{-1,4}+76 \zeta _2 H_{2,0}+32 \zeta _2 H_{2,1}-48 H_{2,3}+12 H_{4,1}-28
   H_{-2,2,1}-56 H_{-1,-3,0}-234 H_{-1,-2,2}  \nonumber \\
   & -264 \zeta _2 H_{-1,-1,-1}+183 \zeta _2
   H_{-1,-1,0}-148 H_{-1,-1,3}-72 \zeta _2 H_{-1,0,0}+78 H_{-1,2,2}+37 H_{-1,3,0}-32
   H_{-1,3,1}  \nonumber \\
   & +32 \zeta _2 H_{0,0,0}-4 H_{2,-2,0}+38 H_{2,1,2}+26 H_{2,2,0}+12
   H_{-2,-1,-1,0}-155 H_{-1,-2,0,0}+6 H_{-1,-1,-2,0}  \nonumber \\
   & +264 H_{-1,-1,-1,2}+10
   H_{-1,-1,2,0}+48 H_{-1,-1,2,1}+76 H_{-1,2,0,0}   +64 H_{-1,2,1,0}-42 H_{2,0,0,0}-12
   H_{2,1,0,0} \nonumber \\
   & +64 H_{2,1,1,0}+156 H_{-1,-1,-1,0,0}-104 H_{-1,-1,0,0,0}+54
   H_{-1,0,0,0,0}\biggl)+8 \left(3 z^2+1\right) \left(8 H_{-3,-1,0}+9
   H_{0,0,0,0,0}\right)  \bigg] \bigg\} \,. 
\end{align}

\begin{align}
I^{*}_{qq}(z) &\,= \frac{1}{(1-z)_+} \bigg[   C_A C_F N_f
   T_F \left(-\frac{1648 \zeta _2}{81}-\frac{1808 \zeta
   _3}{27}+\frac{40 \zeta _4}{3}+\frac{125252}{729}\right) +   C_A^2
   C_F \biggl( -\frac{176}{3} \zeta _3 \zeta _2+\frac{6392 \zeta _2}{81}+\frac{12328 \zeta
   _3}{27} \nonumber \\
   & +\frac{154 \zeta _4}{3}-192 \zeta _5-\frac{297029}{729}\biggl) +  C_F N_f^2
   T_F^2 \left(-\frac{128 \zeta _3}{9}-\frac{7424}{729}\right)+ C_F^2 N_f
   T_F \left(-\frac{608 \zeta _3}{9}-32 \zeta _4+\frac{3422}{27}\right) \bigg] \nonumber \\
   & + C_F N_f^2 T_F^2 \bigg\{-\frac{32}{81}\frac{1}{1-z} \left(37 z^2-24 z+37\right)
   H_{0,0}-\frac{160}{27}\frac{1}{1-z} \left(z^2+1\right) H_{0,0,0}-\frac{32}{81}
  \frac{1}{1-z} H_0 \left(43 z^2-34 z+43\right)  \nonumber \\
  & +\frac{64}{9} (z+1) \zeta
   _3+\frac{64}{729} (157 z-41)\bigg\}    +   C_A C_F N_f T_F \bigg\{ \frac{1}{81}\frac{1}{1-z} \bigg[    8 \left(1322 z^2-882
   z+1247\right) H_{0,0}   \nonumber \\
   &  +144 \left(z^2-6 z+4\right) H_{1,0}-24 H_2 \left(z^2+27
   z-25\right)+8 \left(47 z^2-231 z+175\right) \zeta _2   \bigg] \nonumber \\
   &   +\frac{1}{27}
  \frac{1}{1-z} \bigg[    24 \left(30 z^2-27 z+37\right) H_{1,2}+24 \left(34 z^2-17
   z+26\right) H_{2,0}+16 \left(41 z^2+21 z+98\right) H_{0,0,0}  \nonumber \\
   &  +96 \left(4 z^2-13
   z+4\right) H_{1,0,0}+24 \left(16 z^2+15 z+9\right) H_{1,1,0}-8 H_0 \left(8 z^2+3
   z+20\right) \zeta _2+8 H_3 \left(16 z^2-33 z+28\right)  \nonumber \\
   &  +8 \left(87 z^2-51 z+391\right)
   \zeta _3      \bigg] +   \frac{1}{9}\frac{1}{1-z} \bigg[   -16 \left(2 z^2+3 z+5\right) \zeta _2
   H_{0,0}+16 \left(z^2+1\right) \biggl(-7 \zeta _2 H_{1,0}-4 \zeta _2 H_{1,1}  \nonumber \\
   &  +9
   H_{1,3}-H_{3,1}-4 H_{1,1,2}+3 H_{1,2,0}-3 H_{1,2,1}+H_{2,0,0}+8 H_{0,0,0,0}-8
   H_{1,0,0,0}+5 H_{1,1,0,0}-8 H_{1,1,1,0}-5 H_1 \zeta _3\biggl)  \nonumber \\
   &  +16 \left(z^2+3
   z+4\right) \left(H_{2,2}+H_4\right)+48 \left(4 z^2-z+3\right) H_{3,0}+16 \left(5 z^2-3
   z+2\right) H_{2,1,0}+16 H_2 \left(z^2-3 z-2\right) \zeta _2  \nonumber \\
   &  +16 H_0 \left(4 z^2+9
   z+13\right) \zeta _3+12 \left(21 z^2-14 z-3\right) \zeta _4  \bigg]  -\frac{16}{9} (z+1)
   H_{2,1}-\frac{8 H_{1,1}}{3}  \nonumber \\
   &  +\frac{80}{81}\frac{1}{1-z} H_0 \left(157 z^2-121
   z+168\right)+\frac{8}{9} H_1(18 z-25) \zeta _2+\frac{64}{27} H_1(5
   z+1)-\frac{346}{729} (451 z-89)\bigg\}     \nonumber \\
   & +        C_F^2 N_f T_F \bigg\{\frac{1}{81}\frac{1}{1-z} \bigg[    -16 \left(349 z^2-474 z+503\right)
   H_{0,0}-16 \left(35 z^2-312 z+53\right) H_{1,0}  \nonumber \\
   &  +8 H_2 \left(146 z^2+174
   z+191\right)-72 \left(36 z^2-70 z+41\right) \zeta _2  \bigg] +\frac{1}{27}
  \frac{1}{1-z} \bigg[        -96 \left(13 z^2-6 z+13\right) H_{1,2}  \nonumber \\
   &  -16 \left(88 z^2-72
   z+73\right) H_{2,0}-192 \left(3 z^2-z+3\right) H_{2,1}+8 \left(191 z^2-126
   z-169\right) H_{0,0,0}  \nonumber \\
   &  +16 \left(z^2+138 z+1\right) H_{1,0,0}-192 \left(3 z^2+4
   z+3\right) H_{1,1,0}-16 H_0 \left(16 z^2-24 z+37\right) \zeta _2+16 H_3 \left(8 z^2+12
   z+29\right)  \nonumber \\
   &  -16 \left(344 z^2-138 z+221\right) \zeta _3    \bigg]  +\frac{1}{9}
  \frac{1}{1-z} \bigg[    32 \left(z^2+1\right) \biggl(-2 \zeta _2 H_{0,0}+3 \zeta _2
   H_{1,0}+2 \zeta _2 H_{1,1}-5 H_{1,3}-4 H_{2,2}  \nonumber \\
   &  -7 H_{3,0}-2 H_{3,1}+6 H_{1,1,2}-5
   H_{1,2,0}+4 H_{1,2,1}+H_{2,0,0}-3 H_{2,1,0}+2 H_{2,1,1}+13 H_{1,0,0,0}-4 H_{1,1,0,0}  \nonumber \\
   &  +8
   H_{1,1,1,0}+2 H_2 \zeta _2-22 H_0 \zeta _3-H_1 \zeta _3+3 H_4 \biggl)+16 \left(17
   z^2-5\right) H_{0,0,0,0}-8 \left(59 z^2+23\right) \zeta _4  \bigg] +\frac{32}{9} z
   H_{1,1}  \nonumber \\
   &  -\frac{4}{81}\frac{1}{1-z} H_0 \left(1127 z^2-1628
   z+1470\right)+\frac{224}{9} H_1(1-z) \zeta _2-\frac{104}{27} H_1(z+2)+\frac{1}{81}
   (-3755 z-6511) \bigg\}  \nonumber \\
   & +           C_F^3 \bigg\{ 964 (1-z)+\frac{1}{3} \left(250 z^2-648 z+471\right)\frac{1}{1-z}
   H_0-\frac{2}{3} (1245 z-1204) H_1-\frac{32}{3} (10 z-7) H_{-2} \zeta _2 \nonumber \\
   & +64 z
   H_{-3,0}-64 H_{-2,2}+\frac{8}{3} \left(4 z^2+z+2\right) \frac{z+1}{z}
   H_{-1,0}+ \frac{1}{3}\frac{1}{1-z}  \bigg[     -2 \left(840 z^2-624 z-199\right) H_2 \nonumber \\
   & -2
   \left(16 z^3-502 z^2-36 z+553\right) \zeta _2+8 \left(4 z^3-95 z^2+93 z-27\right)
   H_{0,0}-16 \left(34 z^2-67 z+39\right) H_{1,0}    \bigg] \nonumber \\
   & -\frac{8}{3} (7 z-13)
   H_{1,1}-\frac{64}{3} (10 z-1) H_{-2,-1,0}-\frac{16}{3} (z+1) \left(-39 H_{-1} \zeta
   _2+16 H_{-1,2}-46 H_{-1,-1,0}+53 H_{-1,0,0}\right) \nonumber \\
   &  -\frac{32}{3} (19 z-17)
   H_{1,-2,0}+\frac{1}{3}\frac{1}{1-z} \bigg[   8 \left(89 z^2-130 z+29\right) H_3+8
   \left(z^2+66 z-29\right) H_0 \zeta _2 \nonumber \\
   &  +4 \left(97 z^2-128 z+23\right) H_1 \zeta _2-4
   \left(193 z^2-176 z-83\right) \zeta _3+8 \left(91 z^2-98 z-3\right) H_{-2,0} -24 \left(7 z^2-24 z+1\right) H_{2,0}  \nonumber \\ 
   & -4
   \left(161 z^2-264 z+95\right) H_{1,2}-4 \left(51
   z^2-60 z+5\right) H_{2,1}-4 \left(119 z^2-126 z+36\right) H_{0,0,0}  \nonumber \\
   & -4 \left(91 z^2-314
   z+91\right) H_{1,0,0}     \bigg]  +\frac{4}{3} (133 z-173) H_{1,1,0}+\frac{8}{3} (15 z-13)
   H_{1,1,2}+\frac{32}{3} (z-2) H_{1,2,1}  \nonumber \\
   &  +\frac{1}{3} \frac{1}{(1-z) (z+1)} \bigg[    8 \left(2 z^3-38
   z^2+7 z+23\right) H_2 \zeta _2+8 \left(22 z^3-6 z^2-31 z+9\right) H_{0,0} \zeta _2 \nonumber \\
   &  -8
   \left(38 z^3+2 z^2+z+61\right) H_0 \zeta _3+2 \left(118 z^3-62 z^2-523 z-367\right)
   \zeta _4+8 \left(58 z^3-12 z^2-73 z+9\right) H_{3,0} \nonumber \\
   &  +24 \left(35 z^3+13 z^2-23
   z+7\right) H_{2,0,0}+8 \left(53 z^3-23 z^2-77 z-25\right)
   H_{2,1,0}        \bigg] -\frac{32}{3} (z+1) \biggl( -5 H_{-1} \zeta _3 \nonumber \\
   &  +6 \zeta _2
   H_{-1,-1}-\zeta _2 H_{-1,0}+2 H_{-1,3}-4 H_{-1,-1,2}+4 H_{-1,-1,-1,0}+6
   H_{-1,0,0,0}\biggl) +8 (z+1) \biggl(-H_{-1} \zeta _4-4 \zeta _3 H_{-1,0} \nonumber \\
   &  -4 \zeta _2
   H_{-1,2}+2 \zeta _2 H_{-1,0,0}+2 H_{-1,3,0}+4 H_{-1,2,0,0}-4
   H_{-1,2,1,0}\biggl) +\frac{1}{3}\frac{1}{1-z} \bigg[    -8 \left(17 z^2-26 z-3\right)
   H_4 \nonumber \\
   &  -16 \left(15 z^2+28 z+11\right) H_1 \zeta _3-48 \left(4 z^2-2 z+5\right) \zeta _2
   H_{1,0}+8 \left(93 z^2-164 z+95\right) H_{1,3}+24 \left(7 z^2-2 z-6\right) H_{2,2} \nonumber \\
   &  +8
   \left(25 z^2-22 z+3\right) H_{3,1}-32 \left(21 z^2-26 z+11\right) H_{-2,0,0}+16
   \left(2 z^2-26 z+3\right) H_{1,2,0} \nonumber \\
   &  -12 \left(5 z^2-10 z-1\right) H_{0,0,0,0}-16
   \left(13 z^2-50 z+10\right) H_{1,0,0,0}       \bigg]-\frac{8}{3} (35 z-29)
   H_{1,1,0,0}+\frac{8}{3} (11 z-9) H_{1,1,1,0} \nonumber \\
   &  +\frac{1}{3}\frac{1}{1-z} \bigg[    256
   H_{-4,0} z^2-16 \left(7 z^2+5\right) H_5+16 \left(19 z^2-5\right) H_2 \zeta _3+16
   \left(47 z^2+41\right) \zeta _2 \zeta _3-4 \left(17 z^2+139\right) H_0 \zeta _4 \nonumber \\
   &  -8
   \left(355 z^2+381\right) \zeta _5+48 (3 z-1) (3 z+1) \zeta _3 H_{0,0}+16 \left(5
   z^2-13\right) \zeta _2 H_{2,0}+16 \left(z^2-5\right) \zeta _2 H_{2,1} \nonumber \\
   &  +16 \left(17
   z^2+23\right) H_{2,3}+32 \left(9 z^2+5\right) H_{3,2}+128 \left(2 z^2+1\right)
   H_{4,0}+32 \left(5 z^2+1\right) H_{-3,0,0}+32 \left(13 z^2+7\right) H_{2,1,2}\nonumber \\
   &  +32
   \left(23 z^2+15\right) H_{2,2,0}+32 \left(10 z^2+9\right) H_{3,0,0}+96 \left(3
   z^2+2\right) H_{3,1,0}-96 \left(2 z^2-1\right) H_{2,0,0,0} \nonumber \\
   &  +32 \left(19 z^2+9\right)
   H_{2,1,0,0}+16 \left(3 z^2+1\right) \left(-2 H_3 \zeta _2+H_{4,1}+9
   H_{2,1,1,0}\right)-8 \left(z^2+1\right) \biggl(14 H_{-3} \zeta _2+24 H_{-2,-1} \zeta
   _2 \nonumber \\
   &  -4 H_{-2,0} \zeta _2-20 H_{1,-2} \zeta _2-18 H_{1,2} \zeta _2-8 H_{0,0,0} \zeta
   _2+16 H_{1,1,0} \zeta _2-20 H_{-2} \zeta _3+9 H_1 \zeta _4+28 H_{-3,2}+8 H_{-2,3} \nonumber \\
   &  -68
   \zeta _3 H_{1,0}-36 \zeta _3 H_{1,1}+84 H_{-3,-1,0}-16 H_{-2,-1,2}+8 H_{1,-3,0}+24
   H_{1,-2,2}-96 H_{1,1,3}-42 H_{1,2,2}-76 H_{1,3,0} \nonumber \\
   &  -52 H_{1,3,1}+32 H_{2,-2,0}-32
   H_{2,2,1}-24 H_{3,1,1}+16 H_{-2,-1,-1,0}+24 H_{-2,0,0,0}+8 H_{1,-2,-1,0}+56
   H_{1,-2,0,0} \nonumber \\
   &  -40 H_{1,1,-2,0}-60 H_{1,1,2,0}-68 H_{1,2,0,0}-52 H_{1,2,1,0}-12
   H_{0,0,0,0,0}+56 H_{1,1,0,0,0}-48 H_{1,1,1,0,0}\biggl)    \bigg]\bigg\} \nonumber \\
   &  +  
   C_A^2 C_F \bigg\{  \frac{1136639 z-542581}{1458}+\frac{2}{81} \left(12863 z^2-12422
   z-15080\right) \frac{1}{1-z} H_0-\frac{2}{27} (5369 z-4526) H_1 \nonumber \\
   &   +\frac{80}{3} (z+2)
   H_{-2} \zeta _2-\frac{2}{9} (579 z-431) H_1 \zeta _2-\frac{32}{3} (7 z+4)
   H_{-2,2}+\frac{4}{3} \left(6 z^2+64 z+3\right) \frac{z+1}{z} H_{-1,0} \nonumber \\
   & +\frac{1}{81}
   \frac{1}{1-z} \bigg[    -6 \left(2512 z^2-909 z-1492\right) H_2-2 \left(324 z^3-8471 z^2+4821 z+3605\right) \zeta _2 \nonumber \\
   & +6 \left(108 z^3-6333 z^2+5026 z-2974\right) H_{0,0}+18
   \left(185 z^2-339 z+86\right) H_{1,0}    \bigg]+\frac{52}{3} z H_{1,1}+\frac{4}{9} (11
   z+14) H_{2,1} \nonumber \\
   & -\frac{32}{3} (9 z-2) H_{-2,-1,0}-\frac{160}{3} (1-z)
   H_{-2,0,0}-\frac{8}{3} (z+1) \left(-36 H_{-1} \zeta _2+14 H_{-1,2}-44 H_{-1,-1,0}+45
   H_{-1,0,0}\right) \nonumber \\
   & +\frac{1}{27} \frac{1}{1-z} \bigg[     -2 \left(281 z^2+492 z-100\right)
   H_3+2 \left(1225 z^2-264 z+448\right) H_0 \zeta _2+2 \left(2598 z^2-4506 z-2183\right)
   \zeta _3 \nonumber \\
   & +36 \left(127 z^2-136 z+27\right) H_{-2,0}-6 \left(651 z^2-642 z+527\right)
   H_{1,2}-6 \left(623 z^2-406 z+232\right) H_{2,0} \nonumber \\
   & -30 \left(149 z^2-64 z+103\right)
   H_{0,0,0}-12 \left(340 z^2-847 z+349\right) H_{1,0,0}-6 \left(245 z^2-18 z+309\right)
   H_{1,1,0}      \bigg]  \nonumber \\
   & +\frac{1}{9} \frac{1}{(1-z) (z+1)} \bigg[    8 \left(17 z^3-28 z^2+23 z+50\right)
   H_2 \zeta _2+4 \left(28 z^3-23 z^2+37 z+106\right) H_{0,0} \zeta _2 \nonumber \\
   & -4 \left(98 z^3+59
   z^2-202 z-127\right) H_0 \zeta _3-3 \left(115 z^3+680 z^2+313 z-240\right) \zeta _4-12
   \left(10 z^3+54 z^2+65 z+15\right) H_{3,0} \nonumber \\
   & +4 \left(115 z^3+124 z^2-110 z-83\right)
   H_{2,0,0}-4 \left(4 z^3+31 z^2+76 z+85\right) H_{2,1,0}   \bigg]-\frac{32}{3} (z+1) \biggl(    -6 H_{-1} \zeta _3 \nonumber \\
   & +\frac{15}{2} \zeta _2 H_{-1,-1}-3 \zeta _2 H_{-1,0}+3
   H_{-1,3}+2 H_{-1,-2,0}-7 H_{-1,-1,2}-2 H_{-1,2,0}-H_{-1,2,1}+H_{-1,-1,-1,0} \nonumber \\
   & -2
   H_{-1,-1,0,0}+4 H_{-1,0,0,0}  \biggl)+2 (z+1) \biggl(-H_{-1} \zeta _4-4 \zeta _3
   H_{-1,0}-4 \zeta _2 H_{-1,2}+2 \zeta _2 H_{-1,0,0}+2 H_{-1,3,0} \nonumber \\
   & +4 H_{-1,2,0,0}-4
   H_{-1,2,1,0}\biggl)+ \frac{1}{9} \frac{1}{1-z} \bigg[     -4 \left(59 z^2-39 z+68\right)
   H_4+4 \left(145 z^2-330 z+133\right) H_1 \zeta _3 \nonumber \\
   & -24 \left(18 z^2-25 z-8\right)
   H_{-3,0}-4 \left(13 z^2-66 z+43\right) \zeta _2 H_{1,0}+8 \left(40 z^2-27 z+31\right)
   \zeta _2 H_{1,1} \nonumber \\
   & +12 \left(31 z^2-113 z+40\right) H_{1,3}-8 \left(49 z^2-45 z+49\right)
   H_{2,2}+4 \left(56 z^2-39 z+23\right) H_{3,1} \nonumber \\
   & +48 (2 z-1) (11 z-14) H_{1,-2,0}-4
   \left(z^2-75 z-14\right) H_{1,1,2}-24 \left(9 z^2+3 z+2\right) H_{1,2,0}+12 \left(14
   z^2-z+9\right) H_{1,2,1} \nonumber \\
   & -4 \left(67 z^2+72 z+64\right) H_{0,0,0,0}+4 \left(19 z^2+84
   z+37\right) H_{1,0,0,0}-4 \left(109 z^2-120 z+121\right) H_{1,1,0,0}\nonumber \\
   & +352
   \left(z^2+1\right) H_{1,1,1,0}      \bigg] +\frac{1}{3} \frac{1}{1-z} \bigg[           
   64 \zeta _2
   H_{2,1} z^2-16 \left(z^2+5\right) H_5+8 \left(17 z^2+21\right) H_3 \zeta _2-20
   \left(11 z^2+13\right) H_2 \zeta _3 \nonumber \\
   & +4 \left(49 z^2+95\right) \zeta _2 \zeta _3-4
   \left(31 z^2+42\right) H_0 \zeta _4-4 \left(75 z^2-1\right) \zeta _5-32
   \left(z^2+2\right) H_{-4,0}-4 \left(37 z^2+11\right) \zeta _3 H_{0,0} \nonumber \\
   & +4 \left(35
   z^2+9\right) \zeta _2 H_{2,0}-4 \left(31 z^2+13\right) H_{2,3}+4 \left(z^2+7\right)
   H_{4,1}-16 \left(5 z^2+7\right) H_{-3,0,0}-16 \left(7 z^2+9\right) H_{2,-2,0} \nonumber \\
   & +4
   \left(z^2-17\right) \left(H_{2,1,2}-\zeta _2 H_{0,0,0}\right)+16 \left(2 z^2+1\right)
   H_{2,2,0}+4 \left(15 z^2+13\right) H_{3,0,0}+4 \left(33 z^2+35\right) H_{2,0,0,0} \nonumber \\
   & +4
   \left(49 z^2+31\right) H_{2,1,0,0}+4 \left(13 z^2-17\right) H_{2,1,1,0}+12 \left(5
   z^2+3\right) H_{0,0,0,0,0} -4 \left(z^2+1\right) \biggl(-40 H_{-3} \zeta _2 \nonumber \\
   & +60 H_{-2,-1}
   \zeta _2-24 H_{-2,0} \zeta _2-20 H_{1,-2} \zeta _2-40 H_{1,2} \zeta _2+2 H_{1,0,0}
   \zeta _2-12 H_{1,1,0} \zeta _2-16 H_{1,1,1} \zeta _2-48 H_{-2} \zeta _3 \nonumber \\
   & +21 H_1 \zeta
   _4+64 H_{-3,2}+24 H_{-2,3}+38 \zeta _3 H_{1,0}-12 \zeta _3 H_{1,1}+6 H_{1,4}+12
   H_{3,2}-12 H_{4,0}+48 H_{-3,-1,0} \nonumber \\
   & +16 H_{-2,-2,0}-56 H_{-2,-1,2}-16 H_{-2,2,0}-8
   H_{-2,2,1}+72 H_{1,-3,0}+8 H_{1,-2,2}-32 H_{1,1,3}+16 H_{1,2,2}-44 H_{1,3,0}\nonumber \\
   & -16
   H_{1,3,1}+2 H_{2,2,1}+6 H_{3,1,0}+8 H_{-2,-1,-1,0}-16 H_{-2,-1,0,0}+32 H_{-2,0,0,0}-24
   H_{1,-2,-1,0}+64 H_{1,-2,0,0} \nonumber \\
   & -40 H_{1,1,-2,0}+16 H_{1,1,1,2}-12 H_{1,1,2,0}+8
   H_{1,1,2,1}-30 H_{1,2,0,0}+2 H_{1,2,1,0}-32 H_{1,0,0,0,0}-28 H_{1,1,0,0,0} \nonumber \\
   & -16
   H_{1,1,1,0,0}\biggl)    \bigg]   \bigg\}   +
C_A C_F^2 \bigg\{ \frac{23060 (1-z)}{81}+\frac{1}{81} \left(-76790 z^2+68912 z+30003\right)
   \frac{1}{1-z} H_0  \nonumber \\
   & +\frac{4}{27} (7333 z-6709) H_1-144 H_{-2} \zeta _2+\frac{32}{3} (14
   z+11) H_{-2,2}-\frac{4}{3} (z+8) (16 z+1) \frac{z+1}{z} H_{-1,0}  \nonumber \\
   & +\frac{1}{81}
   \frac{1}{1-z} \bigg[     2 \left(23474 z^2-13731 z-15685\right) H_2+18 \left(96 z^3-2229
   z^2+841 z+1450\right) \zeta _2  \nonumber \\
   & -2 \left(864 z^3-29056 z^2+29454 z-12035\right)
   H_{0,0}-2 \left(1069 z^2+4155 z-704\right) H_{1,0}     \bigg]-\frac{4}{9} (91 z-15)
   H_{1,1}  \nonumber \\
   & +\frac{32}{3} (28 z-5) H_{-2,-1,0}+\frac{8}{3} (z+1) \left(-111 H_{-1} \zeta
   _2+44 H_{-1,2}-134 H_{-1,-1,0}+143 H_{-1,0,0}\right)  \nonumber \\
   & +\frac{1}{27} \frac{1}{1-z}
   \bigg[    -2 \left(1217 z^2-3354 z+1382\right) H_3-2 \left(2291 z^2+690 z-286\right) H_0
   \zeta _2  \nonumber \\
   & -6 \left(1295 z^2-2014 z+671\right) H_1 \zeta _2+2 \left(4376 z^2+6378
   z-3193\right) \zeta _3-36 \left(345 z^2-370 z+51\right) H_{-2,0}  \nonumber \\
   & +6 \left(1631 z^2-1686
   z+1079\right) H_{1,2}+2 \left(4387 z^2-4554 z+1348\right) H_{2,0}+12 \left(225 z^2-137
   z+168\right) H_{2,1}  \nonumber \\
   & +2 \left(3305 z^2-3024 z+2081\right) H_{0,0,0}+4 \left(1810
   z^2-6279 z+1837\right) H_{1,0,0}+6 \left(873 z^2-1010 z+1209\right)
   H_{1,1,0}    \bigg]  \nonumber \\
   & +\frac{1}{9} \frac{1}{(1-z) (z+1)} \bigg[     -4 \left(143 z^3-235 z^2+35
   z+269\right) H_2 \zeta _2-16 \left(22 z^3-35 z^2-35 z+40\right) H_{0,0} \zeta _2  \nonumber \\
   & +4
   \left(733 z^3+355 z^2-479 z+43\right) H_0 \zeta _3+\left(761 z^3+4331 z^2+3482
   z+56\right) \zeta _4  \nonumber \\
   & -4 \left(167 z^3-307 z^2-628 z-82\right) H_{3,0}-4 \left(568
   z^3+406 z^2-389 z-83\right) H_{2,0,0}  \nonumber \\
   & -12 \left(56 z^3-32 z^2-139 z-99\right)
   H_{2,1,0}      \bigg]   +\frac{16}{3} (z+1) \biggl(  -29 H_{-1} \zeta _3+36 \zeta _2 H_{-1,-1}-13
   \zeta _2 H_{-1,0}+14 H_{-1,3}  \nonumber \\
   & +8 H_{-1,-2,0}-32 H_{-1,-1,2}-8 H_{-1,2,0}-4 H_{-1,2,1}+8
   H_{-1,-1,-1,0}-8 H_{-1,-1,0,0}+22 H_{-1,0,0,0}\biggl)  \nonumber \\
   &  +\frac{8}{3} (z+1) \biggl(16 H_5+3
   H_{-1} \zeta _4+12 \zeta _3 H_{-1,0}+12 \zeta _2 H_{-1,2}+4 H_{3,2}-6 \zeta _2
   H_{-1,0,0}-6 H_{-1,3,0}-12 H_{-1,2,0,0} \nonumber \\
   & +12 H_{-1,2,1,0}  \biggl)+\frac{1}{9}
   \frac{1}{1-z} \bigg[       12 \left(45 z^2-66 z+16\right) H_4-4 \left(239 z^2-1008
   z+239\right) H_1 \zeta _3  \nonumber \\
   & +48 \left(24 z^2-31 z-8\right) H_{-3,0}+24 \left(17 z^2-19
   z+28\right) \zeta _2 H_{1,0}-16 \left(29 z^2-9 z+20\right) \zeta _2 H_{1,1}  \nonumber \\
   & -4
   \left(469 z^2-1116 z+517\right) H_{1,3}+8 \left(101 z^2-120 z+125\right) H_{2,2}-16
   \left(25 z^2-18 z-2\right) H_{3,1}  \nonumber \\
   & +48 \left(41 z^2-66 z+31\right) H_{-2,0,0}-144
   \left(21 z^2-38 z+15\right) H_{1,-2,0}+12 \left(z^2-54 z-11\right) H_{1,1,2}  \nonumber \\
   & +4
   \left(161 z^2+96 z+71\right) H_{1,2,0}-4 \left(91 z^2-12 z+61\right) H_{1,2,1}-176
   \left(z^2+1\right) H_{2,1,1}   \nonumber \\
   & -4 \left(157 z^2-162 z-52\right) H_{0,0,0,0}-8 \left(8
   z^2+288 z+35\right) H_{1,0,0,0}+8 \left(47 z^2-3 z+62\right) H_{1,1,0,0}  \nonumber \\
   & -8 \left(73
   z^2+33 z+70\right) H_{1,1,1,0}        \bigg]+\frac{1}{3} \frac{1}{1-z} \bigg[      -48 \left(5
   z^2+7\right) H_3 \zeta _2-24 \left(45 z^2+43\right) \zeta _3 \zeta _2-8 \left(23
   z^2-17\right) H_{2,0} \zeta _2  \nonumber \\
   & -16 \left(7 z^2-1\right) H_{2,1} \zeta _2+8 \left(79
   z^2+107\right) H_2 \zeta _3+4 \left(153 z^2+181\right) H_0 \zeta _4+8 \left(289
   z^2+325\right) \zeta _5-64 \left(z^2-2\right) H_{-4,0}  \nonumber \\
   & +16 \left(29 z^2+15\right) \zeta
   _3 H_{0,0}-8 \left(3 z^2+23\right) H_{2,3}-16 \left(17 z^2+11\right) H_{4,0}+16
   \left(5 z^2+13\right) H_{-3,0,0}  \nonumber \\
   & +32 \left(2 z^2-1\right) \left(H_{4,1}+3 \zeta _2
   H_{0,0,0}\right)+32 \left(11 z^2+13\right) H_{2,-2,0}-48 \left(5 z^2-1\right)
   H_{2,1,2}-8 \left(45 z^2+29\right) H_{2,2,0} \nonumber \\
   & -8 \left(43 z^2+35\right) H_{3,0,0}-8
   \left(27 z^2+47\right) H_{2,0,0,0}-352 \left(2 z^2+1\right) H_{2,1,0,0}-8 \left(35
   z^2-13\right) H_{2,1,1,0}  \nonumber \\
   & -8 \left(7 z^2+3\right) \left(H_{3,1,0}+6
   H_{0,0,0,0,0}\right)+8 \left(z^2+1\right) \biggl(-33 H_{-3} \zeta _2+72 H_{-2,-1} \zeta
   _2-26 H_{-2,0} \zeta _2-30 H_{1,-2} \zeta _2  \nonumber \\
   & -50 H_{1,2} \zeta _2+19 H_{1,0,0} \zeta
   _2+10 H_{1,1,0} \zeta _2-8 H_{1,1,1} \zeta _2-58 H_{-2} \zeta _3+\frac{111 H_1 \zeta
   _4}{2}+78 H_{-3,2}+28 H_{-2,3}+39 \zeta _3 H_{1,0} \nonumber \\
   & -12 \zeta _3 H_{1,1}-11 H_{1,4}+90
   H_{-3,-1,0}+16 H_{-2,-2,0}-64 H_{-2,-1,2}-16 H_{-2,2,0}-8 H_{-2,2,1}+76 H_{1,-3,0} \nonumber \\
   & +20
   H_{1,-2,2}-94 H_{1,1,3}-4 H_{1,2,2}-88 H_{1,3,0}-36 H_{1,3,1}-7 H_{2,2,1}+16
   H_{-2,-1,-1,0}-16 H_{-2,-1,0,0}  \nonumber \\
   & +44 H_{-2,0,0,0}-20 H_{1,-2,-1,0}+92 H_{1,-2,0,0}-60
   H_{1,1,-2,0}+8 H_{1,1,1,2}-36 H_{1,1,2,0}+4 H_{1,1,2,1}-67 H_{1,2,0,0}  \nonumber \\
   & -21
   H_{1,2,1,0}-50 H_{1,0,0,0,0}-44 H_{1,1,1,0,0}\biggl)    \bigg]   \bigg\}  \,.     
\end{align}

\begin{align}
I^{(3)}_{qg}(z) &\, = C_A N_f T_F^2 \bigg\{ \frac{1}{81} \frac{1}{z} \left(-24 \left(154 z^3-123 z^2+3
   z-40\right) H_{1,0}-8 \left(435 z^3-172 z^2-18 z-120\right) \zeta
   _2\right) \nonumber \\
   & +\frac{1}{27} \frac{1}{z} \bigg[   -16 \left(116 z^3-128 z^2+31 z-24\right)
   H_{1,0,0}+32 \left(26 z^3-29 z^2+4 z-6\right) H_{1,1,0} \nonumber \\
   & +16 H_1 \left(82 z^3-88 z^2+23
   z-12\right) \zeta _2+48 \left(30 z^3+7 z^2+18 z-4\right) \zeta _3    \bigg] \nonumber \\
   & +\frac{80}{9}
   \left(2 z^2+2 z+1\right) \left(2 H_{-1,-1,0}+H_{-1} \zeta _2\right)-\frac{32}{9}
   \left(2 z^2+2 z+1\right) \biggl( \frac{3}{2} \zeta _2 H_{-1,-1}-\frac{3}{2} \zeta _2
   H_{-1,0}+3 H_{-3,0} \nonumber \\
   &  -3 H_{-2,-1,0}+\frac{9}{2} H_{-2,0,0}-3 H_{-1,-2,0}+3
   H_{-1,-1,-1,0}-\frac{9}{2} H_{-1,-1,0,0}+4 H_{-1,0,0,0}-\frac{3}{2} H_{-2} \zeta
   _2-\frac{3}{2} H_{-1} \zeta _3 \biggl) \nonumber \\
   &  -\frac{16}{9} \left(2 z^2-2 z+1\right)
   \biggl(    \zeta _2 H_{1,0}+2 \zeta _2 H_{1,1}+2 H_{1,3}-8 H_{1,1,2}-2 H_{1,2,0}-5
   H_{1,2,1}-3 H_{1,0,0,0}+4 H_{1,1,0,0} \nonumber \\
   & -4 H_{1,1,1,0}+5 H_{1,1,1,1}-10 H_1 \zeta
   _3   \biggl)-\frac{16}{9} \left(20 z^2+21 z+10\right) H_{-2,0} -\frac{8}{81} \left(299
   z^2+290 z+85\right) H_{-1,0} \nonumber \\
   & -\frac{8}{81} \left(391 z^2+1407 z+32\right)
   H_{0,0}-\frac{8}{81} \left(290 z^2-326 z+85\right) H_{1,1}+\frac{8}{9} \left(8 z^2+65
   z+12\right) H_{2,0} \nonumber \\
   &  -\frac{16}{27} \left(62 z^2+62 z+25\right) H_{-1,0,0}+\frac{32}{27}
   \left(23 z^2-74 z-4\right) H_{0,0,0}-\frac{32}{27} \left(23 z^2-23 z+10\right)
   \left(H_{1,2}-H_{1,1,1}\right) \nonumber \\
   & +\frac{16}{3} \left(2 z^2+6 z+1\right)
   H_{2,0,0}+\frac{16}{9} z \left(12 H_{3,0}-12 H_{2,1,0}-12 H_2 \zeta _2-19 \zeta
   _4\right)-\frac{8}{9} z \left(5 H_3-2 H_{2,1}\right) \nonumber \\
   & -\frac{32}{9} (12 z-1)
   H_{0,0,0,0} +\frac{8}{243} \frac{1}{z} H_1 \left(668 z^3-1064 z^2+103
   z-36\right)+\frac{8}{9} H_0 \left(8 z^2+27 z+12\right) \zeta _2 \nonumber \\
   & +\frac{8}{243} 
   \left(2280 z^2-8119 z-523\right)H_0+\frac{16}{27} H_2 \left(3 z^2-23 z-3\right)+\frac{4
   \left(43324 z^3-24198 z^2-8697 z-1582\right)}{729 z}  \bigg\}  \nonumber \\
   &+       
   C_F N_f T_F^2 \bigg\{ -\frac{32}{9} \frac{1-z}{z} \left(22 z^2+13 z+4\right)
   \left(H_{1,1,0}+H_1 \zeta _2\right)+\frac{1}{81} \frac{1}{z} \bigg[   32 \left(382
   z^3-403 z^2+131 z-96\right) H_{1,0} \nonumber \\
   & +32 \left(253 z^3-274 z^2+89 z-96\right) \zeta
   _2     \bigg] +\frac{1}{27} \frac{1}{z} \bigg[ -16 \left(184 z^3-28 z^2-133 z-48\right)
   H_{1,0,0} \nonumber \\
   &  -32 \left(100 z^3-133 z^2+77 z+12\right) \zeta _3   \bigg] +\frac{32}{3} \left(4
   z^2+6 z+3\right) \left(\zeta _2 H_{0,0}+H_{3,0}+2 H_{2,0,0}-H_{2,1,0}-H_2 \zeta _2+H_0
   \zeta _3\right) \nonumber \\
   &  -\frac{4}{81} \left(7424 z^2+6559 z-6023\right) H_{0,0}+\frac{8}{81}
   \left(308 z^2-353 z+112\right) H_{1,1}+\frac{4}{27} \left(1344 z^2-1282 z+1805\right)
   H_{0,0,0} \nonumber \\
   &  -\frac{32}{27} \left(23 z^2-23 z+10\right)
   \left(H_{2,1}+H_{1,1,1}\right)-\frac{16}{9} \left(44 z^2+94 z-47\right)
   H_{0,0,0,0} \nonumber \\
   & +\frac{80}{9} \left(2 z^2-2 z+1\right) \left(H_{2,1,1}+2
   H_{1,0,0,0}+H_{1,1,1,1}\right)-\frac{32}{9} z (10 z-9) \left(H_{2,0}+H_0 \zeta
   _2\right)-64 (2 z-1) H_{0,0,0,0,0} \nonumber \\
   & +\frac{8}{243} H_0 \left(17172 z^2+7693
   z+20752\right)+\frac{8}{243} H_1 \left(268 z^2+155 z-355\right)+\frac{224}{81} 
   \left(11 z^2-11 z+4\right) H_2\nonumber \\
   & -\frac{8}{9} \left(104 z^2+106 z+73\right) \zeta
   _4-\frac{441320 z^3+989034 z^2-1294539 z-160736}{1458 z}  \bigg\} \nonumber \\
   &  +  \frac{1}{3} C_F^2 T_F \bigg\{ \frac{1}{8} \left(-5020 z^2+2406 z+2127\right)+\left(-708
   z^2-734 z+235\right) H_0-2 \left(354 z^2+191 z-505\right) H_1 \nonumber \\
   &  -4 \left(151 z^2-343
   z-70\right) H_2+4 \left(24 z^2-11 z+49\right) H_3-4 \left(50 z^2-136 z+5\right) H_4+8
   \left(194 z^2+18 z-35\right) \zeta _2 \nonumber \\
   &  +4 \left(57 z^2-33 z-49\right) H_0 \zeta _2-8
   \left(3 z^2-9 z-10\right) H_1 \zeta _2-16 \left(5 z^2+2 z-2\right) H_2 \zeta _2+64
   \left(7 z^2+2 z+1\right) H_3 \zeta _2 \nonumber \\
   &  +4 \left(757 z^2-758 z+317\right) \zeta _3-96
   (z+1) (2 z+1) H_{-1} \zeta _3+4 \left(398 z^2-344 z+57\right) H_0 \zeta _3 \nonumber \\
   &  -8 (1-z)
   (175 z-73) H_1 \zeta _3+32 \left(39 z^2-34 z+20\right) H_2 \zeta _3-64 \left(23 z^2-5
   z+11\right) \zeta _2 \zeta _3  \nonumber \\
   &  +2 \left(227 z^2+416 z+344\right) \zeta _4+4 \left(348
   z^2-118 z+109\right) H_0 \zeta _4-8 \left(84 z^2-454 z+63\right) \zeta _5+64 \left(16
   z^2+7\right) H_{-4,0} \nonumber \\
   & +16 \left(53 z^2+46 z+47\right) H_{-3,0}+8 \left(63 z^2+98
   z+135\right) H_{-2,0}-64 z (3 z+2) \zeta _2 H_{-2,0} \nonumber \\
   & +4 \left(251 z^2+634 z+395\right)
   H_{-1,0}-8 \left(21 z^2+2 z-13\right) \zeta _2 H_{-1,0}+\left(-1012 z^2-857
   z+15\right) H_{0,0} \nonumber \\
   & +4 \left(158 z^2-168 z+5\right) \zeta _2 H_{0,0}-4 \left(2 z^2-26
   z+77\right) H_{1,0}+16 \left(29 z^2-35 z+15\right) \zeta _2 H_{1,0} \nonumber \\
   & -2 \left(302
   z^2-353 z+110\right) H_{1,1}-8 \left(10 z^2-18 z+17\right) \zeta _2 H_{1,1}+16 \left(6
   z^2-9 z+2\right) H_{1,2} \nonumber \\
   & -8 \left(25 z^2-12 z-7\right) H_{1,3}-4 \left(4 z^2+108
   z-7\right) H_{2,0}+32 \left(9 z^2-10 z+6\right) \zeta _2 H_{2,0}+4 \left(27 z^2-47
   z+35\right) H_{2,1} \nonumber \\
   & -16 \left(6 z^2+6 z-1\right) \zeta _2 H_{2,1}+8 \left(13
   z^2+2\right) H_{2,2}-32 \left(3 z^2-8 z+3\right) H_{2,3}+8 \left(33 z^2-98 z+7\right)
   H_{3,0} \nonumber \\
   & -8 \left(6 z^2-8 z-3\right) H_{3,1}+64 \left(3 z^2-2 z+1\right) H_{3,2}-32
   \left(10 z^2+6 z+5\right) \left(H_{-3} \zeta _2+2 H_{-3,-1,0}\right) \nonumber \\
   & +32 \left(26
   z^2+18 z+11\right) H_{-3,0,0}-64 \left(8 z^2+4 z+3\right) H_{-2,-2,0}-8 \left(36
   z^2+38 z+21\right) \left(H_{-2} \zeta _2+2 H_{-2,-1,0}\right) \nonumber \\
   & +8 \left(112 z^2+138
   z+55\right) H_{-2,0,0}-112 (z+1) (7 z+3) H_{-1,-2,0}-4 \left(63 z^2+96 z+41\right)
   \left(H_{-1} \zeta _2+2 H_{-1,-1,0}\right) \nonumber \\
   & +4 \left(209 z^2+384 z+183\right)
   H_{-1,0,0}+192 z (z+1) H_{-1,2,0}+\left(-792 z^2-694 z+65\right) H_{0,0,0} \nonumber \\
   & -8 \left(4
   z^2-2 z+1\right) \left(13 H_5-43 \zeta _3 H_{0,0}+4 H_{4,0}-4 H_{4,1}-13 \zeta _2
   H_{0,0,0}\right)+32 \left(2 z^2-22 z+17\right) H_{1,-2,0} \nonumber \\
   & +4 (z+1) (11 z-21)
   H_{1,0,0}-8 \left(2 z^2-5 z-6\right) H_{1,1,0}+4 \left(27 z^2-38 z+15\right)
   H_{1,1,1}+8 \left(13 z^2-12 z+8\right) H_{1,1,2} \nonumber \\
   & +8 \left(57 z^2-68 z+23\right)
   H_{1,2,0}-8 \left(6 z^2-4 z-5\right) H_{1,2,1}+64 \left(2 z^2+3\right) H_{2,-2,0}+16
   \left(4 z^2-35 z+2\right) H_{2,0,0} \nonumber \\
   & -8 \left(19 z^2-13\right) H_{2,1,0}-4 (2 z-1) (8
   z+3) H_{2,1,1}+16 \left(2 z^2-6 z+5\right) H_{2,1,2}+32 \left(17 z^2-12 z+5\right)
   H_{2,2,0} \nonumber \\
   & +16 \left(16 z^2-14 z+7\right) H_{2,2,1}-32 \left(16 z^2+14 z+1\right)
   H_{3,0,0}+16 \left(48 z^2-2 z+13\right) H_{3,1,0} \nonumber \\
   & +16 \left(28 z^2-26 z+13\right)
   H_{3,1,1}+32 \left(8 z^2+6 z+3\right) \left(-H_{-2} \zeta _3+\zeta _2 H_{-2,-1}+2
   H_{-2,-1,-1,0}\right) \nonumber \\
   &  -32 \left(24 z^2+10 z+9\right) H_{-2,-1,0,0}+64 \left(3
   z^2+z+1\right) H_{-2,0,0,0}+96 (z+1) (3 z+1) \left(\zeta _2 H_{-1,-1}+2
   H_{-1,-1,-1,0}\right) \nonumber \\
   & -64 (z+1) (14 z+9) H_{-1,-1,0,0}+8 \left(61 z^2+82 z+27\right)
   H_{-1,0,0,0}-8 \left(94 z^2+21 z-4\right) H_{0,0,0,0} \nonumber \\
   & -8 \left(33 z^2-56 z+17\right)
   H_{1,0,0,0}+8 \left(8 z^2-50 z+45\right) H_{1,1,0,0}-8 \left(19 z^2-42 z+26\right)
   H_{1,1,1,0}\nonumber \\
   & -4 \left(16 z^2-20 z+1\right) H_{1,1,1,1}-16 \left(40 z^2-26 z+11\right)
   H_{2,0,0,0}+32 \left(4 z^2+2 z+1\right) H_{2,1,0,0} \nonumber \\
   & -16 \left(4 z^2+14 z-5\right)
   H_{2,1,1,0}+8 \left(60 z^2-62 z+31\right) H_{2,1,1,1}-16 \left(2 z^2+2 z+1\right)
   \biggl(      22 H_{-1} \zeta _4-10 \zeta _2 H_{-1,-2} \nonumber \\
   & -8 \zeta _3 H_{-1,-1}+4 \zeta _3
   H_{-1,0}+12 \zeta _2 H_{-1,2}+22 H_{-1,-3,0}+8 \zeta _2 H_{-1,-1,-1}-6 \zeta _2
   H_{-1,-1,0}-6 H_{-1,3,0} \nonumber \\
   & -20 H_{-1,-2,-1,0}+30 H_{-1,-2,0,0}-16 H_{-1,-1,-2,0}-22
   H_{-1,2,0,0}+12 H_{-1,2,1,0}+16 H_{-1,-1,-1,-1,0} \nonumber \\
   & -24 H_{-1,-1,-1,0,0}+6
   H_{-1,-1,0,0,0}-5 H_{-1,0,0,0,0}   \biggl)-16 \left(24 z^2+4 z+3\right) H_{0,0,0,0,0}-16
   \left(2 z^2-2 z+1\right) \biggl(    \nonumber \\
   & -25 H_1 \zeta _4-39 \zeta _3 H_{1,0}-39 \zeta _3
   H_{1,1}-2 \zeta _2 H_{1,2}+12 H_{1,4}-10 H_{1,-3,0}-12 \zeta _2 H_{1,0,0}-8 \zeta _2
   H_{1,1,0}+3 \zeta _2 H_{1,1,1} \nonumber \\
   &  +2 H_{1,1,3}-6 H_{1,2,2}-2 H_{1,3,0}-3 H_{1,3,1}+4
   H_{1,-2,0,0}-4 H_{1,1,-2,0}-H_{1,1,1,2}-17 H_{1,1,2,0}-7 H_{1,1,2,1} \nonumber \\
   & -5 H_{1,2,0,0}-13
   H_{1,2,1,0}-14 H_{1,2,1,1}+7 H_{1,0,0,0,0}+20 H_{1,1,0,0,0}-3
   H_{1,1,1,0,0}+H_{1,1,1,1,0}-15 H_{1,1,1,1,1}\biggl)\bigg\}  \nonumber \\ 
   & +  \frac{1}{27} C_A^2 T_F \bigg\{ 3 \frac{z+1}{z} \bigg[ 12 \left(60 z^2+9 z+8\right)
   H_{-1,3}-8 \left(112 z^2+53 z+16\right) H_{-1,-1,2}-32 \left(17 z^2+4 z+2\right)
   H_{-1,2,1} \bigg]  \nonumber \\ 
   &  +72 \left(8 z^2+74 z-41\right) H_{-4,0}+24 \left(817 z^2+180
   z+216\right) H_{-3,0}+72 \left(56 z^2+46 z+29\right) H_{-3,2} \nonumber \\ 
   & -4 \left(2426 z^2-30
   z-213\right) H_{-2,0}+72 \left(96 z^2+37 z+31\right) H_{-2,2}+36 \left(100 z^2+110
   z+49\right) H_{-2,3} \nonumber \\ 
   & +\frac{2}{3} \left(126201 z^2+29235 z+20278\right) H_{0,0}-36
   \left(28 z^2+66 z+29\right) H_{2,3}+12 \left(317 z^2-135 z+51\right) H_{3,1} \nonumber \\ 
   & -72
   \left(20 z^2+50 z+19\right) H_{3,2}+108 \left(8 z^2-14 z-3\right) H_{4,1}-144 \left(10
   z^2+22 z+3\right) H_{-3,-1,0} \nonumber \\ 
   & +36 \left(52 z^2+214 z-35\right) H_{-3,0,0}+72 \left(8
   z^2-26 z+13\right) H_{-2,-2,0}-72 \left(78 z^2+37 z+14\right) H_{-2,-1,0} \nonumber \\ 
   & -72 \left(76
   z^2+66 z+39\right) H_{-2,-1,2}+12 \left(1522 z^2+246 z+297\right) H_{-2,0,0}-72
   \left(8 z^2+18 z+3\right) H_{-2,2,0} \nonumber \\ 
   & +108 \left(4 z^2+6 z+3\right) H_{-1,3,0}-4
   \left(7385 z^2-1712 z+1679\right) H_{0,0,0}+144 \left(8 z^2+10 z+7\right)
   H_{2,-2,0} \nonumber \\ 
   & -36 \left(16 z^2-62 z+9\right) H_{2,1,2}+72 \left(24 z^2-62 z-1\right)
   H_{2,2,0}-72 \left(2 z^2-26 z+1\right) H_{2,2,1} \nonumber \\ 
   & +72 \left(6 z^2-100 z+17\right)
   H_{3,0,0}+36 \left(4 z^2-86 z-25\right) H_{3,1,0}-5184 z H_{4,0}+504 (z+1)
   H_{-1,2,0} \nonumber \\ 
   & +144 (8 z+3) H_{3,1,1}+\frac{1}{9} \frac{1}{z} \bigg[   -2 H_0 \left(747175
   z^3+62809 z^2+167620 z+17152\right) \nonumber \\ 
   &  -2 H_1 \left(30151 z^3+24401 z^2-50692
   z-6528\right)  \bigg] +4 H_3 \left(1460 z^2-483 z-321\right)-24 H_4 \left(137 z^2-48
   z+51\right) \nonumber \\ 
   & -36 H_5 \left(72 z^2-58 z+9\right)+\frac{2 \left(2153647 z^3-2175249
   z^2+450714 z-470494\right)}{27 z}                             -144 \left(33 z^2+34 z+16\right) H_{-3} \zeta _2 \nonumber \\ 
   & -36 \left(270 z^2+111 z+76\right) H_{-2}
   \zeta _2+36 \left(20 z^2+18 z-13\right) H_3 \zeta _2-36 \left(136 z^2+242 z+93\right)
   \zeta _3 \zeta _2 \nonumber \\ 
   & +72 \left(70 z^2+66 z+33\right) H_{-2,-1} \zeta _2-72 \left(50 z^2+90
   z+17\right) H_{-2,0} \zeta _2+36 \left(88 z^2-2 z+47\right) H_{2,0} \zeta _2 \nonumber \\ 
   & +288
   \left(8 z^2-z+6\right) H_{2,1} \zeta _2-108 \left(16 z^2+14 z+7\right) H_{-1,0,0}
   \zeta _2+324 (2 z-1) (4 z-1) H_{0,0,0} \zeta _2 \nonumber \\ 
   &  -144 \left(19 z^2+22 z+8\right) H_{-2}
   \zeta _3+108 \left(52 z^2+30 z+27\right) H_2 \zeta _3-36 \left(163 z^2+166 z+83\right)
   H_{-1} \zeta _4 \nonumber \\ 
   &  +9 \left(136 z^2-38 z-425\right) H_0 \zeta _4-36 \left(136 z^2-1206
   z+77\right) \zeta _5   -144 \left(25 z^2+28 z+14\right) \zeta _3 H_{-1,0} \nonumber \\ 
   & -144 (41 z+27)
   \zeta _3 H_{0,0}+\frac{1}{3} \frac{1}{z} \bigg[    -4 \left(98 z^3-12231 z^2-702
   z-1256\right) H_2  \nonumber \\ 
   & -2 \left(49619 z^3-58657 z^2+11376 z-12536\right) \zeta _2+2
   \left(23245 z^3+33484 z^2+17579 z+5436\right) H_{-1,0} \nonumber \\ 
   & -2 \left(43978 z^3-43401
   z^2+9435 z-9490\right) H_{1,0}+2 \left(6748 z^3-12541 z^2+7796 z-918\right)
   H_{1,1}    \bigg]  \nonumber \\ 
   & +\frac{1}{z (z+1)} \bigg[    2 \left(3233 z^4+7511 z^3+5946 z^2+2238
   z+624\right) H_0 \zeta _2 \nonumber \\ 
   & -2 \left(8577 z^4+1403 z^3-6989 z^2-1467 z-1544\right) H_1
   \zeta _2-2 \left(509 z^4-13207 z^3-8142 z^2+2178 z-3288\right) \zeta _3 \nonumber \\ 
   & +6 \left(2059
   z^4+2811 z^3+1118 z^2+556 z+208\right) H_{2,0}+4 \left(7486 z^4+123 z^3-6050 z^2-235
   z-1494\right) H_{1,0,0} \nonumber \\ 
   & -2 \left(8519 z^4-1361 z^3-6983 z^2+805 z-1984\right)
   H_{1,1,0}     \bigg] +\frac{1}{z} \bigg[   4 \left(851 z^3+1062 z^2+246 z+272\right) H_{-1}
   \zeta _2  \nonumber \\ 
   & -8 \left(422 z^3+393 z^2+96 z+116\right) H_{-1,2}-4 \left(92 z^3-1678 z^2+1463
   z-208\right) H_{1,2} \nonumber \\ 
   & -4 \left(445 z^3-861 z^2-111 z-116\right) H_{2,1}+8 \left(7
   z^3+276 z^2+54 z+40\right) H_{-1,-1,0} \nonumber \\ 
   & -4 \left(393 z^3+431 z^2+124 z+484\right)
   H_{-1,0,0}-4 \left(337 z^3-194 z^2+175 z+40\right) H_{1,1,1}    \bigg]  \nonumber \\ 
   & +3 \frac{1}{z (z+1)^2}
   \bigg[   12 \left(102 z^5+456 z^4+671 z^3+398 z^2+103 z+16\right) H_2 \zeta _2 \nonumber \\ 
   & +4
   \left(1757 z^5+2308 z^4-811 z^3-1470 z^2-42 z+48\right) H_0 \zeta _3 \nonumber \\ 
   & +\left(5103
   z^5+22925 z^4+31855 z^3+17187 z^2+5012 z+1840\right) \zeta _4 \nonumber \\ 
   & -4 \left(155 z^5+1756
   z^4+3167 z^3+1830 z^2+426 z+144\right) H_{2,0,0}    \bigg] \nonumber \\ 
   & +3 \frac{1}{(z+1)^2} \bigg[   4
   \left(436 z^4+473 z^3-323 z^2-321 z+30\right) \zeta _2 H_{0,0}+36 \left(19 z^4-37
   z^3-138 z^2-89 z-8\right) H_{3,0} \nonumber \\ 
   & +4 \left(436 z^4+1133 z^3+1024 z^2+393 z+84\right)
   H_{2,1,0}    \bigg]-864 \left(z^2+1\right) H_{-2,-1,-1,0} \nonumber \\ 
   & -36 \left(80 z^2+254 z+5\right)
   H_{-2,-1,0,0}+72 \left(10 z^2+78 z-7\right) H_{-2,0,0,0} \nonumber \\ 
   & +216 \left(8 z^2+10 z+5\right)
   \left(H_{-1,2,0,0}-\zeta _2 H_{-1,2}\right)+72 \left(4 z^2-2 z-1\right)
   H_{-1,2,1,0}-24 \left(108 z^2-345 z-55\right) H_{0,0,0,0} \nonumber \\ 
   & +3 \frac{1}{z} \bigg[     -12
   \left(58 z^3+72 z^2+33 z+8\right) H_{-1} \zeta _3-4 \left(555 z^3-307 z^2+38
   z-176\right) H_1 \zeta _3 \nonumber \\ 
   & +4 \left(298 z^3+414 z^2+189 z+40\right) \zeta _2 H_{-1,-1}-4
   \left(401 z^3+264 z^2-30 z+56\right) \zeta _2 H_{-1,0} \nonumber \\ 
   & -4 \left(287 z^3-389 z^2+97
   z-96\right) \zeta _2 H_{1,0}-8 \left(112 z^3-98 z^2-5 z-20\right) \zeta _2 H_{1,1} \nonumber \\ 
   & +4
   \left(624 z^3-761 z^2+217 z-112\right) H_{1,3}+12 \left(97 z^3-172 z^2-59 z-16\right)
   H_{2,2} \nonumber \\ 
   & -8 \left(205 z^3+195 z^2+39 z+16\right) H_{-1,-2,0}-16 \left(8 z^3+69 z^2-60
   z-8\right) H_{1,-2,0} \nonumber \\ 
   & -4 \left(471 z^3-539 z^2+220 z-64\right) H_{1,1,2}+4 \left(457
   z^3-352 z^2-25 z-48\right) H_{1,2,0} \nonumber \\ 
   & -4 \left(383 z^3-404 z^2+124 z-48\right)
   H_{1,2,1}-16 \left(53 z^3+6 z^2-2\right) H_{2,1,1}+8 \left(74 z^3+84 z^2+51 z+8\right)
   H_{-1,-1,-1,0} \nonumber \\ 
   & -12 \left(286 z^3+245 z^2+16 z+24\right) H_{-1,-1,0,0}+4 \left(647
   z^3+494 z^2+z+48\right) H_{-1,0,0,0} \nonumber \\ 
   & -4 \left(233 z^3-63 z^2-147 z-8\right)
   H_{1,0,0,0}+4 \left(845 z^3-685 z^2-4 z-112\right) H_{1,1,0,0} \nonumber \\ 
   & -4 \left(734 z^3-811
   z^2+233 z-112\right) H_{1,1,1,0}-4 \left(14 z^2-61 z-8\right) H_{1,1,1,1} \bigg]         -36
   \left(52 z^2-118 z-11\right) H_{2,0,0,0} \nonumber \\ 
   & -36 \left(20 z^2+198 z+43\right)
   H_{2,1,0,0}+36 \left(20 z^2+94 z+27\right) H_{2,1,1,0}+144 (4 z+1) H_{2,1,1,1} \nonumber \\ 
   & +72
   \left(2 z^2+2 z+1\right) \biggl(28 \zeta _2 H_{-1,-2}+4 \zeta _3 H_{-1,-1}+10
   H_{-1,4}-16 H_{-2,2,1}-24 H_{-1,-3,0}-24 H_{-1,-2,2} \nonumber \\ 
   & -18 \zeta _2 H_{-1,-1,-1}+18 \zeta
   _2 H_{-1,-1,0}-11 H_{-1,-1,3}+10 H_{-1,2,2}-10 H_{-1,3,1}+8 H_{-1,-2,-1,0}-16
   H_{-1,-2,0,0} \nonumber \\ 
   & +2 H_{-1,-1,-2,0}+18 H_{-1,-1,-1,2}+16 H_{-1,-1,2,1}+4 H_{-1,2,1,1}+7
   H_{-1,-1,-1,0,0}-3 H_{-1,-1,0,0,0}+13 H_{-1,0,0,0,0}\biggl) \nonumber \\ 
   & +432 (15 z-4)
   H_{0,0,0,0,0}-18 \left(2 z^2-2 z+1\right) \biggl(57 H_1 \zeta _4-24 \zeta _2
   H_{1,-2}-120 \zeta _3 H_{1,0}+172 \zeta _3 H_{1,1}-4 \zeta _2 H_{1,2} \nonumber \\ 
   & +4 H_{1,4}-56
   H_{1,-3,0}+16 H_{1,-2,2}-44 \zeta _2 H_{1,0,0}+32 \zeta _2 H_{1,1,0}+16 \zeta _2
   H_{1,1,1}-64 H_{1,1,3}-28 H_{1,2,2}-76 H_{1,3,0} \nonumber \\ 
   & -20 H_{1,3,1}-16 H_{1,-2,-1,0}+140
   H_{1,1,1,2}-8 H_{1,1,2,0}+80 H_{1,1,2,1}-116 H_{1,2,0,0}-4 H_{1,2,1,0}+40
   H_{1,2,1,1} \nonumber \\ 
   & +44 H_{1,0,0,0,0}+76 H_{1,1,0,0,0}-52 H_{1,1,1,0,0}+172 H_{1,1,1,1,0}-120
   H_{1,1,1,1,1}\biggl)     \bigg\}  \nonumber \\ 
   & +  \frac{1}{27} C_A C_F T_F \bigg\{ \frac{920124 z^3-57058 z^2-834293 z-116208}{72
   z}+\frac{1}{9} \left(-26970 z^2-32212 z-13165\right) H_0  \nonumber \\ 
   & +\frac{2}{9} \left(48075
   z^3+15479 z^2-60229 z-2285\right) \frac{1}{z} H_1-\frac{8}{3} \left(2672 z^2+3701
   z+2576\right) H_2 \nonumber \\ 
   & +12 \left(179 z^2-278 z+134\right) H_3-12 \left(631 z^2+78 z-6\right)
   H_4+144 \left(8 z^2+16 z+11\right) H_5 \nonumber \\ 
   & -144 \left(28 z^2-14 z+25\right) H_{-4,0}-72
   \left(63 z^2+84 z+139\right) H_{-3,0}+144 \left(44 z^2+10 z+23\right) H_{-3,2} \nonumber \\ 
   & +36
   \left(70 z^2+194 z-339\right) H_{-2,0}+216 \left(16 z^2+28 z+15\right) H_{-2,2}+144
   \left(47 z^2+30 z+23\right) H_{-2,3} \nonumber \\ 
   & +18 \left(125 z^2-824 z-929\right) H_{-1,0}+72
   \left(28 z^2+82 z+55\right) H_{-1,2}+216 (z+1) (13 z+10) H_{-1,3} \nonumber \\ 
   & +\frac{1}{3}
   \left(-47946 z^2+26249 z+1745\right) H_{0,0}+4 \left(694 z^2-881 z-2\right) H_{2,1}-72
   \left(14 z^2-7 z-18\right) H_{2,2} \nonumber \\ 
   & +144 \left(3 z^2-40 z+9\right) H_{2,3}+12 \left(172
   z^2+186 z+3\right) H_{3,1}+144 \left(4 z^2-14 z+1\right) H_{3,2}+576 (1-z) H_{4,0} \nonumber \\ 
   & +72
   \left(4 z^2-10 z-3\right) H_{4,1}-144 \left(10 z^2-10 z+1\right) H_{-3,-1,0}+144
   \left(25 z^2+8 z+4\right) H_{-3,0,0} \nonumber \\ 
   & +144 \left(8 z^2-22 z+3\right) H_{-2,-2,0}+216
   \left(z^2+20 z+6\right) H_{-2,-1,0}-288 \left(33 z^2+14 z+17\right) H_{-2,-1,2} \nonumber \\ 
   & -36
   \left(45 z^2+79\right) H_{-2,0,0}+144 \left(4 z^2+22 z+3\right) H_{-2,2,0}+1152 z
   H_{-2,2,1}+72 \left(27 z^2+92 z+53\right) H_{-1,-2,0} \nonumber \\ 
   & -72 \left(35 z^2+64 z+36\right)
   H_{-1,-1,0}-432 (z+1) (8 z+11) H_{-1,-1,2}+36 \left(27 z^2-98 z-133\right)
   H_{-1,0,0} \nonumber \\ 
   & +72 \left(5 z^2-18 z-14\right) H_{-1,2,0}+144 (z-7) (z+1) H_{-1,2,1}-432
   \left(3 z^2+2 z+1\right) H_{-1,3,0} \nonumber \\ 
   & +\left(1836 z^2+11222 z+299\right) H_{0,0,0}-288
   \left(9 z^2-34 z+22\right) H_{1,-2,0}+\frac{1}{z} \bigg[   4 \left(349 z^3-1290
   z^2+1056 z-16\right) H_{1,2} \nonumber \\ 
   & +8 \left(21 z^3+74 z^2+20 z+46\right) H_{1,1,1}   \bigg]-288
   \left(2 z^2-4 z+5\right) H_{2,-2,0}+12 \left(176 z^2+128 z-37\right) H_{2,1,1} \nonumber \\ 
   & +72
   \left(76 z^2-34 z+35\right) H_{2,1,2}-288 \left(2 z^2+7 z+1\right) H_{2,2,0}+72
   \left(18 z^2-46 z+11\right) H_{2,2,1} \nonumber \\ 
   & -72 \left(10 z^2-22 z+17\right) H_{3,0,0}-144
   \left(31 z^2+4 z+9\right) H_{3,1,0}-288 \left(2 z^2+3 z+2\right) H_{3,1,1} \nonumber \\ 
   &           -72 \left(98 z^2+10 z+47\right) H_{-3} \zeta _2-108 \left(31 z^2+36 z+24\right) H_{-2}
   \zeta _2-36 \left(91 z^2+228 z+146\right) H_{-1} \zeta _2 \nonumber \\ 
   & -432 \left(17 z^2-10
   z+5\right) H_3 \zeta _2+72 \left(128 z^2-6 z+77\right) \zeta _3 \zeta _2+216 \left(38
   z^2+18 z+21\right) H_{-2,-1} \zeta _2 \nonumber \\ 
   & -72 \left(102 z^2+54 z+59\right) H_{-2,0} \zeta
   _2+108 \left(31 z^2+60 z+34\right) H_{-1,-1} \zeta _2-36 \left(93 z^2+254 z+143\right)
   H_{-1,0} \zeta _2 \nonumber \\ 
   & +432 \left(8 z^2+6 z+3\right) H_{-1,2} \zeta _2-72 \left(8 z^2-86
   z+25\right) H_{2,0} \zeta _2-288 \left(z^2-11 z+6\right) H_{2,1} \zeta _2 \nonumber \\ 
   & -72 \left(52
   z^2+46 z+23\right) H_{-1,0,0} \zeta _2-144 \left(8 z^2+14 z+11\right) H_{0,0,0} \zeta
   _2-144 \left(42 z^2+18 z+23\right) H_{-2} \zeta _3 \nonumber \\ 
   & -72 \left(37 z^2+66 z+32\right)
   H_{-1} \zeta _3+72 \left(20 z^2-38 z-31\right) H_2 \zeta _3 \nonumber \\ 
   & +3 \left(-1601 z^5-8772
   z^4-15458 z^3-10812 z^2-2351 z+192\right) \frac{1}{z (z+1)^2} \zeta _4+18 \left(454 z^2+442
   z+221\right) H_{-1} \zeta _4  \nonumber \\ 
   &  -18 \left(148 z^2-130 z+167\right) H_0 \zeta _4+36
   \left(304 z^2-974 z-49\right) \zeta _5-72 \left(70 z^2+82 z+41\right) \zeta _3
   H_{-1,0} \nonumber \\ 
   & +72 \left(4 z^2-66 z-33\right) \zeta _3 H_{0,0}+\frac{1}{3} \frac{1}{z}
   \bigg[   2 \left(11111 z^3-7057 z^2+12248 z-1296\right) \zeta _2 \nonumber \\ 
   & -2 \left(6190 z^3-5051
   z^2-1007 z-404\right) H_{1,0}+4 \left(564 z^3+905 z^2-1894 z+661\right)
   H_{1,1}     \bigg] \nonumber \\ 
   & +\frac{1}{z+1} \bigg[ 2 \left(2239 z^3+1765 z^2+180 z+708\right)
   H_{2,0}-6 \left(367 z^3-1173 z^2-1254 z+268\right) H_0 \zeta _2\bigg] \nonumber \\ 
   & +\frac{1}{z (z+1)}
   \bigg[    -2 \left(59 z^4-1351 z^3+513 z^2+1999 z-32\right) H_1 \zeta _2 \nonumber \\ 
   & -2 \left(7677
   z^4-4303 z^3-16010 z^2-5458 z-1536\right) \zeta _3-4 \left(710 z^4-563 z^3-575 z^2+736
   z+92\right) H_{1,0,0} \nonumber \\ 
   & +2 \left(2087 z^4-763 z^3-1863 z^2+655 z-224\right)
   H_{1,1,0}     \bigg]   \nonumber \\ 
   &   +3 \frac{1-z}{z} \bigg[ 4 \left(7 z^2+49 z+16\right) \zeta _2
   H_{1,1}-96 \left(17 z^2+2\right) H_{1,2,1}\bigg]  \nonumber \\ 
   &  +3 \frac{1}{(z+1)^2} \bigg[   12 \left(37
   z^4+36 z^3-72 z^2-104 z-39\right) H_2 \zeta _2+4 \left(388 z^4+1034 z^3+889 z^2+228
   z-6\right) H_{0,0} \zeta _2  \nonumber \\ 
   &   +4 \left(1132 z^4+3764 z^3+4405 z^2+2046 z+255\right) H_0
   \zeta _3-4 \left(613 z^4+494 z^3-758 z^2-546 z+84\right) H_{3,0}  \nonumber \\ 
   &  -12 \left(199 z^4+282
   z^3+45 z^2+40 z+72\right) H_{2,0,0}+12 \left(57 z^4+76 z^3-24 z^2-48 z-11\right)
   H_{2,1,0}       \bigg]  \nonumber \\ 
   &  -144 \left(18 z^2+2 z+5\right) H_{-2,-1,-1,0}-72 \left(64 z^2+70
   z+27\right) H_{-2,-1,0,0}+72 \left(90 z^2+70 z+37\right) H_{-2,0,0,0}  \nonumber \\ 
   &  -216 \left(z^2+16
   z+10\right) H_{-1,-1,-1,0}+36 (5 z+8) (9 z+10) H_{-1,-1,0,0}-36 \left(13 z^2-26
   z-9\right) H_{-1,0,0,0}  \nonumber \\ 
   &  -288 \left(7 z^2+4 z+2\right) H_{-1,2,0,0}+864 \left(5 z^2+4
   z+2\right) H_{-1,2,1,0}-12 \left(948 z^2-230 z+121\right) H_{0,0,0,0}  \nonumber \\ 
   &  +3 \frac{1}{z}
   \bigg[      -24 \left(88 z^3-187 z^2+97 z+8\right) H_1 \zeta _3-4 \left(617 z^3-492 z^2+159
   z-32\right) \zeta _2 H_{1,0}  \nonumber \\ 
   &  +4 \left(329 z^3-72 z^2-81 z-32\right) H_{1,3}+8 \left(17
   z^3-87 z^2+69 z-8\right) H_{1,1,2}-4 \left(173 z^3-204 z^2+114 z+16\right) H_{1,2,0}  \nonumber \\ 
   &  -4
   \left(203 z^3-340 z^2+239 z+8\right) H_{1,0,0,0}-4 \left(373 z^3-522 z^2+174
   z-16\right) H_{1,1,0,0}  \nonumber \\ 
   &  +4 \left(277 z^3-552 z^2+273 z-16\right) H_{1,1,1,0}+8 \left(24
   z^3-23 z^2-29 z-4\right) H_{1,1,1,1}       \bigg]+144 \left(7 z^2-24 z-2\right)
   H_{2,0,0,0}  \nonumber \\ 
   &  -144 \left(7 z^2-14 z-3\right) H_{2,1,0,0}+144 \left(47 z^2-14 z+10\right)
   H_{2,1,1,0}-216 \left(20 z^2-18 z+11\right) H_{2,1,1,1}  \nonumber \\ 
   &  +72 \left(2 z^2+2 z+1\right)
   \biggl(48 \zeta _2 H_{-1,-2}+44 \zeta _3 H_{-1,-1}+20 H_{-1,4}+32 H_{-1,-3,0}-60
   H_{-1,-2,2}-62 \zeta _2 H_{-1,-1,-1}  \nonumber \\ 
   &  +58 \zeta _2 H_{-1,-1,0}-53 H_{-1,-1,3}+10
   H_{-1,2,2}-6 H_{-1,3,1}-24 H_{-1,-2,-1,0}-22 H_{-1,-2,0,0}-10 H_{-1,-1,-2,0}  \nonumber \\ 
   &  +78
   H_{-1,-1,-1,2}-4 H_{-1,-1,2,0}-4 H_{-1,2,1,1}+32 H_{-1,-1,-1,-1,0}+41
   H_{-1,-1,-1,0,0}-51 H_{-1,-1,0,0,0}  \nonumber \\ 
   &  +15 H_{-1,0,0,0,0}\biggl)+288 (8 z+3)
   H_{0,0,0,0,0}+18 \left(2 z^2-2 z+1\right) \biggl(H_1 \zeta _4-24 \zeta _2 H_{1,-2}-80
   \zeta _3 H_{1,0}-20 \zeta _3 H_{1,1}  \nonumber \\ 
   &  -68 \zeta _2 H_{1,2}+44 H_{1,4}-8 H_{1,-3,0}-16
   H_{1,-2,2}-68 \zeta _2 H_{1,0,0}-32 \zeta _2 H_{1,1,0}+40 \zeta _2 H_{1,1,1}+24
   H_{1,1,3}+44 H_{1,2,2}  \nonumber \\ 
   &  -52 H_{1,3,0}+52 H_{1,3,1}-80 H_{1,-2,-1,0}+48 H_{1,-2,0,0}-32
   H_{1,1,-2,0}+132 H_{1,1,1,2}-24 H_{1,1,2,0}+24 H_{1,1,2,1} \nonumber \\ 
   &  -92 H_{1,2,0,0}+12
   H_{1,2,1,0}-72 H_{1,2,1,1}-36 H_{1,0,0,0,0}+68 H_{1,1,0,0,0}-28 H_{1,1,1,0,0} \nonumber \\ 
   &  +180
   H_{1,1,1,1,0}-240 H_{1,1,1,1,1}\biggl)     \bigg\}  \,.               
\end{align}

\subsection{Numerical fitting of the analytical results}
\label{sec:numer}

Eq.~\eqref{eq:twoloopCoef} of appendix.~\ref{sec:beam} contains Harmonic Polylogarithms up to weight $5$. To facilitate straightforward numerical implementation, we provide in this appendix a numerical fitting for the coefficient functions. Following Ref.~\cite{Moch:2017uml}, we use the following elementary functions to fit the results,
\begin{align}
\label{eq:Lzdefinition}
L_z \equiv \ln z\,,\,L_{\bar{z}} \equiv \ln (1-z )\,, \, \bar{z}\equiv 1-z \,. 
\end{align}
For two loop and three loop coefficient functions, we fit the exact results in the region $10^{-6} < z <1-10^{-6}$ (Numerical evaluation of HPLs are made with the \texttt{HPL} package~\cite{Maitre:2005uu}), and we have set the color factor to numerical values in QCD, i.e. 
\begin{align}
C_F = \frac{4}{3}\,, C_A = 3\,, T_f = \frac{1}{2} \,. 
\end{align}
In more detail, we subtract the $z \to 0$ and $z \to 1$ limits up to next-to-next-to-leading power ($z^1 $ and $(1-z)^1$). Then we fit the remaining terms in the region $10^{-6} < z <1-10^{-6}$. Combining the two parts, the fitted results can achieve an accuracy better than $10^{-3}$ for $0<z<1$. We shown below the numerical fitting with six significant digits. The full numerical fitting is attached as ancillary files with the arXiv submission. The two loop scale independent coefficient functions are given by
\begin{align}
\label{eq:TwoloopNumerical}
I^{(2)}_{qq'}(z) = &  -4.56035\, +z^3 \left(-0.0170296 L_z^3-0.143469 L_z^2+1.21562
   L_z-3.60403\right) +z^2 \big(0.000306717 L_z^3-1.77306 L_z^2\nonumber  \\ 
   &+5.65477
   L_z+3.57046\big)  +z \left(0.444444 L_z^3 -0.666667 L_z^2-5.33333
   L_z+1.89369\right)+0.444444 L_z^3  \nonumber \\
   &-0.666667 L_z^2+2.66667
   L_z -0.00131644 z^5+0.0563783 z^4+1.33333 \bar{z}+\frac{2.64517}{z} \,,   \nonumber \\
  I^{(2)}_{q\bar{q}}(z) = & I^{(2)}_{qq'}(z)+ z^3 \left(23.8756 L_z^3-68.5281 L_z^2+391.31 L_z-479.112\right)+z^2
   \big(1.73989 L_z^3+29.5744 L_z^2+207.751 L_z \nonumber \\
   &+533.913\big)+z
   \left(-0.148148 L_z^3-0.888889 L_z^2-1.33333
   L_z+6.98894\right)+0.148148 L_z^3-1.33333 L_z\nonumber \\
   &+1.66959 z^5-60.5553
   z^4-0.444444 \bar{z}-2.90423 \,, \nonumber \\
  I^{(2)}_{qq}(z) = & I^{(2)}_{qq'}(z)+  \left(5.53086 N_f+14.9267\right) \frac{1}{(\bar{z})_+}+N_f
   \bigg\{z^3 \left(-0.0532042 L_z^2+1.92031 L_z-4.39249\right) \nonumber \\
   &+z^2
   \left(0.939547 L_z^2+3.50359 L_z+1.65854\right)+z \left(0.444444
   L_z^2+1.48148 L_z+2.36991\right)+0.444444 L_z^2\nonumber \\
   &+1.48148
   L_z+0.0178399 z^5-0.246399 z^4+4.24691 \bar{z}-7.90123 \bigg\}+z^3
   \big(3.49597 L_z^3-18.6432 L_z^2+60.163 L_z\nonumber \\
   &-48.6244\big)+z^2
   \left(-2.49636 L_z^3-11.0306 L_z^2-7.51243
   L_z+59.7912\right)+\left(\bar{z}\right)^3 \left(3.19726 L_{\bar{z}}-1.32635
   L_{\bar{z}}^2\right)\nonumber \\
   &-7.11111 L_{\bar{z}}^2+\left(\bar{z}\right)^2 \left(13.4628
   L_{\bar{z}}-2.37726 L_{\bar{z}}^2\right)+22.2222 L_{\bar{z}}+\bar{z} \left(3.55556
   L_{\bar{z}}^2-17.7778 L_{\bar{z}}-0.105144\right)\nonumber \\
   &+z \left(-0.740741 L_z^3-10.
   L_z^2-11.5556 L_z+1.87655\right)-0.740741 L_z^3-2. L_z^2-8.
   L_z\nonumber \\
   &+0.070919 z^5-2.2589 z^4-10.2974 \,, \nonumber \\
   I^{(2)}_{qg}(z) = & -52.3982\, +z^3 \left(-0.5403 L_z^3+15.2935 L_z^2+21.5425
   L_z-103.137\right)+z^2 \big(-1.00863 L_z^3-20.4656 L_z^2\nonumber \\
   &-24.7319
   L_z+200.419\big)+0.555556 L_{\bar{z}}^3+\left(\bar{z}\right)^3
   \left(-2.91634 L_{\bar{z}}^3+2.52056 L_{\bar{z}}^2-54.8176
   L_{\bar{z}}\right)\nonumber \\
   &+\left(\bar{z}\right)^2 \left(0.982654 L_{\bar{z}}^3+2.72223
   L_{\bar{z}}^2-15.0644 L_{\bar{z}}\right)-1.66667 L_{\bar{z}}+\bar{z} \big(-1.11111
   L_{\bar{z}}^3-4.66667 L_{\bar{z}}^2+5. L_{\bar{z}}\nonumber \\
   &+58.9092\big)+z \left(2.44444
   L_z^3+11.6667 L_z^2+6.66667 L_z-53.6197\right)+0.777778
   L_z^3-1.16667 L_z^2\nonumber \\
   &+11.3333 L_z+4.0827 z^5-16.7693
   z^4+\frac{5.95164}{z} \,.
\end{align}

The fitted results for the three-loop coefficient functions in Eq.~\eqref{eq:Idecomp} are 
\begin{align}
\label{eq:ThreeLoopCoef}
I_{d33}(z)&\, = 32. \bigg\{  -z^3 \left(-19156.8 L_z^5+199731.
   L_z^4-2.18614\times 10^6 L_z^3+1.2083\times 10^7
   L_z^2-4.5734\times 10^7 L_z+7.52147\times
   10^7\right) \nonumber 
   \\& - z^2 \left(-538.817 L_z^5-25014.2
   L_z^4-500678. L_z^3-5.37744\times 10^6
   L_z^2-3.09137\times 10^7 L_z-7.59345\times
   10^7\right)\nonumber 
   \\& - \left(\bar{z}\right)^3 \left(1023.79
   L_{\bar{z}}^2+121.796 L_{\bar{z}}\right)- \left(\bar{z}\right)^2
   \left(-93.2504 L_{\bar{z}}^2-876.881 L_{\bar{z}}\right)\nonumber 
   \\& - \bar{z}
   \left(0.0241557 L_{\bar{z}}^2-0.278902
   L_{\bar{z}}-0.582336\right)- z \left(-1.64493
   L_z^3+3.75242 L_z^2+0.397826
   L_z-774.514\right) \nonumber 
   \\& -0.0333333 L_z^5+0.0833333
   L_z^4-0.132844 L_z^3-0.339443 L_z^2-12.6418
   L_z+8847.28 z^5-729429. z^4-13.0667 \bigg\}  \,,  \\ 
 I^*_{qq'}(z)&\, = N_f \bigg\{-50.3634\, +z^3 \left(0.117123 L_z^4-0.314883 L_z^3+3.53761
   L_z^2-10.541 L_z+18.5347\right)+z^2 \big(-0.0026012 L_z^4\nonumber \\
   &+0.703183
   L_z^3+0.199736 L_z^2+5.14781 L_z-24.9854\big)+\left(\bar{z}\right)^3
   \left(-0.227266 L_{\bar{z}}^3-0.0305807 L_{\bar{z}}^2-3.18523
   L_{\bar{z}}\right)\nonumber \\
   &+\left(\bar{z}\right)^2 \left(-0.000912305 L_{\bar{z}}^3+0.865032
   L_{\bar{z}}^2+1.85446 L_{\bar{z}}\right)+\bar{z} \big(-0.0987654 L_{\bar{z}}^3-0.493827
   L_{\bar{z}}^2-4.21399 L_{\bar{z}}\nonumber \\
   &-8.73525\big)+z \big(-0.246914
   L_z^4-1.61317 L_z^3-8.77153 L_z^2-12.0035
   L_z+51.7382\big)-0.246914 L_z^4-1.61317 L_z^3\nonumber \\
   &-10.5493
   L_z^2-30.9665 L_z+0.102137 z^5-0.909082
   z^4+\frac{5.88282}{z}\bigg\}+307.912\, +z^3 \big(1.80474
   L_z^5+1.58169 L_z^4 \nonumber \\
   &+92.844 L_z^3+24.4101 L_z^2+724.086
   L_z+168.701\big)+z^2 \big(-0.00854607 L_z^5-4.88703 L_z^4-33.0209
   L_z^3\nonumber \\
   &+188.744 L_z^2-49.3191 L_z+147.078\big)+\left(\bar{z}\right)^3
   \left(-2.52551 L_{\bar{z}}^4+13.1597 L_{\bar{z}}^3-119.84 L_{\bar{z}}^2+333.889
   L_{\bar{z}}\right)\nonumber \\
   &+\left(\bar{z}\right)^2 \left(0.0720812 L_{\bar{z}}^4+1.43763
   L_{\bar{z}}^3+24.7149 L_{\bar{z}}^2+192.782 L_{\bar{z}}\right)+\bar{z} \big(0.246914
   L_{\bar{z}}^4+0.54321 L_{\bar{z}}^3+6.03113 L_{\bar{z}}^2\nonumber \\
   &+38.031
   L_{\bar{z}}+123.709\big)+z \left(1.27407 L_z^5+2.34568 L_z^4+4.50092
   L_z^3-70.7153 L_z^2-270.973 L_z+167.376\right)\nonumber \\
   &-0.592593
   L_z^5+6.79012 L_z^4-46.4127 L_z^3+86.421 L_z^2-470.887
   L_z\nonumber \\
   &+\frac{-78.9847 L_z-466.384}{z}+10.7916 z^5-335.475 z^4  \,, \\ 
I^*_{qq}(z) &\,= \left(-9.09324 N_f^2+154.257 N_f+140.136\right)
  \frac{1}{(\bar{z})_+}+N_f^2 \bigg\{z^3 \left(0.666009
   L_z^3-4.08064 L_z^2+13.5087 L_z-13.4503\right)\nonumber \\
   &+z^2 \left(-0.610414
   L_z^3-2.30736 L_z^2+0.622374 L_z+20.7035\right)+z \big(-0.329218
   L_z^3-0.855967 L_z^2-1.18519 L_z \nonumber \\
   &-6.71567\big)-0.329218
   L_z^3-2.43621 L_z^2-5.66255 L_z-0.000436414 z^5-0.405417 z^4-14.1598
   \bar{z}+15.8093\bigg\}\nonumber \\
   &+N_f \bigg\{z^3 \left(-11.0536 L_z^4+57.1168
   L_z^3-406.207 L_z^2+1183.43 L_z-2303.63\right)+z^2 \big(3.38122
   L_z^4+37.927 L_z^3\nonumber \\
   & +212.612 L_z^2+982.488 L_z+2012.66\big)+2.107
   L_{\bar{z}}^3+\left(\bar{z}\right)^3 \left(0.189723 L_{\bar{z}}^3+1.80938
   L_{\bar{z}}^2-2.62339 L_{\bar{z}}\right)\nonumber \\
   &+10.3704 L_{\bar{z}}^2+\left(\bar{z}\right)^2
   \left(0.69507 L_{\bar{z}}^3+5.12831 L_{\bar{z}}^2-15.0284
   L_{\bar{z}}\right)-10.4458 L_{\bar{z}}+\bar{z} \big(-1.0535 L_{\bar{z}}^3-7.60494
   L_{\bar{z}}^2\nonumber \\
   &+48.0993 L_{\bar{z}}+439.611\big)+z \left(0.855967
   L_z^4+12.6639 L_z^3+16.4326 L_z^2+43.9956
   L_z+164.293\right)+0.855967 L_z^4\nonumber \\
   &+11.9396 L_z^3+59.1809
   L_z^2+187.752 L_z-1.34066 z^5+82.5779 z^4-312.745\bigg\}+z^3
   \big(15.3681 L_z^5-12.7492 L_z^4\nonumber \\
   &+195.91 L_z^3+1220.54 L_z^2-4589.86
   L_z+14892.1\big)+z^2 \big(-2.96846 L_z^5-62.4209 L_z^4-385.992
   L_z^3-2295.84 L_z^2\nonumber \\
   &-8061.05 L_z-14175.7\big)-34.7654
   L_{\bar{z}}^3+\left(\bar{z}\right)^3 \left(-9.48494 L_{\bar{z}}^3-52.1748
   L_{\bar{z}}^2+111.502 L_{\bar{z}}\right)-5.09037 L_{\bar{z}}^2\nonumber \\
   &+\left(\bar{z}\right)^2
   \left(-11.021 L_{\bar{z}}^3-85.3807 L_{\bar{z}}^2+413.664
   L_{\bar{z}}\right)+637.843 L_{\bar{z}}+\bar{z} \big(-7.90123 L_{\bar{z}}^3+9.5085
   L_{\bar{z}}^2-487.943 L_{\bar{z}}\nonumber \\
   &-2438.69\big)+z \left(-0.301235
   L_z^5-13.037 L_z^4-60.4295 L_z^3+90.2398 L_z^2+400.357
   L_z+265.017\right)-0.301235 L_z^5\nonumber \\
   &-8.69136 L_z^4-69.1867
   L_z^3-286.991 L_z^2-913.865 L_z+9.73661 z^5-868.539 z^4+1033.69 \,, \\ 
 I^*_{q\bar{q}}(z)& \, = \frac{9 N_f}{4} \bigg\{z^3 \left(-37.3313 L_z^4+69.5328 L_z^3-1189.55 L_z^2+2193.6
   L_z-4962.46\right)+z^2 \big(-0.0538545 L_z^4+8.68065 L_z^3\nonumber \\
   &+179.886
   L_z^2+1428.28 L_z+4299.55\big)+0.101532 \left(\bar{z}\right)^3
   L_{\bar{z}}+0.0172219 \left(\bar{z}\right)^2 L_{\bar{z}}+\bar{z} \left(0.395062
   L_{\bar{z}}+0.460905\right)\nonumber \\
   &+z \left(0.131687 L_z^4+0.877915 L_z^3+4.86623
   L_z^2+11.488 L_z-9.51143\right)-0.131687 L_z^4-0.943759
   L_z^3-3.41767 L_z^2\nonumber \\
   &-4.49137 L_z-16.6625 z^5+688.059
   z^4+1.02498\bigg\}+ \frac{9}{4} \bigg\{ z^3 \big(-1385.38 L_z^5+15741.4 L_z^4-165914.
   L_z^3+944859. L_z^2\nonumber \\
   &-3.53957\times 10^6 L_z+5.88567\times
   10^6\big)+z^2 \big(-42.7578 L_z^5-1977.89 L_z^4-39497.9
   L_z^3-422824. L_z^2\nonumber \\
   &-2.42091\times 10^6 L_z-5.92341\times
   10^6\big)+3.3635 \left(\bar{z}\right)^3 L_{\bar{z}}-1.78893
   \left(\bar{z}\right)^2 L_{\bar{z}}+\bar{z} \left(-4.98979 L_{\bar{z}}-13.046\right)\nonumber \\
   &+z
   \left(0.0148148 L_z^5-1.48148 L_z^4-13.6105 L_z^3-83.3483
   L_z^2-64.0087 L_z+105.151\right)-0.0148148 L_z^5\nonumber \\
   &+1.08642
   L_z^4+10.911 L_z^3+17.5783 L_z^2+6.83186 L_z-246.269 z^5+37870.3
   z^4+15.737 \bigg\} \,,   \\
 I^{(3)}_{qg}(z)&\, = N_f \bigg\{532.389\, +z^3 \left(3.23607 L_z^5-4.68217 L_z^4+120.483
   L_z^3-119.874 L_z^2+453.908 L_z+442.25\right)\nonumber \\
   & +z^2 \left(-0.0565211
   L_z^5-3.10462 L_z^4-20.1426 L_z^3-202.066 L_z^2-561.287
   L_z+424.055\right)-0.154321 L_{\bar{z}}^4\nonumber \\
   &-0.823045
   L_{\bar{z}}^3+\left(\bar{z}\right)^3 \left(2.61221 L_{\bar{z}}^4-9.9784
   L_{\bar{z}}^3+116.968 L_{\bar{z}}^2-267.94 L_{\bar{z}}\right)+1.48131
   L_{\bar{z}}^2\nonumber \\
   &+\left(\bar{z}\right)^2 \left(-0.378008 L_{\bar{z}}^4-4.9419
   L_{\bar{z}}^3-38.9532 L_{\bar{z}}^2-203.044 L_{\bar{z}}\right)+22.2518 L_{\bar{z}}+\bar{z}
   \big(0.308642 L_{\bar{z}}^4+2.38683 L_{\bar{z}}^3\nonumber \\
   &+3.85219 L_{\bar{z}}^2-18.9149
   L_{\bar{z}}-275.431\big)+z \big(-0.355556 L_z^5-3.65432 L_z^4-21.6379
   L_z^3-75.7833 L_z^2-71.9808 L_z \nonumber \\
   &-1040.7\big)+0.177778 L_z^5+1.2716
   L_z^4+14.2634 L_z^3+67.2339 L_z^2+216.898 L_z-2.00607 z^5-287.428
   z^4+\frac{11.9546}{z}\bigg\}\nonumber \\
   &-3636.55\, +z^3 \left(-259.099
   L_z^5+3952.96 L_z^4-34085. L_z^3+221224. L_z^2-768027.
   L_z+594756.\right)\nonumber \\
   &+z^2 \left(-10.1315 L_z^5-463.996 L_z^4-8964.71
   L_z^3-91774.6 L_z^2-536653. L_z+96171.9\right)-0.925926
   L_{\bar{z}}^5\nonumber \\
   &+2.36111 L_{\bar{z}}^4+14.151 L_{\bar{z}}^3+\left(\bar{z}\right)^3
   \left(-264.815 L_{\bar{z}}^5+1791.21 L_{\bar{z}}^4-23536.1 L_{\bar{z}}^3+107809.
   L_{\bar{z}}^2-431667. L_{\bar{z}}\right)\nonumber \\
   &-34.2113 L_{\bar{z}}^2+\left(\bar{z}\right)^2
   \left(-5.38358 L_{\bar{z}}^5-180.755 L_{\bar{z}}^4-3771.6 L_{\bar{z}}^3-43136.2
   L_{\bar{z}}^2-262007. L_{\bar{z}}\right)-103.526 L_{\bar{z}}\nonumber \\
   &+\bar{z} \left(1.85185
   L_{\bar{z}}^5+1.08025 L_{\bar{z}}^4-37.8328 L_{\bar{z}}^3-107.055 L_{\bar{z}}^2+85.0903
   L_{\bar{z}}+2491.42\right)+z \big(10.2642 L_z^5\nonumber \\
   &+65.8827 L_z^4+175.497
   L_z^3-49.2838 L_z^2-1754.58 L_z-692466.\big)-1.98519 L_z^5+5.08025
   L_z^4-171.088 L_z^3\nonumber \\
   &-103.238 L_z^2-2331.02 L_z+\frac{-177.716
   L_z-1109.28}{z}+1274.6 z^5+4328.89 z^4 \,.    
\end{align}

\end{document}